\documentclass[letterpaper,english,notitlepage,superscriptaddress,nofootinbib]{revtex4-2}
\usepackage[T1]{fontenc}
\usepackage{textcomp}
\usepackage[latin9]{inputenc}
\usepackage{xcolor}
\usepackage{babel}
\usepackage{amsmath}
\usepackage{amssymb}
\usepackage{graphicx}
\usepackage[pdfusetitle,
 bookmarks=true,bookmarksnumbered=false,bookmarksopen=false,
 breaklinks=false,pdfborder={0 0 1},backref=false,colorlinks=false]
 {hyperref}



\begin{document}
\title{Photoproduction of $\eta_{c}\gamma$ pairs in CGC framework}
\author{M. Siddikov}
\affiliation{Departamento de Física, Universidad Técnica Federico Santa María,~~~~~~~\\
 y Centro Científico - Tecnológico de Valparaíso, Casilla 110-V, Valparaíso,
Chile}
\author{I. Zemlyakov}
\affiliation{Departamento de Física, Universidad Técnica Federico Santa María,~~~~~~~\\
 y Centro Científico - Tecnológico de Valparaíso, Casilla 110-V, Valparaíso,
Chile}
\author{M. Roa}
\affiliation{Facultad de Ingeniería, Laboratorio DataScience, Universidad de Playa
Ancha, ~\\
Leopoldo Carvallo 270, Valparaíso, Chile,~~~~~~ ~\\
}
\affiliation{Centro de Estudios Avanzados, Universidad de Playa Ancha, Traslaviña
450, Viña del Mar, Chile~~\\
 }
\author{S. Valdebenito}
\affiliation{Departamento de Física, Universidad Técnica Federico Santa María,~~~~~~~\\
 y Centro Científico - Tecnológico de Valparaíso, Casilla 110-V, Valparaíso,
Chile}
\begin{abstract}
In this paper we analyzed the exclusive photoproduction of $\eta_{c}\gamma$
pairs in the Color Glass Condensate framework. We found that the cross-section
of this process is sensitive only to the forward dipole scattering
amplitude, and thus could be used as a new tool for analysis of this
fundamental nonperturbative object. Using the phenomenological parametrizations
of this object, we estimated numerically the production cross-section
and counting rates in the kinematics of the ongoing and forthcoming
experiments at LHC and future Electron Ion Collider. We found that
the cross-section is sufficiently large for dedicated experimental
study. We also estimated the role of this process as a potential background
to $\eta_{c}$ photoproduction, which is conventionally considered
as a gateway for studies of odderons. We found that the contribution
of $\eta_{c}\gamma$ (with undetected photon) is on par with expected
contributions of odderons in the kinematics of small momentum transfer
$|t|\lesssim1$ GeV$^{2}$, though decreases rapidly at larger $|t|$.
Finally, we also calculated the feed-down contribution (from radiative
decays of other charmonia) and found that a sizable correction comes
from $J/\psi\to\eta_{c}\gamma$ decays. This contribution remains
pronounced even at relatively large $|t|$ and potentially can impose
constraint on detectability of odderons via $\eta_{c}$ photoproduction. 
\end{abstract}
\maketitle

\section{Introduction}

In the high-energy collisions, due to onset of saturation effects,
the dynamics of all the hadronic processes is conventionally described
in the Color Glass Condensate (CGC) framework~\cite{GLR,McLerran:1993ni,McLerran:1993ka,McLerran:1994vd,MUQI,MV,gbw01:1,Kopeliovich:2002yv,Kopeliovich:2001ee,Gelis:2010nm,Iancu:2003uh},
which naturally incorporates the saturation effects and provides a
phenomenologically successful description of both hadron-hadron and
lepton-hadron collisions~\cite{Kovchegov:1999yj,Kovchegov:2006vj,Balitsky:2008zza,Kovchegov:2012mbw,Balitsky:2001re,Cougoulic:2019aja,Aidala:2020mzt,Ma:2014mri}.
The cross-sections of physical processes in this framework are expressed
in terms of the forward multipole scattering amplitudes ($n$-point
correlators of Wilson lines), which have a probabilistic interpretation
and present important physical characteristics of the target. Technically,
these correlators in the dilute limit may be related to the multigluon
distribution functions used in collinear and $k_{T}$ factorization
picture. The behavior of these multipole correlators is theoretically
constrained by the evolution equations~\cite{Balitsky:2008zza} and
the asymptotic behavior in different asymptotic regimes. At present
it is not possible to extract these objects from the first principles,
and for this reason the available parametrizations inevitably rely
on phenomenological analysis of the experimental data. However, it
has been realized that different models can provide a reasonable description
of the same data, thus opening a quest for new channels which could
be used in the future precision studies of the forward dipole amplitudes.

The electro- and photoproduction of heavy mesons for a long time has
played an important role for studies of the gluonic field of the target
almost since their discovery~\cite{Korner:1991kf,Neubert:1993mb}.
The modern NRQCD framework allows to describe systematically the hadronization
of the produced $\bar{Q}Q$ pairs into final state quarkonia~\cite{Bodwin:1994jh,Maltoni:1997pt,Cho:1995ce,Cho:1995vh,Baranov:2002cf,Baranov:2007dw,Baranov:2016clx,Brambilla:2008zg,Baranov:2015laa,Baranov:2011ib,Feng:2015cba,Brambilla:2010cs}.
While conventionally the exclusive photoproduction of $J/\psi$ has
been widely used for studies of the hadronic structure, it provides
only limited information about the dipole cross-section. Due to experimental
advances and exceptionally high luminosity of the ongoing photoproduction
experiments at LHC in ultraperipheral collisions~\cite{Mangano:2017tke,Agostini:2020fmq,Abada:2019lih}
and future experiments at the Electron Ion Collider~\cite{Accardi:2012qut,AbdulKhalek:2021gbh,Burkert:2022hjz},
nowadays it is possible to study various $2\to3$ processes~\cite{GPD2x3:9,GPD2x3:8,Siddikov:2024blb,GPD2x3:7,GPD2x3:6,GPD2x3:5,GPD2x3:4,GPD2x3:3,GPD2x3:2,GPD2x3:1,Duplancic:2022wqn,Qiu:2024mny,Qiu:2023mrm,Deja:2023ahc,Siddikov:2022bku,Siddikov:2023qbd,GPD2x3:10,GPD2x3:11}
which could provide a more detailed information about the target.
While most of the previous studies focused on production of light
mesons and in the moderate-energy kinematics, we believe that the
$2\to3$ processes can be also studied in high-energy (small-$x_{B}$)
kinematics using quarkonia-photon pair production. Unfortunately,
the process $\gamma p\to J/\psi\,\gamma p$ has too small cross-section
because it requires $C$-odd exchanges in the $t$-channel. However,
the process $\gamma p\to\eta_{c}\,\gamma p$ can present an interesting
tool for analysis of the forward dipole amplitudes. Previously this
channel has been studied in~\cite{HarlandLang:2018ytk}, assuming
that it proceeds via a radiative decay of higher excited states (mainly
$\gamma p\to J/\psi p\to\eta_{c}\gamma\,p$). Though this mechanism
has a relatively large cross-section, it is controlled by the $J/\psi$
photoproduction cross-section and thus does not bring any new constraints
for dipole scattering amplitudes. In this paper we will focus on the
associated production of $\eta_{c}\gamma$ pairs with large invariant
mass, $M_{\gamma\eta_{c}}\ge3.5-4\,{\rm GeV}^{2}$, where the feed-down
contributions are negligible, and at high photon-proton energies,
where the dipole approach is applicable. This kinematic regime can
be studied in high-energy $pA$ and $AA$ collisions in ultraperipheral
kinematics at LHC, as well as electron-proton collisions at the future
Electron Ion Collider (EIC)~\cite{Accardi:2012qut,AbdulKhalek:2021gbh,Burkert:2022hjz},
the Large Hadron electron Collider (LHeC)~\cite{AbelleiraFernandez:2012cc},
and the Future Circular Collider (FCC-he)~\cite{Mangano:2017tke,Agostini:2020fmq,Abada:2019lih}.

We also need to mention that the $\eta_{c}\gamma$ pair photoproduction
(with undetected photon) presents interest as a potential background
to the $\eta_{c}$ photoproduction $\gamma p\to\eta_{c}p$, which
has been considered as one of the most promising channels for studies
of odderons~\cite{Odd3,Odd4,Odd1,Bartels:2001hw,Odd4,Odd5,Odd6,Odd7}.
Since the $\eta_{c}\gamma$ pair photoproduction does not require
$C$-odd exchanges in $t$-channel, it can constitute a sizable background
which sets the limits on detectability of odderons via $\eta_{c}$
photoproduction at future experiments.

The paper is structured as follows. Below in Section~\ref{sec:Formalism}
we briefly describe the kinematics of the process, the main components
of the CGC framework and then present the theoretical results for
the photoproduction of heavy quarkonia pairs in the CGC approach.
In Section~\ref{sec:Numer} we provide numerical estimates using
the phenomenological parametrizations of dipole amplitudes. In that
section we also make comparison with odderonic contribution found
in~\cite{Benic:2023} and the $\eta_{c}\gamma$ production from $J/\psi\to\eta_{c}\gamma$
radiative decays. Finally, in Section~\ref{sec:Conclusions} we draw
conclusions.

\section{Theoretical framework}

\label{sec:Formalism} As we mentioned in the introduction, the previous
studies of the meson-photon photoproduction were based on partonic
picture, which is valid at moderate energies~\cite{GPD2x3:8,GPD2x3:9,Siddikov:2024blb}
and can obtain sizable corrections from higher twists and diagrams
of higher order in strong coupling $\alpha_{s}$. At high energies
it is more appropriate to describe the interaction in the Color Glass
Condensate (CGC) framework~\cite{GLR,McLerran:1993ni,McLerran:1993ka,McLerran:1994vd,MUQI,MV,gbw01:1,Kopeliovich:2002yv,Kopeliovich:2001ee,Gelis:2010nm}.
Furthermore, we have to pay attention to the hierarchy of scales.
In what follows we will consider the mass of the meson $M_{\eta_{c}}$
and the invariant mass $M_{\gamma\eta_{c}}$ as hard scales of the
same order. Since in electroproduction and ultraperipheral hadroproduction
experiments the spectrum of equivalent photons is dominated by quasi-real
photons, we'll focus on the photoproduction by transversely polarized
photons with zero virtuality $Q=0$. In the following subsections~\ref{subsec:Kinematics},~\ref{subsec:Derivation},~\ref{subsec:Formalism}
we briefly introduce the main kinematic variables used for description
of the process and present the amplitude of the process in the Color
Glass Condensate framework.

\subsection{Kinematic of the process}

\label{subsec:Kinematics} We will perform our evaluations in the
photon-proton collision frame, where the incoming photon and proton
move in the direction of axis $z$, though our final result (the invariant
cross-sections) do not depend on the choice of the frame. In contrast
to earlier studies of $\gamma\rho,\,\gamma\pi$ production in~\cite{GPD2x3:8,GPD2x3:9},
we no longer disregard the mass of the meson $M_{\eta_{c}}$ but rather
consider it as a hard scale on par with the invariant mass $M_{\gamma\eta_{c}}$,
as was suggested in~\cite{Siddikov:2024blb}. Both in electroproduction
and ultraperipheral hadroproduction the spectrum of equivalent photons
is dominated by quasi-real photons, so we'll focus on the photoproduction
by transversely polarized photons with zero virtuality $Q=0$. We
will use a notation $q$ for the momentum of the incoming photon,
$P_{{\rm in}},P_{{\rm out}}$ for the momenta of the proton before
and after interaction, $k$ for the momentum of the emitted (outgoing)
photon, and $p_{\eta_{c}}$ for the momentum of produced $\eta_{c}$
meson. In this frame the light-cone decomposition of the particles'
momenta may be written as~\cite{GPD2x3:8,GPD2x3:9} 
\begin{align}
q^{\mu} & =p^{\mu}\label{eq:q}\\
P_{{\rm in}}^{\mu} & =\left(1+\xi\right)n^{\mu}+\frac{m_{N}^{2}}{s(1+\xi)}p^{\mu},\qquad P_{{\rm out}}^{\mu}=\left(1-\xi\right)n^{\mu}+\frac{m_{N}^{2}+\Delta_{\perp}^{2}}{s(1-\xi)}p^{\mu}+\Delta_{\perp}^{\mu},\\
p_{\eta_{c}}^{\mu} & =\alpha_{\eta_{c}}p^{\mu}+\frac{\left(\boldsymbol{p}_{\perp}+\boldsymbol{\Delta}_{\perp}/2\right)^{2}+M_{\eta_{c}}^{2}}{\alpha_{\eta_{c}}s}n^{\mu}\underbrace{-\boldsymbol{p}_{\perp}-\frac{\boldsymbol{\Delta}_{\perp}}{2}}_{\boldsymbol{p}_{\perp}^{\eta_{c}}},\\
k^{\mu} & =\left(1-\alpha_{\eta_{c}}\right)p^{\mu}+\frac{\left(\boldsymbol{p}_{\perp}-\boldsymbol{\Delta}_{\perp}/2\right)^{2}}{\left(1-\alpha_{\eta_{c}}\right)s}n^{\mu}+\underbrace{\boldsymbol{p}_{\perp}-\frac{\boldsymbol{\Delta}_{\perp}}{2}}_{\boldsymbol{k}_{\gamma}^{\perp}},\label{eq:k}
\end{align}
where the basis light-cone vectors $p^{\mu},\,n^{\mu}$ are defined
as 
\begin{equation}
p^{\mu}=\frac{\sqrt{s}}{2}\left(1,0,0,1\right),\qquad n^{\mu}=\frac{\sqrt{s}}{2}\left(1,0,0,-1\right),\qquad p\cdot n=\frac{s}{2}.
\end{equation}
The parameter $\xi$, which characterizes a longitudinal momentum
transfer to the proton is very small in the high-energy kinematics,
$\xi\sim M_{\gamma\eta_{c}}^{2}/s\ll1$, and for this reason in many
expressions may be disregarded. In what follows we will use the invariant
Mandelstam variables 
\begin{align}
S_{\gamma N} & \equiv W^{2}=\left(q+P_{{\rm in}}\right)^{2}=s\left(1+\xi\right)+m_{N}^{2},\\
t & =\left(P_{{\rm out}}-P_{{\rm in}}\right)^{2}=-\frac{1+\xi}{1-\xi}\Delta_{\perp}^{2}-\frac{4\xi^{2}m_{N}^{2}}{1-\xi^{2}}.\label{eq:tDep}
\end{align}
From Eq.~(\ref{eq:tDep}) we can see that at given $\xi$, the invariant
momentum transfer $t$ is bound by 
\[
t\le t_{{\rm min}}=-\frac{4\xi^{2}m_{N}^{2}}{1-\xi^{2}}.
\]
We will also use the variables 
\begin{align}
 & u'=\left(p_{\eta_{c}}-q\right)^{2},\qquad t'=\left(k-q\right)^{2},\qquad M_{\gamma\eta_{c}}^{2}=\left(k+p_{\eta_{c}}\right)^{2}\label{eq:uPrimetPrime}
\end{align}
which are related as 
\begin{align}
 & -u'-t'=M_{\gamma\eta_{c}}^{2}-M_{\eta_{c}}^{2}-t.\label{eq:Constr}
\end{align}
Using definitions~(\ref{eq:uPrimetPrime}) and light-cone decomposition~(\ref{eq:q}-\ref{eq:k}),
it is possible to show that 
\begin{equation}
u'=M_{\eta_{c}}^{2}-2q\cdot p_{\eta_{c}}=M_{\eta_{c}}^{2}-\frac{\left(\boldsymbol{p}_{\perp}+\boldsymbol{\Delta}_{\perp}/2\right)^{2}+M_{\eta_{c}}^{2}}{\alpha_{\eta_{c}}}
\end{equation}
\begin{equation}
t'=-2q\cdot k=-\frac{\left(\boldsymbol{p}_{\perp}-\boldsymbol{\Delta}_{\perp}/2\right)^{2}}{\left(1-\alpha_{\eta_{c}}\right)}
\end{equation}
so 
\begin{equation}
\boldsymbol{p}_{\perp}\cdot\boldsymbol{\Delta}_{\perp}=\frac{\alpha_{\eta_{c}}\left(M_{\gamma\eta_{c}}^{2}-t\right)+t'-M_{\eta_{c}}^{2}}{2}=\frac{\bar{\alpha}_{\eta_{c}}\left(t-M_{\gamma\eta_{c}}^{2}\right)-u'}{2}\label{eq:pDelta}
\end{equation}
Using the exact identities~~\cite{GPD2x3:8,GPD2x3:9} 
\begin{align}
\bar{\alpha}_{\eta_{c}} & =\frac{1}{2\xi s}\left(-u'-\frac{2\xi m_{N}^{2}}{s\left(1-\xi^{2}\right)}\left(-u'+M_{\eta_{c}}^{2}\right)\right)+\frac{2\xi m_{N}^{2}}{s\left(1-\xi^{2}\right)},\qquad2\xi s=M_{\gamma\eta_{c}}^{2}-t
\end{align}
we may reduce~(\ref{eq:pDelta}) to the form 
\begin{equation}
\boldsymbol{p}_{\perp}\cdot\boldsymbol{\Delta}_{\perp}=\frac{2\xi m_{N}^{2}}{\left(1-\xi^{2}\right)}\,\frac{t'}{s}\label{eq:pDelta-1}
\end{equation}
As we will show below, the absolute values of the transverse vectors
$\boldsymbol{p}_{\perp}$ and $\boldsymbol{\Delta}_{\perp}$ are largely
fixed by the invariant mass $M_{\gamma\eta_{c}}$ and the momentum
transfer $t$, for this reason the variables $t,t',M_{\gamma\eta_{c}}$
allow to fix completely the kinematics of the process and even the
relative orientation of these two vectors (up to global rotation in
the transverse plane). In the high-energy kinematics $W\gg M_{\eta_{c}},M_{\gamma\eta_{c}}\gtrsim p_{\perp},\Delta_{\perp}$,
so in view of the smallness of the variable $\xi\ll1$, the ratio
$t'/s\sim\mathcal{O}\left(\xi\right)$ and thus the scalar product~(\ref{eq:pDelta})
is very small, $\boldsymbol{p}_{\perp}\cdot\boldsymbol{\Delta}_{\perp}\sim\mathcal{O}\left(\xi^{2}\right)$,
which implies that the angle between the vectors $\boldsymbol{p}_{\perp},\boldsymbol{\Delta}_{\perp}$
is close to $\pi/2$.

The polarization vectors of the incoming and outgoing real photons
with momenta $\boldsymbol{k}$ are chosen in the light-cone gauge
as 
\begin{equation}
\varepsilon_{T}^{(\lambda=\pm1)}(\boldsymbol{k})=\left(0,\,\frac{\boldsymbol{\varepsilon}_{\lambda}\cdot\boldsymbol{k}_{\perp}}{k^{+}},\boldsymbol{\varepsilon}_{\lambda}\right),\qquad\boldsymbol{\varepsilon}_{\lambda}=\frac{1}{\sqrt{2}}\left(\begin{array}{c}
1\\
i\lambda
\end{array}\right).\label{eq:PolVector}
\end{equation}

The parametrization~~(\ref{eq:q}-\ref{eq:k}) implicitly implements
various kinematic constraints on momenta of the produced particles
which follow from onshellness of final state particles and energy-momentum
conservation. In the generalized Bjorken kinematics, the variables
$m_{N},\,\left|\Delta_{\perp}\right|,t$ are negligibly small, whereas
all the other variables are parametrically large, $\sim M_{\eta_{c}}$.
In this kinematics it is possible to simplify the light-cone decomposition~(\ref{eq:q}-\ref{eq:k})
and obtain approximate relations of the Mandelstam variables with
variables $\alpha_{\eta_{c}},\boldsymbol{p}_{\perp}$ as 
\begin{align}
-t' & \approx\alpha_{\eta_{c}}M_{\gamma\eta_{c}}^{2}-M_{\eta_{c}}^{2},\qquad-u'\approx\left(1-\alpha_{\eta_{c}}\right)M_{\gamma\eta_{c}}^{2},\label{eq:KinApprox}\\
\boldsymbol{p}_{\perp}^{2} & =\bar{\alpha}_{\eta_{c}}\left[\alpha_{\eta_{c}}M_{\gamma\eta_{c}}^{2}-M_{\eta_{c}}^{2}\right]\approx-\bar{\alpha}_{\eta_{c}}t',\qquad M_{\gamma\eta_{c}}^{2}\approx2s\xi
\end{align}

The physical constraint for real photon momenta 
\begin{equation}
t'=\left(q-k\right)^{2}=-2q\cdot k=-2\,\left|\boldsymbol{q}\right|\left|\boldsymbol{k}\right|\left(1-\cos\theta_{\boldsymbol{q},\,\boldsymbol{k}}\right)\le0,
\end{equation}
implies that the variable $\alpha_{\eta_{c}}$ is bound by $\alpha_{\eta_{c}}\ge M_{\eta_{c}}^{2}/M_{\gamma\eta_{c}}^{2}$,
so for the variable $u'$ we get 
\begin{equation}
-u'\le\left(-u'\right)_{{\rm max}}=M_{\gamma\eta_{c}}^{2}-M_{\eta_{c}}^{2}-t.
\end{equation}
We also may check that the pairwise invariant masses 
\begin{align*}
\left(P_{{\rm out}}+p_{\eta_{c}}\right)^{2} & \approx M_{\eta_{c}}^{2}+s\alpha_{\eta_{c}}\left(1-\xi\right)=t'+s\alpha_{\eta_{c}},\qquad\left(P_{{\rm out}}+k\right)^{2}\approx s\left(1-\alpha_{\eta_{c}}\right)\left(1-\xi\right),
\end{align*}
remain large, which shows that the produced $\eta_{c}$ and $\gamma$
are well-separated kinematically from the recoil proton.

In what follows we prefer to use as independent variables $t,t',M_{\gamma\eta_{c}}$
since, as we will see below, the cross-section decreases homogeneously
as a function of these variables. The photoproduction cross-section
in terms of these variables may be represented as 
\begin{equation}
\frac{d\sigma_{\gamma p\to\eta_{c}\gamma p}}{dt\,dt'\,dM_{\gamma\eta_{c}}}\approx\frac{\left|\mathcal{A}_{\gamma p\to\eta_{c}\gamma p}^{(\lambda,\sigma)}\right|^{2}}{128\pi^{3}M_{\gamma\eta_{c}}}\label{eq:Photo}
\end{equation}
where $\mathcal{A}_{\gamma p\to\eta_{c}\gamma p}^{(\lambda,\sigma)}$
is the amplitude of the process, which in general depends on helicities
$\lambda$ and $\sigma$ of the incoming and outgoing photons. For
the unpolarized cross-section, summation over spins of photons in
the final state and averaging over helicities in the initial state
yields explicitly 
\begin{equation}
\left|\mathcal{A}_{\gamma p\to\eta_{c}\gamma p}^{({\rm unpolarized)}}\right|^{2}\equiv\frac{1}{2}\sum_{\lambda=\pm1}\sum_{\sigma=\pm1}\left|\mathcal{A}_{\gamma p\to\eta_{c}\gamma p}^{(\sigma,\,\lambda)}\right|^{2}=\left|\mathcal{A}_{\gamma p\to\eta_{c}\gamma p}^{(+,+)}\right|^{2}+\left|\mathcal{A}_{\gamma p\to\eta_{c}\gamma p}^{(+,-)}\right|^{2}.\label{eq:PolAmps}
\end{equation}
The electroproduction cross-section in the small-$Q$ kinematics gets
the dominant contribution from events with single-photon exchange
between leptonic and hadronic parts, and may be represented as~\cite{Weizsacker:1934,Williams:1935,Budnev:1975poe}
\begin{equation}
\frac{d\sigma_{ep\to eM_{1}M_{2}p}}{d\ln W^{2}dQ^{2}\,dt\,dt'\,dM_{\gamma\eta_{c}}}\approx\frac{\alpha_{{\rm em}}}{\pi\,Q^{2}}\,\left(1-y+\frac{y^{2}}{2}-(1-y)\frac{Q_{{\rm min}}^{2}}{Q^{2}}\right)\frac{d\sigma_{\gamma p\to M_{1}M_{2}p}}{dt\,dt'\,dM_{\gamma\eta_{c}}},\label{eq:LTSep}
\end{equation}
where $Q_{{\rm min}}^{2}=m_{e}^{2}y^{2}/\left(1-y\right)$, $m_{e}$
is the mass of the electron and $y$ is the fraction of the electron
energy which passes to the virtual photon (the so-called inelasticity);
it may be related to the invariant energy $\sqrt{s_{ep}}$ of the
electron-proton collision as 
\begin{equation}
y=\frac{W^{2}+Q^{2}-m_{N}^{2}}{s_{ep}-m_{N}^{2}}.
\end{equation}

\subsection{High energy scattering in CGC picture}

\label{subsec:Derivation} In heavy charmonia production processes,
the charm mass $m_{c}$ serves as a characteristic hard scale which
controls the strength of interaction of heavy quarks with the gluonic
field. For infinitely heavy quarks formally we can develop a systematic
expansion in the strong coupling $\alpha_{s}\left(m_{c}\right)\ll1$.
However, in the small-$x$ (high energy) kinematics, when the gluon
fields are enhanced and reach values $A_{\mu}^{a}\sim1/\alpha_{s}$,
such perturbative approach becomes unjustified for description of
the interaction with the target. In CGC picture this interaction is
analyzed in the eikonal approximation and is described by a Wilson
line $U(\boldsymbol{x}_{\perp})$~\cite{GLR,McLerran:1993ni,McLerran:1993ka,McLerran:1994vd,MUQI,MV,gbw01:1,Kopeliovich:2002yv,Kopeliovich:2001ee,Gelis:2010nm,Iancu:2003uh}
\begin{equation}
U\left(\boldsymbol{x}_{\perp}\right)=P\exp\left(ig\int dx^{-}A_{a}^{+}\left(x^{-},\,\boldsymbol{x}_{\perp}\right)t^{a}\right),\label{eq:Wilson}
\end{equation}
where $\boldsymbol{x}_{\perp}$ is the impact parameter (transverse
coordinate) of the parton, $t_{a}$ are the color group generators
in the corresponding representation (fundamental or adjoint), $A_{\mu}^{a}(x)=-\frac{1}{\nabla_{\perp}^{2}}\rho_{a}(x^{-},\,\boldsymbol{x}_{\perp})$
is the gluonic field of the target, and $\rho_{a}$ is the density
of the color charges inside the target. According to classical CGC
picture~\cite{McLerran:1993ni,McLerran:1993ka,McLerran:1994vd},
the probability of different color charge configurations $\rho(x^{-},\boldsymbol{x}_{\perp})$
in the target is controlled by the weight functional $W[\rho]$, and
physical observables (amplitudes) require averaging over all possible
configurations $\rho(x^{-},\boldsymbol{x}_{\perp})$, namely 
\begin{equation}
\left\langle \mathcal{O}\right\rangle =\int\mathcal{D}\rho\,W[\rho]\,\mathcal{O}[\rho],\label{eq:Ave}
\end{equation}
where $\mathcal{O[\rho]}$ is the corresponding amplitude found for
a fixed distribution of charges $\rho(x^{-},\boldsymbol{x}_{\perp})$,
and the angular brackets $\langle...\rangle$ imply the above-mentioned
averaging. The explicit evaluation of the integral $\mathcal{D}\rho$
over all configurations of color sources can be realized only for
some simple forms of $W[\rho]$ (e.g. for gaussian). Fortunately,
for some high energy processes it is possible to express the physical
amplitudes in terms of universal (process-independent) correlators
of Wilson-lines, such as dipole or quadrupole scattering amplitudes,
which represent fundamental characteristics of the target and can
be extracted from phenomenological analysis.

\begin{figure}
\includegraphics[width=5cm]{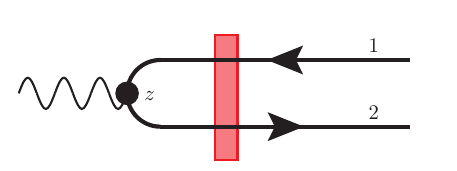}

\caption{The diagram which describes photoproduction of the heavy quark pair
in CGC picture in the leading order over $\alpha_{s}$. The subscript
letter $z$ is the coordinate of the interaction vertex in the configuration
space. The red block represents the interaction with the target (shockwave).}\label{fig:CGCBasic}
\end{figure}

For example, the leading order contribution to inclusive heavy quark
pair photoproduction is described by the diagram shown in the Figure~~\ref{fig:CGCBasic}
(see for details~\cite{Ayala:2017rmh,Caucal:2021ent,Caucal:2022ulg}
), and the amplitude of such process is given by 
\begin{equation}
\mathcal{A}^{a}=-ig\int d^{4}z\,\bar{u}\left(p_{q},\,z\right)\gamma^{\mu}\varepsilon_{\mu}\left(k\right)v\left(p_{\bar{q}},\,z\right),\label{eq:amp_pert}
\end{equation}
where $z$ represents the coordinate of the interaction point in the
configuration space, and the final state quark and antiquark fields
are given by

\begin{align}
\bar{u}\left(p_{q},\,z\right) & =\frac{1}{2}\left(\frac{p_{q}^{+}}{2\pi}\right)\int d^{2}\boldsymbol{x}_{q}e^{ip_{q}^{+}\left(z^{-}-\frac{\left(\boldsymbol{x}_{q}-\boldsymbol{z}\right)^{2}}{2z^{+}}\right)-i\boldsymbol{p}_{q}\cdot\boldsymbol{x}_{q}+\frac{iz^{+}}{2p_{q}^{+}}m^{2}}\times\label{eq:SWubar}\\
 & \times\left(\frac{i}{z^{+}}\right)\bar{u}_{p}\gamma^{+}\left[U\left(\boldsymbol{x}_{q}\right)\theta\left(-z^{+}\right)+\theta\left(z^{+}\right)\right]\left(\gamma^{-}-\frac{\hat{\boldsymbol{x}}_{q}-\hat{\boldsymbol{z}}}{z^{+}}+\frac{m}{p_{q}^{+}}\right),\nonumber 
\end{align}
\begin{align}
v\left(p_{\bar{q}},\,z\right) & =\frac{1}{2}\left(\frac{p_{\bar{q}}^{+}}{2\pi}\right)\int d^{2}\boldsymbol{x}_{\bar{q}}e^{ip_{\bar{q}}^{+}\left(z^{-}-\frac{\left(\boldsymbol{x}_{\bar{q}}-\boldsymbol{z}\right)^{2}}{2z^{+}}\right)-i\boldsymbol{p}_{\bar{q}}\cdot\boldsymbol{x}_{\bar{q}}+\frac{iz^{+}}{2p_{\bar{q}}^{+}}m^{2}}\times\label{eq:SWv}\\
 & \times\left(\frac{i}{z^{+}}\right)\left(\gamma^{-}-\frac{\hat{\boldsymbol{x}}_{\bar{q}}-\hat{\boldsymbol{z}}}{z^{+}}-\frac{m}{p_{\bar{q}}^{+}}\right)\left[U^{\dagger}\left(\boldsymbol{x}_{\bar{q}}\right)\theta\left(-z^{+}\right)+\theta\left(z^{+}\right)\right]\gamma^{+}v_{p_{\bar{q}}}\nonumber 
\end{align}
respectively. The Heaviside functions $\theta\left(\pm z^{+}\right)$
in (\ref{eq:SWubar}-\ref{eq:SWv}) imply that corresponding Wilson
lines $U,\,U^{\dagger}$ should be taken into account only if the
colored quark or antiquark was formed before the interaction with
the target. After averaging over color charges~(\ref{eq:Ave}), the
interaction of the dipole with the target can be described in the
nonperturbative $S$-matrix element~\cite{Gelis:2010nm,Kovchegov:2012mbw,Iancu:2003uh}

\begin{equation}
S_{2}\left(Y,\,\boldsymbol{x}_{q},\,\boldsymbol{x}_{\bar{q}}\right)=\frac{1}{N_{c}}\left\langle {\rm tr}\left(U\left(\boldsymbol{\boldsymbol{x}}_{q}\right)U^{\dagger}\left(\boldsymbol{x}_{\bar{q}}\right)\right)\right\rangle _{Y},\label{eq:S_matrix}
\end{equation}
where $Y$ is the dipole rapidity in the target rest frame. Conventionally,
the phenomenological parametrizations of $S_{2}\left(Y,\,\boldsymbol{x}_{q},\,\boldsymbol{x}_{\bar{q}}\right)$
are given in terms of the dipole scattering amplitude $\mathcal{N}(x,\,\boldsymbol{r},\,\boldsymbol{b})$,
which is related to $S_{2}\left(Y,\,\boldsymbol{x}_{q},\,\boldsymbol{x}_{\bar{q}}\right)$
as 
\begin{equation}
\mathcal{N}\left(x,\,\boldsymbol{r},\,\boldsymbol{b}\right)=1-S_{2}\left(Y=\ln\left(\frac{1}{x}\right),\,\boldsymbol{x}_{q},\,\boldsymbol{x}_{\bar{q}}\right),\label{eq:NS}
\end{equation}
where the variable $\boldsymbol{r}\equiv\boldsymbol{x}_{q}-\boldsymbol{x}_{\bar{q}}$
is the transverse size of the dipole, and $\boldsymbol{b}\equiv\alpha_{q}\,\boldsymbol{x}_{q}+\alpha_{\bar{q}}\boldsymbol{x}_{\bar{q}}$
is the transverse position of the dipole's center of mass. However,
photoproduction alone is not sufficient to fix completely the dipole
amplitude $\mathcal{N}$, and for this reason the phenomenological
parametrizations of $\mathcal{N}(x,\,\boldsymbol{r},\,\boldsymbol{b})$
usually rely on analysis of various experimental channels. The inclusive
production channels in view of optical theorem are sensitive only
to the imaginary part of the amplitude. In the exclusive channels,
the real part of the amplitude may be restored from imaginary part:
as was proven in~\cite{Bronzan:1974jh,Gotsman:1992ui}, if the amplitude
scales with energy as $\sim s^{\alpha}\sim x^{-\alpha}$, then the
ratio of the real and imaginary parts is given by 
\begin{equation}
\beta\equiv\frac{{\rm Re\,\mathcal{A}}}{{\rm Im}\,\mathcal{A}}=\tan\left(\frac{\pi\alpha}{2}\right),
\end{equation}
so the full amplitude $\mathcal{A}$ of any process and its absolute
value in this limit may be rewritten via the imaginary part just adding
additional multiplicative factor, namely 
\begin{equation}
\mathcal{A}=\left(\beta+i\right)\text{{\rm Im}}\,\mathcal{A},\qquad\left|\mathcal{A}\right|^{2}=\left(1+\beta^{2}\right)\left|{\rm Im}\mathcal{A}\right|^{2}.\label{eq:reA}
\end{equation}
As was discussed in~\cite{ske}, in exclusive processes due to unequal
sharing of the $t$-channel momentum between the gluons which form
the shock wave, the amplitude is enhanced by the so-called ``skewedness
factor'' $R_{g}$ given by 
\begin{equation}
R_{g}(\gamma)=\frac{2^{2\gamma+3}}{\sqrt{\pi}}\frac{\Gamma(\gamma+5/2)}{\Gamma(\gamma+4)},\quad\text{where}\quad\gamma\equiv\frac{\partial\ln\left[xg(x,\mu^{2})\right]}{\partial\ln(1/x)}\approx{\rm const}.\label{eq:Rg}
\end{equation}
and $g\left(x,\mu^{2}\right)$ is the gluon PDF~\footnote{In the physically relevant region of small $\gamma\le0.5$, the factor
$R_{g}(\gamma)$ may be approximated by a simpler expression $R_{g}(\gamma)\approx2.4^{\gamma}$.}. The derivation of~(\ref{eq:Rg}) provided in~\cite{ske} crucially
relies on the dominance of the two-gluon exchange in $t$-channel
and validity of the collinear factorization approach, and both assumptions
eventually may become invalid in the deeply saturated regime $x\lll1$.
However, at moderate values of $x\gtrsim10^{-3}$, where we plan to
make predictions, the expression~(\ref{eq:Rg}) provides a reasonable
estimate of the skewedness effect, and together with contributions
of the real part~(\ref{eq:reA}), allows to improve the phenomenological
description of various exclusive processes (see e.g.~\cite{Kowalski:2003hm,watt:bcgc,watt2007,Kowalski:2008sa,RESH}
and references therein).

\subsection{Photoproduction of $\eta_{c}\gamma$ pairs}

\label{subsec:Formalism} 
\begin{figure}
\includegraphics[width=5cm]{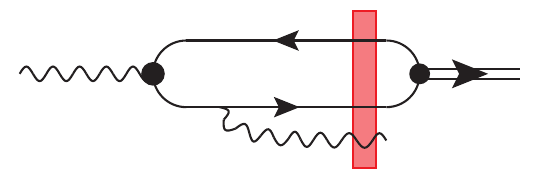}\includegraphics[width=5cm]{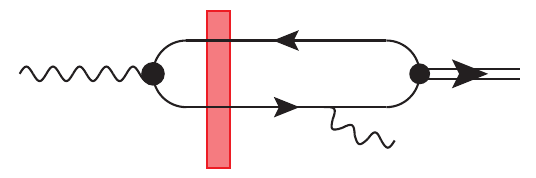}

\includegraphics[width=5cm]{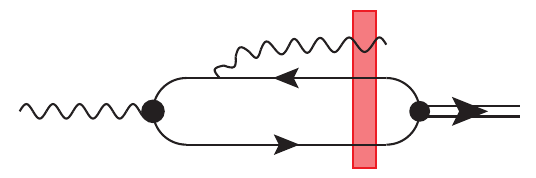}\includegraphics[width=5cm]{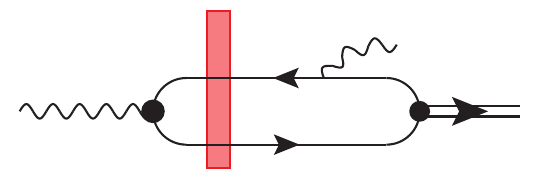}

\caption{The diagrams which describe the $\eta_{c}\gamma$ pair photoproduction
in CGC picture in the leading order in $\alpha_{s}$. Here and in
what follows the double line with arrow in the right part of each
diagram corresponds to produced $\eta_{c}$ meson, and the red block
represents the interaction with the target (shock wave). The diagrams
in the lower row are charge conjugate of the diagrams in the upper
row. For each intermediate quark propagator, which does not cross
the shock wave, an instantaneous contribution should be taken into
account.}\label{fig:CGCBasic-1}
\end{figure}

The exclusive photoproduction of $\eta_{c}\gamma$
pairs in general may proceed via the diagrams shown in the Figure~\ref{fig:CGCBasic-1}.
Formally, the interaction of the photon with the shock wave may be
disregarded as $\mathcal{O}\left(\alpha_{{\rm em}}\right)$-correction,
and for this reason we may represent the amplitude of the process
as 
\begin{equation}
\mathcal{A}_{\gamma p\to\eta_{c}\gamma p}=\mathcal{A}_{1}+\mathcal{A}_{2}\label{eq:ASum}
\end{equation}

where $\mathcal{A}_{1}$ and $\mathcal{A}_{2}$ correspond to the
contributions of the diagrams with photon emission before and after
interaction with the shock wave respectively (left and right columns
in the Figure~\ref{fig:CGCBasic-1} respectively). As was demonstrated
in~\cite{Ayala:2016lhd,Ayala:2017rmh}, the corresponding exclusive
amplitudes $\mathcal{A}_{1},\,\mathcal{A}_{2}$ in general may be
represented as convolutions of the dipole amplitude $\mathcal{N}\left(x,\,\boldsymbol{r},\,\boldsymbol{b}\right)$
with wave functions of the initial and final states, which will be
calculated explicitly below. In what follows we will use subindices
0,1,2 for the kinematic variables related to quark, antiquark and
photon when they interact with the shockwave. Namely, we'll use the
variables 
\begin{equation}
z_{0}\equiv\frac{k_{Q}^{+}}{q^{+}},\quad z_{1}\equiv\frac{k_{\bar{Q}}^{+}}{q^{+}},\quad z_{2}\equiv\frac{k_{\gamma}^{+}}{q^{+}}=1-\alpha_{\eta_{c}},\label{eq:lcFractions}
\end{equation}
for the fractions of the light-cone momentum $q^{+}$ of the incoming
photon which are carried by the produced fermions and emitted photon;
similarly, $\boldsymbol{r}_{0},\,\boldsymbol{r}_{1},\,\boldsymbol{r}_{2}$
are the transverse coordinates of the $Q,\bar{Q}$ and $\gamma$.
For the diagrams in the left column of the Figure~\ref{fig:CGCBasic-1},
the conservation of the plus component of the light-cone momentum
implies that the variables $z_{0},z_{1},z_{2}$ are bound by 
\begin{equation}
z_{0}+z_{1}+z_{2}=1,\qquad z_{0}+z_{1}=\alpha_{\eta_{c}},
\end{equation}
whereas for the diagrams in the right column 
\begin{equation}
z_{0}+z_{1}=1.
\end{equation}

\subsubsection{Evaluation of the amplitude $\mathcal{A}_{1}$}

\begin{figure}
\includegraphics[width=5cm]{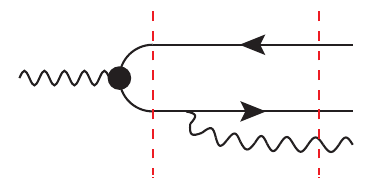}\includegraphics[width=5cm]{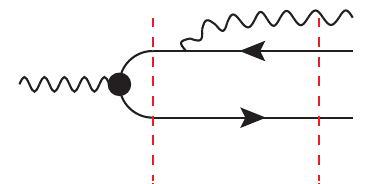}

\includegraphics[width=5cm]{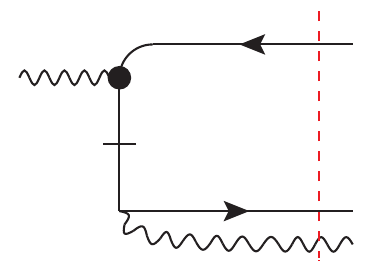}\includegraphics[width=5cm]{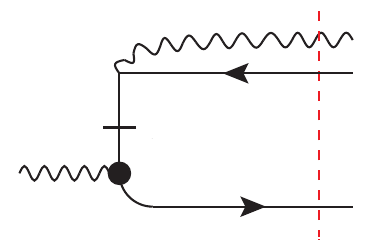}

\caption{The leading order diagrams which contribute to the wave function
of the $\bar{q}q\gamma$ Fock state in the photon (amplitude of the
$\gamma\to\gamma q\bar{q}$ subprocess). The diagrams in the lower
row represent contribution of the instantaneous quark propagator.
The vertical dashed lines denote the light-cone denominators of the
corresponding wave functions in momentum space (see details in Appendix~\ref{sec:WFsAndOverlaps}).}\label{fig:Diags-2}
\end{figure}

As we can see from the diagrams in the left column of the Figure~\ref{fig:CGCBasic-1},
the amplitude $\mathcal{A}_{1}$ may be represented as a convolution
of the wave functions of the dipole amplitude $\mathcal{N}\left(x,\,\boldsymbol{r},\,\boldsymbol{b}\right)$,
the wave function $\Phi_{\eta_{c}}$of the produced $\eta_{c}$ meson,
and the wave function of the $\bar{Q}Q\gamma$ Fock component of the
incoming photon (the amplitude of the subprocess $\gamma\to\bar{Q}Q\gamma$),
namely

\begin{align}
\mathcal{A}_{1} & =\int_{0}^{\alpha_{\eta_{c}}}dz_{0}\prod_{k=1}^{3}\left(d^{2}\boldsymbol{r}_{k}\right)\,\,\Phi_{\eta_{c}}^{(h,\bar{h})\dagger}\left(\frac{z_{0}}{z_{0}+z_{1}},\,\boldsymbol{r}_{10}\right)\,\Psi_{\gamma\to\gamma\bar{Q}Q}^{(\lambda,\sigma,h,\bar{h})}\left(z_{0},\,z_{1}=\alpha_{\eta_{c}}-z_{0},\,z_{2}\equiv\bar{\alpha}_{\eta_{c}},\,\boldsymbol{r}_{0},\,\boldsymbol{r}_{1},\,\boldsymbol{r}_{2}\right)\times\label{eq:A}\\
 & \times\mathcal{N}\left(x,\,\boldsymbol{r}_{10},\,\,\boldsymbol{b}_{10}\right)\exp\left[-i\boldsymbol{p}_{\perp}^{\eta_{c}}\cdot\left(\boldsymbol{b}_{10}-\frac{\bar{\alpha}_{\eta_{c}}}{\alpha_{\eta_{c}}}\boldsymbol{r}_{\gamma}\right)-i\boldsymbol{k}_{\gamma}^{\perp}\cdot\left(\boldsymbol{r}_{\gamma}+\boldsymbol{b}_{10}\right)\right],\nonumber 
\end{align}
where $\lambda,\sigma$ are helicities of the incoming and outgoing
photons; $h,\bar{h}$ are helicities of the produced quark and antiquark,
and we introduced shorthand notations $\boldsymbol{r}_{10}=\boldsymbol{r}_{1}-\boldsymbol{r}_{0}$
for the relative distance between quark-antiquark, $\boldsymbol{b}_{10}=\left(z_{0}\boldsymbol{r}_{0}+z_{1}\boldsymbol{r}_{1}\right)/\left(z_{0}+z_{1}\right)$
for the center-of-mass position (impact parameter) of the quark-antiquark
pair, and $\boldsymbol{r}_{\gamma}=\boldsymbol{r}_{2}-\boldsymbol{b}_{10}$
for the distance between the emitted gluon and the center of mass
of the dipole. The upper limit of the integral over $z_{0}$ is below
unity because part of the incoming photon's momentum is carried away
by the emitted photon. The value of $x$ in the argument of the dipole
cross-section $\mathcal{N}$ is given by $x\approx M_{\gamma\eta_{c}}^{2}/W^{2}$.
The evaluation of the wave function $\Psi_{\gamma\to\gamma\bar{Q}Q}^{(\lambda,\sigma,h,\bar{h})}$
in the leading order in the coupling $\alpha_{s}$ literally repeats
a similar evaluation of the $\gamma\to\bar{Q}Qg$ wave function, discussed
in detail in~\cite{Lappi:2016oup,Hanninen:2017ddy,Beuf:2021qqa,Beuf:2022ndu}
(see e.g. evaluation of the diagrams $j,k,\ell,m$ in~~\cite{Beuf:2022ndu}).
Technically, it requires evaluation of the diagrams shown in the Figure~\ref{fig:Diags-2}
using the light-cone rules from~\cite{Lepage:1980fj}. The final
result of this evaluation is given by (see details in Appendix~\ref{sec:WFsAndOverlaps})

\begin{align}
 & \Psi_{\gamma\to\gamma\bar{Q}Q}^{(\lambda,\sigma,h,\bar{h})}\left(z_{0},\,z_{1},\,z_{2},\,\boldsymbol{r}_{0},\,\boldsymbol{r}_{1},\,\boldsymbol{r}_{2}\right)=8\pi\alpha_{{\rm em}}e_{f}^{2}\sqrt{z_{0}z_{1}}\times\label{eq:WFPsi}\\
 & \left\{ -\frac{1}{z_{0}+z_{2}}\left[\delta_{h,-\bar{h}}\left(\left(2z_{0}+z_{2}\right)\left(2z_{1}-1\right)+\lambda\sigma z_{2}\right)\boldsymbol{\varepsilon}_{\lambda}^{k}\boldsymbol{\varepsilon}_{\sigma}^{*i}-\epsilon_{h,\bar{h}}^{\perp}\boldsymbol{\varepsilon}_{\lambda}^{k}\boldsymbol{\varepsilon}_{\sigma}^{*i}\left(-\sigma z_{2}\left(2z_{1}-1\right)+\lambda\left(2z_{0}+z_{2}\right)\right)\right]\mathcal{I}_{\mathopen{{\color{purple}(}}{\color{purple}1}\mathclose{{\color{purple})}}}^{ik}\right.+\nonumber \\
 & -\frac{1}{z_{1}+z_{2}}\left[\delta_{h,-\bar{h}}\left(\left(2z_{1}+z_{2}\right)\left(2z_{0}-1\right)-\lambda\sigma z_{2}\right)\boldsymbol{\varepsilon}_{\lambda}^{k}\boldsymbol{\varepsilon}_{\sigma}^{*i}+\epsilon_{h,\bar{h}}^{\perp}\boldsymbol{\varepsilon}_{\lambda}^{k}\boldsymbol{\varepsilon}_{\sigma}^{*i}\left(-\sigma z_{2}\left(2z_{0}-1\right)+\lambda\left(2z_{1}+z_{2}\right)\right)\right]\mathcal{I}_{\mathopen{{\color{purple}(}}{\color{purple}{\rm 2}}\mathclose{{\color{purple})}}}^{ik}+\nonumber \\
 & +\frac{z_{0}z_{2}}{\left(z_{0}+z_{2}\right)^{2}}\left(\delta_{h,-\bar{h}}+\lambda\,\epsilon_{h,\bar{h}}^{\perp}\right)\delta_{\lambda,\sigma}\mathcal{J}_{\mathopen{{\color{purple}(}}{\color{purple}{\rm 1}}\mathclose{{\color{purple})}}}-\frac{z_{1}z_{2}}{\left(z_{1}+z_{2}\right)^{2}}\left(\delta_{h,-\bar{h}}-\lambda\epsilon_{h,\bar{h}}^{\perp}\right)\delta_{\lambda,\sigma}\mathcal{J}_{\mathopen{{\color{purple}(}}{\color{purple}{\rm 2}}\mathclose{{\color{purple})}}}+\nonumber \\
 & -\frac{m/q^{+}}{z_{0}+z_{2}}\left[\left(2z_{0}+z_{2}\right)(-\lambda\sqrt{2})\boldsymbol{\varepsilon}_{\sigma}^{*i}\delta_{2h,\lambda}\delta_{2\bar{h},\lambda}+z_{2}\left(\sigma\sqrt{2}\boldsymbol{\varepsilon}_{\lambda}^{i}\delta_{2h,-\sigma}\delta_{2\bar{h},-\sigma}-\delta_{\lambda,\sigma}\left(\epsilon_{h,-\bar{h}}\delta_{i,1}-i\delta_{h,\bar{h}}\delta_{i,2}\right)\right)\right]\hat{\mathcal{I}}_{\mathopen{{\color{purple}(}}{\color{purple}{\rm 1}}\mathclose{{\color{purple})}}}^{i}+\nonumber \\
 & +\frac{mz_{2}^{2}/q^{+}}{\left(z_{0}+z_{2}\right)^{2}}\left[\left(2z_{1}-1\right)\boldsymbol{\varepsilon}_{\lambda}^{k}\left(-\sigma\sqrt{2}\delta_{2h,-\sigma}\delta_{2\bar{h},-\sigma}\right)-\left(\boldsymbol{\varepsilon}_{\sigma}^{*k}\left(\lambda\sqrt{2}\delta_{2h,\lambda}\delta_{2\bar{h},\lambda}\right)-\delta_{\lambda,\sigma}\left(\epsilon_{h,-\bar{h}}\delta_{k,1}-i\delta_{h,\bar{h}}\delta_{k,2}\right)\right)\right]\mathcal{\hat{I}}_{\mathopen{{\color{purple}(}}{\color{purple}{\rm 1}}\mathclose{{\color{purple})}}}^{k}+\nonumber \\
 & -\frac{m^{2}z_{2}^{2}/q^{+}}{\left(z_{0}+z_{2}\right)^{2}}\,\delta_{\lambda,\,\sigma}\left[\delta_{h,-\bar{h}}+\frac{\lambda}{q^{+}}\frac{2z_{1}-1}{z_{1}\left(1-z_{1}\right)}\left[m\epsilon_{h,\bar{h}}-\delta_{h,\bar{h}}\left(\partial_{r_{0x}}-ih\partial_{r_{0y}}\right)\right]\right]\mathcal{I}_{\mathopen{{\color{purple}(}}{\color{purple}{\rm 1}}\mathclose{{\color{purple})}}}\nonumber \\
 & -\frac{m/q^{+}}{z_{1}+z_{2}}\left[\left(\left(2z_{1}+z_{2}\right)\boldsymbol{\varepsilon}_{\sigma}^{*i}\lambda\sqrt{2}\delta_{2h,\lambda}\delta_{2\bar{h},\lambda}-z_{2}\left(\boldsymbol{\varepsilon}_{\lambda}^{i}\sigma\sqrt{2}\delta_{2h,-\sigma}\delta_{2\bar{h},-\sigma}-\delta_{\lambda,\sigma}\boldsymbol{\varepsilon}_{\sigma}^{*i}\lambda\sqrt{2}\delta_{2h,\lambda}\delta_{2\bar{h},\lambda}\right)\right)\right]\hat{\mathcal{I}}_{\mathopen{{\color{purple}(}}{\color{purple}{\rm 2}}\mathclose{{\color{purple})}}}^{i}+\nonumber \\
 & -\frac{z_{2}^{2}\,m/q^{+}}{\left(z_{1}+z_{2}\right)^{2}}\left[\left(\left(2z_{0}-1\right)\boldsymbol{\varepsilon}_{\lambda}^{k}\sigma\sqrt{2}\delta_{2h,-\sigma}\delta_{2\bar{h},-\sigma}+\left(\boldsymbol{\varepsilon}_{\sigma}^{*k}\lambda\sqrt{2}\delta_{2h,\lambda}\delta_{2\bar{h},\lambda}-\delta_{\lambda,\sigma}\left(\epsilon_{h,-\bar{h}}\delta_{k,1}-i\delta_{h,\bar{h}}\delta_{k,2}\right)\right)\right)\right]\mathcal{\hat{I}}_{\mathopen{{\color{purple}(}}{\color{purple}{\rm 2}}\mathclose{{\color{purple})}}}^{k}+\nonumber \\
 & \left.+\frac{m^{2}z_{2}^{2}/q^{+}}{\left(z_{1}+z_{2}\right)^{2}}\,\delta_{\lambda,\,\sigma}\left[\delta_{h,-\bar{h}}-\frac{\lambda}{q^{+}}\frac{2z_{1}-1}{z_{1}\left(1-z_{1}\right)}\left[m\epsilon_{h,\bar{h}}-\delta_{h,\bar{h}}\left(\partial_{r_{1x}}-ih\partial_{r_{1y}}\right)\right]\right]\mathcal{I}_{\mathopen{{\color{purple}(}}{\color{purple}{\rm 2}}\mathclose{{\color{purple})}}}\right\} ,\nonumber 
\end{align}
where $\epsilon_{h,\bar{h}}$ is the 2-dimensional antisymmetric Levi-Civita
symbol (s$\epsilon_{+,-}=-\epsilon_{-,+}=1$), the functions $\mathcal{I}_{\mathopen{{\color{purple}(}}{\color{purple}1}\mathclose{{\color{purple})}}},\,\mathcal{I}_{\mathopen{{\color{purple}(}}{\color{purple}2}\mathclose{{\color{purple})}}},\,\,\mathcal{J}_{\mathopen{{\color{purple}(}}{\color{purple}1}\mathclose{{\color{purple})}}},\,\mathcal{J}_{\mathopen{{\color{purple}(}}{\color{purple}2}\mathclose{{\color{purple})}}}$
with different upper indices (e.g. $\mathcal{I},\,\mathcal{I}^{i},\,\hat{\mathcal{I}}^{i},\,\mathcal{I}^{ij}$)
are defined as~\footnote{Technically, the inferior indices $(1,2)$ of the functions $\mathcal{I},\,\mathcal{J}$
distinguish contributions of the diagrams with photon emission from
the quark or antiquark.} 
\begin{align}
\mathcal{I}_{\mathopen{{\color{purple}(}}{\color{purple}1}\mathclose{{\color{purple})}}} & =\mathcal{I}\left(\boldsymbol{b}_{\mathopen{{\color{purple}(}}{\color{purple}1}\mathclose{{\color{purple})}}},\,\boldsymbol{r}_{\mathopen{{\color{purple}(}}{\color{purple}1}\mathclose{{\color{purple})}}},\,\omega_{\mathopen{{\color{purple}(}}{\color{purple}1}\mathclose{{\color{purple})}}},\,\bar{Q}_{\mathopen{{\color{purple}(}}{\color{purple}1}\mathclose{{\color{purple})}}},\,\lambda_{\mathopen{{\color{purple}(}}{\color{purple}1}\mathclose{{\color{purple})}}}\right),\qquad\mathcal{I}_{\mathopen{{\color{purple}(}}{\color{purple}2}\mathclose{{\color{purple})}}}=\mathcal{I}\left(\boldsymbol{b}_{\mathopen{{\color{purple}(}}{\color{purple}2}\mathclose{{\color{purple})}}},\,\boldsymbol{r}_{\mathopen{{\color{purple}(}}{\color{purple}2}\mathclose{{\color{purple})}}},\,\omega_{\mathopen{{\color{purple}(}}{\color{purple}2}\mathclose{{\color{purple})}}},\,\bar{Q}_{\mathopen{{\color{purple}(}}{\color{purple}2}\mathclose{{\color{purple})}}},\,\lambda_{\mathopen{{\color{purple}(}}{\color{purple}2}\mathclose{{\color{purple})}}}\right),\label{eq:I1}\\
\mathcal{J}_{\mathopen{{\color{purple}(}}{\color{purple}1}\mathclose{{\color{purple})}}} & =\mathcal{J}\left(\boldsymbol{b}_{\mathopen{{\color{purple}(}}{\color{purple}1}\mathclose{{\color{purple})}}},\,\boldsymbol{r}_{\mathopen{{\color{purple}(}}{\color{purple}1}\mathclose{{\color{purple})}}},\,\omega_{\mathopen{{\color{purple}(}}{\color{purple}1}\mathclose{{\color{purple})}}},\,\bar{Q}_{\mathopen{{\color{purple}(}}{\color{purple}1}\mathclose{{\color{purple})}}},\,\lambda_{\mathopen{{\color{purple}(}}{\color{purple}1}\mathclose{{\color{purple})}}}\right),\qquad\mathcal{J}_{\mathopen{{\color{purple}(}}{\color{purple}2}\mathclose{{\color{purple})}}}=\mathcal{J}\left(\boldsymbol{b}_{\mathopen{{\color{purple}(}}{\color{purple}2}\mathclose{{\color{purple})}}},\,\boldsymbol{r}_{\mathopen{{\color{purple}(}}{\color{purple}2}\mathclose{{\color{purple})}}},\,\omega_{\mathopen{{\color{purple}(}}{\color{purple}2}\mathclose{{\color{purple})}}},\,\bar{Q}_{\mathopen{{\color{purple}(}}{\color{purple}2}\mathclose{{\color{purple})}}},\,\lambda_{\mathopen{{\color{purple}(}}{\color{purple}2}\mathclose{{\color{purple})}}}\right),\label{eq:J2}
\end{align}
\begin{align}
 & \boldsymbol{b}_{\mathopen{{\color{purple}(}}{\color{purple}{\rm 1}}\mathclose{{\color{purple})}}}=\left(\frac{z_{0}\boldsymbol{r}_{0}+\boldsymbol{r}_{2}z_{2}}{z_{0}+z_{2}}-\boldsymbol{r}_{1}\right), & \boldsymbol{b}_{\mathopen{{\color{purple}(}}{\color{purple}{\rm 2}}\mathclose{{\color{purple})}}}=-\left(\frac{z_{1}\boldsymbol{r}_{1}+\boldsymbol{r}_{2}z_{2}}{z_{1}+z_{2}}-\boldsymbol{r}_{0}\right),\qquad\boldsymbol{r}_{\mathopen{{\color{purple}(}}{\color{purple}1}\mathclose{{\color{purple})}}}=\boldsymbol{r}_{2}-\boldsymbol{r}_{0},\qquad\boldsymbol{r}_{\mathopen{{\color{purple}(}}{\color{purple}2}\mathclose{{\color{purple})}}}=\boldsymbol{r}_{2}-\boldsymbol{r}_{1},
\end{align}
\begin{align}
 & \omega_{\mathopen{{\color{purple}(}}{\color{purple}{\rm 1}}\mathclose{{\color{purple})}}}=\frac{q^{+}k_{0}^{+}k_{2}^{+}}{k_{1}^{+}\left(k_{0}^{+}+k_{2}^{+}\right)^{2}},\qquad\omega_{\mathopen{{\color{purple}(}}{\color{purple}{\rm 2}}\mathclose{{\color{purple})}}}=\frac{q^{+}k_{1}^{+}k_{2}^{+}}{k_{0}^{+}\left(k_{1}^{+}+k_{2}^{+}\right)^{2}},\qquad & \lambda_{\mathopen{{\color{purple}(}}{\color{purple}1}\mathclose{{\color{purple})}}}=\frac{k_{1}^{+}k_{2}^{+}}{q^{+}k_{0}^{+}},\qquad\lambda_{\mathopen{{\color{purple}(}}{\color{purple}{\rm 2}}\mathclose{{\color{purple})}}}=\frac{k_{0}^{+}k_{2}^{+}}{q^{+}k_{1}^{+}},\label{eq:LambdaDef}
\end{align}
$\bar{Q}_{\mathopen{{\color{purple}(}}{\color{purple}{\rm 1}}\mathclose{{\color{purple})}}}=\bar{Q}_{\mathopen{{\color{purple}(}}{\color{purple}{\rm 2}}\mathclose{{\color{purple})}}}=0$
for onshell photons, and the functions $\mathcal{I},\,\mathcal{J}$
without subscript indices (1,2) in the right-hand side of~(\ref{eq:I1}-\ref{eq:J2})
are defined as 
\begin{align}
\mathcal{I}\left(\boldsymbol{b},\,\boldsymbol{r},\,\omega,\,\bar{Q},\,\lambda\right) & =\int\frac{d^{2}\boldsymbol{P}}{\left(2\pi\right)^{2}}\frac{d^{2}\boldsymbol{K}}{\left(2\pi\right)^{2}}\frac{e^{i\boldsymbol{P}\cdot\boldsymbol{b}}e^{i\boldsymbol{K}\cdot\boldsymbol{r}}}{\left(P^{2}+\bar{Q}^{2}+m^{2}\right)\left(K^{2}+\omega\left(P^{2}+\bar{Q}^{2}+m^{2}+\lambda\,m^{2}\right)\right)},\label{eq:Idef}
\end{align}
\begin{align}
\hat{\mathcal{I}}^{\mathfrak{a}}\left(\boldsymbol{b},\,\boldsymbol{r},\,\omega,\,\bar{Q},\,\lambda\right) & =-i\partial_{b^{\mathfrak{a}}}\mathcal{I}\left(\boldsymbol{b},\,\boldsymbol{r},\,\omega,\,\lambda\right)=\\
 & =\int\frac{d^{2}\boldsymbol{P}}{\left(2\pi\right)^{2}}\frac{d^{2}\boldsymbol{K}}{\left(2\pi\right)^{2}}\frac{P^{\mathfrak{a}}e^{i\boldsymbol{P}\cdot\boldsymbol{b}}e^{i\boldsymbol{K}\cdot\boldsymbol{r}}}{\left(P^{2}+\bar{Q}^{2}+m^{2}\right)\left(K^{2}+\omega\left(P^{2}+\bar{Q}^{2}+m^{2}+\lambda\,m^{2}\right)\right)}\nonumber 
\end{align}

\begin{align}
\mathcal{I}^{\mathfrak{a}}\left(\boldsymbol{b},\,\boldsymbol{r},\,\omega,\,\bar{Q},\,\lambda\right) & =-i\partial_{r^{\mathfrak{a}}}\mathcal{I}\left(\boldsymbol{b},\,\boldsymbol{r},\,\omega,\,\lambda\right)=\\
 & =\int\frac{d^{2}\boldsymbol{P}}{\left(2\pi\right)^{2}}\frac{d^{2}\boldsymbol{K}}{\left(2\pi\right)^{2}}\frac{K^{\mathfrak{a}}e^{i\boldsymbol{P}\cdot\boldsymbol{b}}e^{i\boldsymbol{K}\cdot\boldsymbol{r}}}{\left(P^{2}+\bar{Q}^{2}+m^{2}\right)\left(K^{2}+\omega\left(P^{2}+\bar{Q}^{2}+m^{2}+\lambda\,m^{2}\right)\right)},\nonumber 
\end{align}
\begin{align}
\mathcal{I}^{\mathfrak{a},\mathfrak{b}}\left(\boldsymbol{b},\,\boldsymbol{r},\,\omega,\,\bar{Q},\,\lambda\right) & =-\partial_{b^{\mathfrak{a}}}\partial_{r^{\mathfrak{b}}}\mathcal{I}\left(\boldsymbol{b},\,\boldsymbol{r},\,\omega,\,\lambda\right)=\\
 & =\int\frac{d^{2}\boldsymbol{P}}{\left(2\pi\right)^{2}}\frac{d^{2}\boldsymbol{K}}{\left(2\pi\right)^{2}}\frac{P^{\mathfrak{a}}K^{\mathfrak{b}}e^{i\boldsymbol{P}\cdot\boldsymbol{b}}e^{i\boldsymbol{K}\cdot\boldsymbol{r}}}{\left(P^{2}+\bar{Q}^{2}+m^{2}\right)\left(K^{2}+\omega\left(P^{2}+\bar{Q}^{2}+m^{2}+\lambda\,m^{2}\right)\right)},\nonumber 
\end{align}
\begin{align}
\mathcal{J}\left(\boldsymbol{b},\,\boldsymbol{r},\,\omega,\,\bar{Q},\,\lambda\right) & =\left(-\Delta_{b}+\bar{Q}^{2}+m^{2}\right)\mathcal{I}\left(\boldsymbol{b},\,\boldsymbol{r},\,\omega,\,\lambda\right)=\label{eq:Jdef}\\
 & =\int\frac{d^{2}\boldsymbol{P}}{\left(2\pi\right)^{2}}\frac{d^{2}\boldsymbol{K}}{\left(2\pi\right)^{2}}\frac{e^{i\boldsymbol{P}\cdot\boldsymbol{b}}e^{i\boldsymbol{K}\cdot\boldsymbol{r}}}{\left(K^{2}+\omega\left(P^{2}+\bar{Q}^{2}+m^{2}+\lambda\,m^{2}\right)\right)}.\nonumber 
\end{align}
The contributions which include $\mathcal{I}_{\mathopen{{\color{purple}(}}{\color{purple}1}\mathclose{{\color{purple})}}},\,\mathcal{J}_{\mathopen{{\color{purple}(}}{\color{purple}1}\mathclose{{\color{purple})}}}$
and $\mathcal{I}_{\mathopen{{\color{purple}(}}{\color{purple}2}\mathclose{{\color{purple})}}},\,\mathcal{J}_{\mathopen{{\color{purple}(}}{\color{purple}2}\mathclose{{\color{purple})}}}$
correspond to the charge conjugate diagrams, where the photon emission
happens from the quark or antiquark, respectively. Due to this symmetry,
it is possible to simplify significantly the evaluation of the amplitude~(\ref{eq:A}).
In what follows we will replace the dummy integration variable $z_{0}$
with a variable $\zeta$ defined as a fraction of the photon's momentum
carried by the secondary photon and the active parton (quark or antiquark
which emitted it). Technically, this implies that we should define
the transformation as 
\begin{equation}
z_{0}=\zeta-\bar{\alpha}_{\eta_{c}},\qquad z_{1}=1-\zeta,\qquad z_{2}=1-\alpha_{\eta_{c}}
\end{equation}
for the terms which are proportional to $\mathcal{I}_{\mathopen{{\color{purple}(}}{\color{purple}1}\mathclose{{\color{purple})}}},\,\mathcal{J}_{\mathopen{{\color{purple}(}}{\color{purple}1}\mathclose{{\color{purple})}}}$
and as 
\begin{equation}
z_{0}=1-\zeta,\qquad z_{1}=\zeta-\bar{\alpha}_{\eta_{c}},\qquad z_{2}=1-\alpha_{\eta_{c}}
\end{equation}
for the terms which include $\mathcal{I}_{\mathopen{{\color{purple}(}}{\color{purple}2}\mathclose{{\color{purple})}}},\,\mathcal{J}_{\mathopen{{\color{purple}(}}{\color{purple}2}\mathclose{{\color{purple})}}}$.
This substitution allows to reduce the amplitude~(\ref{eq:A}) to
the form

\begin{align}
\mathcal{A} & =\int_{\alpha_{\eta_{c}}}^{1}d\zeta\,d^{2}\boldsymbol{r}_{10}\,\,d^{2}\boldsymbol{b}_{10}\,\,d^{2}\boldsymbol{\mathfrak{r}}_{2}\,\,\mathcal{N}\left(x,\,\boldsymbol{r}_{10},\,\,\boldsymbol{b}_{10}\right)\exp\left[-i\boldsymbol{p}_{\perp}^{\eta_{c}}\cdot\left(\boldsymbol{b}_{10}-\frac{\bar{\alpha}_{\eta_{c}}}{\alpha_{\eta_{c}}}\boldsymbol{\mathfrak{r}}_{2}\right)-i\boldsymbol{k}_{\gamma}^{\perp}\cdot\left(\boldsymbol{\mathfrak{r}}_{2}+\boldsymbol{b}_{10}\right)\right]\times\label{eq:Ajl-1}\\
 & \times\Phi_{\eta_{c}}^{(h,\bar{h})\dagger}\left(\frac{\zeta-\bar{\alpha}_{\eta_{c}}}{\alpha_{\eta_{c}}},\,\boldsymbol{r}_{10}\right)\Psi_{{\rm red}.}^{(\lambda,\sigma,h,\bar{h})}\left(\boldsymbol{\mathfrak{b}}=-\frac{\left(\zeta-\bar{\alpha}_{\eta_{c}}\right)\boldsymbol{r}_{10}}{\alpha_{\eta_{c}}\zeta}+\frac{\bar{\alpha}_{\eta_{c}}\boldsymbol{\mathfrak{r}}_{2}}{\zeta},\,\boldsymbol{\mathfrak{r}}=\boldsymbol{\mathfrak{r}}_{2}+\frac{\bar{\zeta}}{\alpha_{\eta_{c}}}\boldsymbol{r}_{10}\right),\nonumber 
\end{align}
where for the sake of convenience we introduced the relative transverse
distance $\boldsymbol{\mathfrak{r}}_{2}=\boldsymbol{r}_{2}-\boldsymbol{b}_{10}$
and defined the reduced wave function 
\begin{align}
 & \Psi_{{\rm red.}}^{(\lambda,\sigma,h,\bar{h})}\left(\zeta,\,\bar{\alpha}_{\eta_{c}},\,\boldsymbol{\mathfrak{b}},\,\boldsymbol{\mathfrak{r}}\right)=\frac{8\pi\alpha_{{\rm em}}e^{2}e_{f}^{2}\sqrt{\bar{\zeta}\left(\zeta-\bar{\alpha}_{\eta_{c}}\right)}}{\zeta}\times\label{eq:WFPsi-1-1}\\
 & \left\{ -2\left[\delta_{h,-\bar{h}}\left(\left(2\zeta-\bar{\alpha}_{\eta_{c}}\right)\left(1-2\zeta\right)\right)\boldsymbol{\varepsilon}_{\lambda}^{k}\boldsymbol{\varepsilon}_{\sigma}^{*i}\right]\mathcal{I}^{ik}\left(\boldsymbol{\mathfrak{b}},\,\boldsymbol{\mathfrak{r}},\,\omega,\,\lambda\right)+\frac{2\left(\zeta-\bar{\alpha}_{\eta_{c}}\right)\bar{\alpha}_{\eta_{c}}}{\zeta}\lambda\,\delta_{\lambda,\sigma}\epsilon_{h,\bar{h}}^{\perp}\mathcal{J}\left(\boldsymbol{\mathfrak{b}},\,\boldsymbol{\mathfrak{r}},\,\omega,\,\lambda\right)\right.\nonumber \\
 & -\frac{m}{q^{+}}\hat{\mathcal{I}}^{k}\left(\boldsymbol{\mathfrak{b}},\,\boldsymbol{\mathfrak{r}},\,\omega,\,\lambda\right)\bar{\alpha}_{\eta_{c}}\delta_{\lambda,\sigma}\left(-\epsilon_{h,-\bar{h}}\delta_{k,1}+i\delta_{h,\bar{h}}\delta_{k,2}-\lambda\sqrt{2}\delta_{2h,\lambda}\delta_{2\bar{h},\lambda}\boldsymbol{\varepsilon}_{\sigma}^{*k}\right)+\nonumber \\
 & +\frac{2\bar{\alpha}_{\eta_{c}}^{2}}{\zeta}\frac{m}{q^{+}}\mathcal{\hat{I}}^{k}\left(\boldsymbol{\mathfrak{b}},\,\boldsymbol{\mathfrak{r}},\,\omega,\,\lambda\right)\times\nonumber \\
 & \times\left[\left(\left(1-2\zeta\right)\boldsymbol{\varepsilon}_{\lambda}^{k}\left(-\sigma\sqrt{2}\delta_{2h,-\sigma}\delta_{2\bar{h},-\sigma}\right)-\left(\boldsymbol{\varepsilon}_{\sigma}^{*k}\lambda\sqrt{2}\delta_{2h,\lambda}\delta_{2\bar{h},\lambda}+\delta_{\lambda,\sigma}\left(-\epsilon_{h,-\bar{h}}\delta_{k,1}+i\delta_{h,\bar{h}}\delta_{k,2}\right)\right)\right)\right]\nonumber \\
 & -\left.\frac{\bar{\alpha}_{\eta_{c}}^{2}}{\zeta}\left(\frac{m}{q^{+}}\right)^{2}\,\lambda\delta_{\lambda,\,\sigma}\frac{1-2\zeta}{\zeta\bar{\zeta}}\left[m\epsilon_{h,\bar{h}}-\delta_{h,\bar{h}}\left(\partial_{\mathfrak{r}_{x}}-ih\partial_{\mathfrak{r}_{y}}\right)\right]\mathcal{I}\left(\boldsymbol{\mathfrak{b}},\,\boldsymbol{\mathfrak{r}},\,\omega,\,\lambda\right)\right\} ,\nonumber 
\end{align}
with parameters $\omega,\,\lambda$ given by 
\begin{align}
 & \omega=\frac{\left(\zeta-\bar{\alpha}_{\eta_{c}}\right)\bar{\alpha}_{\eta_{c}}}{\bar{\zeta}\,\zeta^{2}},\quad\lambda=\frac{\bar{\zeta}\,\bar{\alpha}_{\eta_{c}}}{\zeta-\bar{\alpha}_{\eta_{c}}}.
\end{align}
The general result~(\ref{eq:Ajl-1}-\ref{eq:WFPsi-1-1}) can be significantly
simplified in helicity basis. Precisely, for the amplitudes $\mathcal{A}_{1}^{(+,+)}$,
$\mathcal{A}_{1}^{(+,-)}$ with different helicities of the final
state photon we may obtain compact expressions 
\begin{align}
\mathcal{A}_{1}^{(+,+)} & =\int_{\bar{\alpha}_{\eta_{c}}}^{1}d\zeta\,d^{2}\boldsymbol{r}_{10}\,\,d^{2}\boldsymbol{b}_{10}\,\,d^{2}\boldsymbol{r}_{2}\,\,\mathcal{N}\left(x,\,\boldsymbol{r}_{10},\,\,\boldsymbol{b}_{10}\right)\exp\left[-i\boldsymbol{p}_{\perp}^{\eta_{c}}\cdot\left(\boldsymbol{b}_{10}-\frac{\bar{\alpha}_{\eta_{c}}}{\alpha_{\eta_{c}}}\boldsymbol{r}_{2}\right)-i\boldsymbol{k}_{\gamma}^{\perp}\cdot\left(\boldsymbol{r}_{2}+\boldsymbol{b}_{10}\right)\right]\times\label{eq:APP-1}\\
 & \times\frac{8i\pi m\bar{\kappa}\,\alpha_{\eta_{c}}^{2}}{\sqrt{\bar{\zeta}\left(\zeta-\bar{\alpha}_{\eta_{c}}\right)}}\times\left\{ \left[m_{c}^{2}\bar{\alpha}_{\eta_{c}}^{2}\frac{\mathcal{I}_{\mathopen{{\color{purple}(}}{\color{purple}{\rm 1}}\mathclose{{\color{purple})}}}}{\zeta^{2}}-\frac{\left(\zeta-\bar{\alpha}_{\eta_{c}}\right)\bar{\alpha}_{\eta_{c}}}{2\zeta^{2}}\mathcal{J}_{\mathopen{{\color{purple}(}}{\color{purple}1}\mathclose{{\color{purple})}}}\right]\Phi_{\eta_{c}}\left(\frac{\zeta-\bar{\alpha}_{\eta_{c}}}{\alpha_{\eta_{c}}},\,\boldsymbol{r}_{10}\right)\right.-\nonumber \\
 & \qquad-\left.\frac{i\,\bar{\alpha}_{\eta_{c}}^{2}\hat{\mathcal{I}}_{\mathopen{{\color{purple}(}}{\color{purple}1}\mathclose{{\color{purple})}}}^{(+)}e^{-i\phi_{10}}}{\zeta}\partial_{r}\Phi_{\eta_{c}}\left(\frac{\zeta-\bar{\alpha}_{\eta_{c}}}{\alpha_{\eta_{c}}},\,\boldsymbol{r}_{10}\right)\right\} ,\nonumber 
\end{align}
\begin{align}
\mathcal{A}_{1}^{(+,-)} & =\int_{\bar{\alpha}_{\eta_{c}}}^{1}d\zeta\,d^{2}\boldsymbol{r}_{10}\,\,d^{2}\boldsymbol{b}_{10}\,\,d^{2}\boldsymbol{r}_{2}\,\,\mathcal{N}\left(x,\,\boldsymbol{r}_{10},\,\,\boldsymbol{b}_{10}\right)\exp\left[-i\boldsymbol{p}_{\perp}^{\eta_{c}}\cdot\left(\boldsymbol{b}_{10}-\frac{\bar{\alpha}_{\eta_{c}}}{\alpha_{\eta_{c}}}\boldsymbol{r}_{2}\right)-i\boldsymbol{k}_{\gamma}^{\perp}\cdot\left(\boldsymbol{r}_{2}+\boldsymbol{b}_{10}\right)\right]\times\label{eq:AMM-1}\\
 & \times\frac{8\pi m_{c}\bar{\kappa}\,\alpha_{\eta_{c}}^{2}\sqrt{\bar{\zeta}}}{\zeta^{2}\sqrt{\zeta-\bar{\alpha}_{\eta_{c}}}}e^{-i\,\phi_{10}}\,\hat{\mathcal{I}}_{\mathopen{{\color{purple}(}}{\color{purple}1}\mathclose{{\color{purple})}}}^{(-)}\,\partial_{r}\Phi_{\eta_{c}}\left(\frac{\zeta-\bar{\alpha}_{\eta_{c}}}{\alpha_{\eta_{c}}},\,\boldsymbol{r}_{10}\right),\nonumber 
\end{align}
where $\bar{\kappa}=e^{2}e_{f}^{2}/\pi=4\,\alpha_{{\rm em}}e_{c}^{2}=16\alpha_{{\rm em}}/9$,
and the superscript indices $(+),\,(-)$ of functions $\mathcal{I}^k, \hat{\mathcal{I}}^k$ imply contraction of the corresponding free index  with polarization vector of corresponding photon helicity,
\begin{equation}
\mathcal{I}^{(\pm)}=\mathcal{I}^{(1)}\pm i\mathcal{I}^{(2)}=\sqrt{2}\,\sum_{\mathfrak{a}}\varepsilon_{(\pm)}^{\mathfrak{a}}\mathcal{I}^{(\mathfrak{a})},\qquad
\hat{\mathcal{I}}^{(\pm)}=\hat{\mathcal{I}}^{(1)}\pm i\hat{\mathcal{I}}^{(2)}=\sqrt{2}\,\sum_{\mathfrak{a}}\varepsilon_{(\pm)}^{\mathfrak{a}}\hat{\mathcal{I}}^{(\mathfrak{a})}
\end{equation}
The wave function $\Phi_{\eta_{c}}$ of the $\eta_{c}$ meson is a
nonperturbative object whose parametrization will be specified below
in section~\ref{sec:Numer}.

\subsubsection{Evaluation of the amplitude $\mathcal{A}_{2}$}

The evaluation of the amplitude $\mathcal{A}_{2}$ largely follows
the same steps as for the amplitude $\mathcal{A}_{1}$ (see details
in Appendix~\ref{sec:WFsAndOverlaps}). As we can see from the diagrams
in the right column of the Figure~\ref{fig:CGCBasic-1}, the amplitude
$\mathcal{A}_{2}$ may be represented as a convolution of the dipole
amplitude $\mathcal{N}\left(x,\,\boldsymbol{r},\,\boldsymbol{b}\right)$,
the photon wave function $\Psi_{\gamma\to\bar{Q}Q}$ and the amplitude
(``wave function'') of the $\bar{Q}Q\to\eta_{c}\gamma$ subprocess,
\begin{align}
\mathcal{A}_{1} & =\int_{0}^{\alpha_{\eta_{c}}}dz_{0}\prod_{k=1}^{3}\left(d^{2}\boldsymbol{r}_{k}\right)\,\Psi_{\bar{Q}Q\to\gamma\,\eta_{c}}^{(\sigma,h,\bar{h})\dagger}\left(z_{0},\,z_{1}=\alpha_{\eta_{c}}-z_{0},\,z_{2}\equiv\bar{\alpha}_{\eta_{c}},\,\boldsymbol{r}_{0},\,\boldsymbol{r}_{1},\,\boldsymbol{r}_{2}\right)\psi_{\gamma\to\bar{Q}Q}^{(\lambda,\,h,\bar{h})}\left(\frac{z_{0}}{z_{0}+z_{1}},\,\boldsymbol{r}_{10}\right)\times\label{eq:A-2}\\
 & \times\mathcal{N}\left(x,\,\boldsymbol{r}_{10},\,\,\boldsymbol{b}_{10}\right)\exp\left[-i\boldsymbol{p}_{\perp}^{\eta_{c}}\cdot\left(\boldsymbol{b}_{10}-\frac{\bar{\alpha}_{\eta_{c}}}{\alpha_{\eta_{c}}}\boldsymbol{r}_{\gamma}\right)-i\boldsymbol{k}_{\gamma}^{\perp}\cdot\left(\boldsymbol{r}_{\gamma}+\boldsymbol{b}_{10}\right)\right],\nonumber 
\end{align}
where $\Psi_{\bar{Q}Q\to\gamma\,\eta_{c}}^{(\sigma,h,\bar{h})}$ is
the amplitude of the $\bar{Q}Q\to\gamma\,\eta_{c}$ subprocess which
will be specified below, $\psi_{\gamma\to\bar{Q}Q}^{(\lambda,\,h,\bar{h})\dagger}$
is the photon wave function which in the leading order in $\alpha_{s}$
is given by~\cite{Bjorken:1970ah,Dosch:1996ss} 
\begin{align}
\psi_{\gamma\to\bar{Q}Q}^{(\lambda,\,h,\bar{h})}\left(\zeta,\,r\right) & =\frac{\sqrt{2}}{2\pi}ee_{f}\,\left[-i\lambda e^{-i\lambda\phi_{r}}\left(\zeta\delta_{h,\lambda}\delta_{\bar{h},-\lambda}-(1-\zeta)\delta_{h,-\lambda}\delta_{\bar{h},\lambda}\right)mK_{1}\left(mr\right)+m\delta_{h,\lambda}\delta_{\bar{h},\lambda}K_{0}\left(mr\right)\right],
\end{align}
and all other notations have been defined in the previous section
under Eq.~(\ref{eq:A}).

\begin{figure}
\includegraphics[width=6cm]{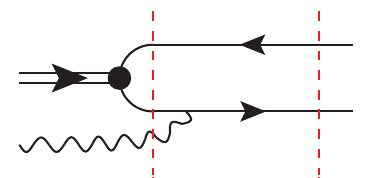}\includegraphics[width=6cm]{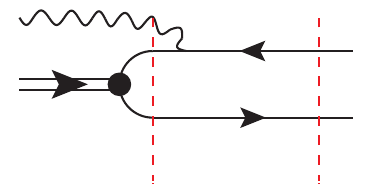}

\includegraphics[width=6cm]{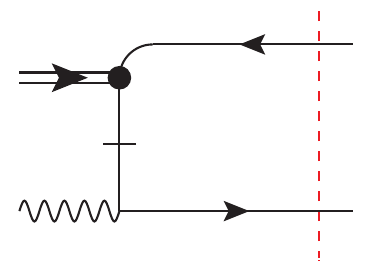}\includegraphics[width=6cm]{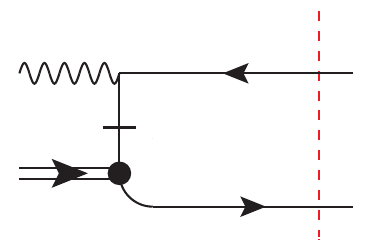}

\caption{The leading order diagrams which contribute to the quarkonium photodissociation
subprocess $\gamma\eta_{c}\to q\bar{q}$ (as explained in Appendix~\ref{sec:WFsAndOverlaps},
they are related to the amplitudes of the $\eta_{c}\to\gamma q\bar{q}$
by a mere change of a sign of the photon momentum $k_{\gamma}^{\mu}$).
The vertical dashed lines denote the light-cone denominators of the
corresponding wave functions in momentum space.}\label{fig:Diags}
\end{figure}

The evaluation of the amplitude $\Psi_{\bar{Q}Q\to\gamma\,\eta_{c}}^{(\sigma,h,\bar{h})}$
in the leading order in the coupling $\alpha_{s}$ resembles a similar
evaluation of the wave function $\Psi_{\gamma\to\gamma\bar{Q}Q}^{(\lambda,\sigma,h,\bar{h})}$
discussed in previous section and requires evaluation of the diagrams
shown in the Figure~\ref{fig:Diags} using the light-cone rules from~\cite{Lepage:1980fj}.
The final result of this evaluation is given by (see details in Appendix~\ref{sec:WFsAndOverlaps})

\begin{align}
\Psi_{\bar{Q}Q\to\gamma\,\eta_{c}}^{(\sigma,h,\bar{h})} & =-iee_{f}\sqrt{2z_{0}z_{1}}\times\label{eq:PsiQQEtac}\\
 & \left[\delta_{\sigma,-}\left\{ \delta_{h,-}\delta_{\bar{h},-}\left[-\frac{1}{\bar{Z}_{1}^{2}}\left(Z_{0}\bar{Z}_{1}\hat{I}_{\mathopen{{\color{purple}(}}{\color{purple}{\rm 1}}\mathclose{{\color{purple})}}}^{(+-)}+m^{2}Z_{2}^{2}I_{\mathopen{{\color{purple}(}}{\color{purple}{\rm 1}}\mathclose{{\color{purple})}}}\right)+\frac{1}{\bar{Z}_{0}^{2}}\left(Z_{1}\bar{Z}_{0}\hat{I}_{\mathopen{{\color{purple}(}}{\color{purple}2}\mathclose{{\color{purple})}}}^{(+-)}-m^{2}Z_{2}^{2}I_{\mathopen{{\color{purple}(}}{\color{purple}2}\mathclose{{\color{purple})}}}\right)\right.\right.\right.\nonumber \\
 & \qquad\qquad\left.+\frac{Z_{0}Z_{2}}{\bar{Z}_{1}^{2}}J_{\mathopen{{\color{purple}(}}{\color{purple}1}\mathclose{{\color{purple})}}}+\frac{Z_{1}Z_{2}}{\bar{Z}_{0}^{2}}J_{\mathopen{{\color{purple}(}}{\color{purple}2}\mathclose{{\color{purple})}}}\right]+\delta_{h,-}\delta_{\bar{h},+}\left[m\frac{1}{\bar{Z}_{1}^{2}}\left(Z_{2}^{2}\hat{I}_{\mathopen{{\color{purple}(}}{\color{purple}{\rm 1}}\mathclose{{\color{purple})}}}^{(-)}-Z_{0}\bar{Z}_{1}I_{\mathopen{{\color{purple}(}}{\color{purple}1}\mathclose{{\color{purple})}}}^{(-)}\right)+mI_{\mathopen{{\color{purple}(}}{\color{purple}{\rm 2}}\mathclose{{\color{purple})}}}^{(-)}\right]\nonumber \\
 & \quad\left.+\delta_{h,+}\delta_{\bar{h},-}\left[mI_{\mathopen{{\color{purple}(}}{\color{purple}{\rm 1}}\mathclose{{\color{purple})}}}^{(-)}-m\frac{1}{\bar{Z}_{0}^{2}}\left(Z_{2}^{2}\hat{I}_{\mathopen{{\color{purple}(}}{\color{purple}{\rm 2}}\mathclose{{\color{purple})}}}^{(-)}+Z_{1}\bar{Z}_{0}I_{\mathopen{{\color{purple}(}}{\color{purple}{\rm 2}}\mathclose{{\color{purple})}}}^{(-)}\right)\right]+\delta_{h,+}\delta_{\bar{h},+}\left[-I_{\mathopen{{\color{purple}(}}{\color{purple}1}\mathclose{{\color{purple})}}}^{(--)}+I_{\mathopen{{\color{purple}(}}{\color{purple}{\rm 2}}\mathclose{{\color{purple})}}}^{(--)}\right]\right\} \nonumber \\
+\delta_{\sigma,-} & \left\{ \left[-\frac{1}{\bar{Z}_{1}^{2}}\left(Z_{0}\bar{Z}_{1}\hat{I}_{\mathopen{{\color{purple}(}}{\color{purple}{\rm 1}}\mathclose{{\color{purple})}}}^{(-+)}+m^{2}Z_{2}^{2}I_{\mathopen{{\color{purple}(}}{\color{purple}1}\mathclose{{\color{purple})}}}\right)+\frac{1}{\bar{Z}_{0}^{2}}\left(Z_{1}\bar{Z}_{0}\hat{I}_{\mathopen{{\color{purple}(}}{\color{purple}{\rm 2}}\mathclose{{\color{purple})}}}^{(-+)}-m^{2}Z_{2}^{2}I_{\mathopen{{\color{purple}(}}{\color{purple}{\rm 2}}\mathclose{{\color{purple})}}}\right)+\frac{Z_{0}Z_{2}}{\bar{Z}_{1}^{2}}J_{\mathopen{{\color{purple}(}}{\color{purple}{\rm 2}}\mathclose{{\color{purple})}}}+\frac{Z_{1}Z_{2}}{\bar{Z}_{0}^{2}}J_{\mathopen{{\color{purple}(}}{\color{purple}{\rm 2}}\mathclose{{\color{purple})}}}\right]\right.\nonumber \\
 & \quad+\delta_{h,+}\delta_{\bar{h},-}\left[-m\frac{1}{\bar{Z}_{1}^{2}}\left(Z_{2}^{2}\hat{I}_{\mathopen{{\color{purple}(}}{\color{purple}1}\mathclose{{\color{purple})}}}^{(+)}-Z_{0}\bar{Z}_{1}I_{\mathopen{{\color{purple}(}}{\color{purple}1}\mathclose{{\color{purple})}}}^{(+)}\right)-mI_{\mathopen{{\color{purple}(}}{\color{purple}{\rm 2}}\mathclose{{\color{purple})}}}^{(+)}\right]\nonumber \\
 & \quad\left.\left.+\delta_{h,-}\delta_{\bar{h},+}\left[-mI_{\mathopen{{\color{purple}(}}{\color{purple}{\rm 1}}\mathclose{{\color{purple})}}}^{(+)}+m\frac{1}{\bar{Z}_{0}^{2}}\left(Z_{2}^{2}\hat{I}_{\mathopen{{\color{purple}(}}{\color{purple}{\rm 2}}\mathclose{{\color{purple})}}}^{(+)}+Z_{1}\bar{Z}_{0}I_{\mathopen{{\color{purple}(}}{\color{purple}{\rm 2}}\mathclose{{\color{purple})}}}^{(+)}\right)\right]+\delta_{h,-}\delta_{\bar{h},-}\left[I_{\mathopen{{\color{purple}(}}{\color{purple}{\rm 2}}\mathclose{{\color{purple})}}}^{(++)}-I_{\mathopen{{\color{purple}(}}{\color{purple}1}\mathclose{{\color{purple})}}}^{(++)}\right]\right\} \right],\nonumber 
\end{align}
where for the sake of brevity we introduced the light-cone fractions
defined with respect to momentum $p_{\eta_{c}}^{+}$ of the produced
$\eta_{c}$-meson, namely 
\begin{equation}
Z_{0}\equiv\frac{k_{Q}^{+}}{p_{\eta_{c}}^{+}},\quad Z_{1}\equiv\frac{k_{\bar{Q}}^{+}}{p_{\eta_{c}}^{+}},\quad Z_{2}\equiv\frac{k_{\gamma}^{+}}{p_{\eta_{c}}^{+}},\quad Z_{0}+Z_{1}-Z_{2}=1,\label{eq:lcFractions-1-1}
\end{equation}
and the functions $I,I^{(\mathfrak{a})},\,\hat{I}^{(\mathfrak{a})},\,I^{(\mathfrak{a,b})},\,J$
differ from the functions $\mathcal{I},\,\mathcal{I}^{(\mathfrak{a})},\,\hat{\mathcal{I}}^{(\mathfrak{a})},\,\mathcal{I}^{(\mathfrak{a,b})},\,\mathcal{J}$
defined in~(\ref{eq:Idef}-\ref{eq:Jdef}) only by inclusion of the
Fourier image of the distribution amplitude $\phi_{\eta_{c}}\left(z,\,\boldsymbol{P}\right)$
in the integrand in the right-hand side, 
\begin{align}
I\left(\boldsymbol{b},\,\boldsymbol{r},\,\omega,\,\bar{Q},\,\lambda\right) & =\int\frac{d^{2}\boldsymbol{P}}{\left(2\pi\right)^{2}}\frac{d^{2}\boldsymbol{K}}{\left(2\pi\right)^{2}}\frac{e^{i\boldsymbol{P}\cdot\boldsymbol{b}}e^{i\boldsymbol{K}\cdot\boldsymbol{r}}\phi_{\eta_{c}}\left(z,\,\boldsymbol{P}\right)}{\left(P^{2}+\bar{Q}^{2}+m^{2}\right)\left(K^{2}+\omega\left(P^{2}+\bar{Q}^{2}+m^{2}+\lambda\,m^{2}\right)\right)},\label{eq:Idef-1}
\end{align}
\begin{align}
\hat{I}^{\mathfrak{a}}\left(\boldsymbol{b},\,\boldsymbol{r},\,\omega,\,\bar{Q},\,\lambda\right) & =-i\partial_{b^{\mathfrak{a}}}I\left(\boldsymbol{b},\,\boldsymbol{r},\,\omega,\,\lambda\right)=\\
 & =\int\frac{d^{2}\boldsymbol{P}}{\left(2\pi\right)^{2}}\frac{d^{2}\boldsymbol{K}}{\left(2\pi\right)^{2}}\frac{P^{\mathfrak{a}}\phi_{\eta_{c}}\left(z,\,\boldsymbol{P}\right)e^{i\boldsymbol{P}\cdot\boldsymbol{b}}e^{i\boldsymbol{K}\cdot\boldsymbol{r}}}{\left(P^{2}+\bar{Q}^{2}+m^{2}\right)\left(K^{2}+\omega\left(P^{2}+\bar{Q}^{2}+m^{2}+\lambda\,m^{2}\right)\right)}\nonumber 
\end{align}

\begin{align}
I^{\mathfrak{a}}\left(\boldsymbol{b},\,\boldsymbol{r},\,\omega,\,\bar{Q},\,\lambda\right) & =-i\partial_{r^{\mathfrak{a}}}I\left(\boldsymbol{b},\,\boldsymbol{r},\,\omega,\,\lambda\right)=\\
 & =\int\frac{d^{2}\boldsymbol{P}}{\left(2\pi\right)^{2}}\frac{d^{2}\boldsymbol{K}}{\left(2\pi\right)^{2}}\frac{K^{\mathfrak{a}}\phi_{\eta_{c}}\left(z,\,\boldsymbol{P}\right)e^{i\boldsymbol{P}\cdot\boldsymbol{b}}e^{i\boldsymbol{K}\cdot\boldsymbol{r}}}{\left(P^{2}+\bar{Q}^{2}+m^{2}\right)\left(K^{2}+\omega\left(P^{2}+\bar{Q}^{2}+m^{2}+\lambda\,m^{2}\right)\right)},\nonumber 
\end{align}
\begin{align}
I^{\mathfrak{a},\mathfrak{b}}\left(\boldsymbol{b},\,\boldsymbol{r},\,\omega,\,\bar{Q},\,\lambda\right) & =-\partial_{b^{\mathfrak{a}}}\partial_{r^{\mathfrak{b}}}\mathcal{I}\left(\boldsymbol{b},\,\boldsymbol{r},\,\omega,\,\lambda\right)=\\
 & =\int\frac{d^{2}\boldsymbol{P}}{\left(2\pi\right)^{2}}\frac{d^{2}\boldsymbol{K}}{\left(2\pi\right)^{2}}\frac{P^{\mathfrak{a}}K^{\mathfrak{b}}\phi_{\eta_{c}}\left(z,\,\boldsymbol{P}\right)e^{i\boldsymbol{P}\cdot\boldsymbol{b}}e^{i\boldsymbol{K}\cdot\boldsymbol{r}}}{\left(P^{2}+\bar{Q}^{2}+m^{2}\right)\left(K^{2}+\omega\left(P^{2}+\bar{Q}^{2}+m^{2}+\lambda\,m^{2}\right)\right)},\nonumber 
\end{align}
\begin{align}
J\left(\boldsymbol{b},\,\boldsymbol{r},\,\omega,\,\bar{Q},\,\lambda\right) & =\left(-\Delta_{b}+\bar{Q}^{2}+m^{2}\right)\mathcal{I}\left(\boldsymbol{b},\,\boldsymbol{r},\,\omega,\,\lambda\right)=\label{eq:Jdef-1}\\
 & =\int\frac{d^{2}\boldsymbol{P}}{\left(2\pi\right)^{2}}\frac{d^{2}\boldsymbol{K}}{\left(2\pi\right)^{2}}\frac{\phi_{\eta_{c}}\left(z,\,\boldsymbol{P}\right)e^{i\boldsymbol{P}\cdot\boldsymbol{b}}e^{i\boldsymbol{K}\cdot\boldsymbol{r}}}{\left(K^{2}+\omega\left(P^{2}+\bar{Q}^{2}+m^{2}+\lambda\,m^{2}\right)\right)}.\nonumber 
\end{align}
However, we should adjust $q^{+}\to p_{\eta_{c}}^{+}$ in~(\ref{eq:LambdaDef})
or equivalently replace $z_{a}\to Z_{a}$. The evaluation of the convolution
is straightforward and yields 
\begin{align}
 & \Psi_{\bar{Q}Q\to\gamma\,\eta_{c}}^{(\sigma,h,\bar{h})\dagger}\,\psi_{\gamma\to\bar{Q}Q}^{(\lambda,\,h,\bar{h})}\left(z_{0},z_{1},\boldsymbol{r}_{0},r_{1},r_{2}\right)=\\
 & =\kappa m\delta_{\lambda,\sigma}\left\{ iK_{0}\left(\varepsilon r\right)\left[\frac{2Z_{0}\bar{Z}_{1}\boldsymbol{\varepsilon}_{\lambda}^{k}\boldsymbol{\varepsilon}_{\sigma}^{*i}I_{\mathopen{{\color{purple}(}}{\color{purple}{\rm 1}}\mathclose{{\color{purple})}}}^{ki}+m^{2}Z_{2}^{2}I_{\mathopen{{\color{purple}(}}{\color{purple}{\rm 1}}\mathclose{{\color{purple})}}}}{\bar{Z}_{1}^{2}}-\frac{2Z_{1}\bar{Z}_{0}\boldsymbol{\varepsilon}_{\lambda}^{k}\boldsymbol{\varepsilon}_{\sigma}^{*i}I_{\mathopen{{\color{purple}(}}{\color{purple}1}\mathclose{{\color{purple})}}}^{ki}-m^{2}Z_{2}^{2}I_{\mathopen{{\color{purple}(}}{\color{purple}2}\mathclose{{\color{purple})}}}}{\bar{Z}_{0}^{2}}-\right.\right.\nonumber \\
 & \qquad\qquad\qquad\qquad-\left.\frac{Z_{0}Z_{2}}{\bar{Z}_{1}^{2}}J_{\mathopen{{\color{purple}(}}{\color{purple}1}\mathclose{{\color{purple})}}}-\frac{Z_{1}Z_{2}}{\bar{Z}_{0}^{2}}J_{\mathopen{{\color{purple}(}}{\color{purple}2}\mathclose{{\color{purple})}}}\right]\nonumber \\
 & +\left.\frac{\varepsilon K_{1}\left(\varepsilon r\right)}{Z_{0}+Z_{1}}e^{i\lambda\phi_{r}}\left[-I_{\mathopen{{\color{purple}(}}{\color{purple}1}\mathclose{{\color{purple})}}}^{(-\lambda)}\left(\frac{Z_{1}\bar{Z}_{1}+Z_{0}^{2}}{\bar{Z}_{1}}\right)+I_{\mathopen{{\color{purple}(}}{\color{purple}2}\mathclose{{\color{purple})}}}^{(-\lambda)}\left(\frac{Z_{0}\bar{Z}_{0}+Z_{1}^{2}}{\bar{Z}_{0}}\right)+Z_{2}^{2}\left(\frac{Z_{0}}{\bar{Z}_{1}^{2}}\hat{I}_{\mathopen{{\color{purple}(}}{\color{purple}1}\mathclose{{\color{purple})}}}^{(-\lambda)}+\frac{Z_{1}}{\bar{Z}_{0}^{2}}\hat{I}_{\mathopen{{\color{purple}(}}{\color{purple}2}\mathclose{{\color{purple})}}}^{(-\lambda)}\right)\right]\right\} \nonumber \\
 & +\kappa m\delta_{\lambda,-\sigma}\left\{ e^{i\lambda\phi_{r}}\varepsilon K_{1}\left(\varepsilon r\right)\left[\frac{I_{\mathopen{{\color{purple}(}}{\color{purple}2}\mathclose{{\color{purple})}}}^{(\lambda)}}{Z_{0}+Z_{1}}\frac{Z_{1}}{\bar{Z}_{0}}-\frac{I_{\mathopen{{\color{purple}(}}{\color{purple}1}\mathclose{{\color{purple})}}}^{(\lambda)}}{Z_{0}+Z_{1}}\frac{Z_{0}}{\bar{Z}_{1}}+\frac{Z_{2}^{2}}{Z_{0}+Z_{1}}\left(\frac{Z_{1}}{\bar{Z}_{1}^{2}}\hat{I}_{\mathopen{{\color{purple}(}}{\color{purple}1}\mathclose{{\color{purple})}}}^{(\lambda)}+\frac{Z_{0}}{\bar{Z}_{0}^{2}}\hat{I}_{\mathopen{{\color{purple}(}}{\color{purple}2}\mathclose{{\color{purple})}}}^{(\lambda)}\right)\right]\right.\nonumber \\
 & \qquad\qquad\qquad\qquad+\left.\frac{i\lambda K_{0}\left(\varepsilon r\right)}{m}\left[I_{\mathopen{{\color{purple}(}}{\color{purple}2}\mathclose{{\color{purple})}}}^{(\lambda,\lambda)}-I_{\mathopen{{\color{purple}(}}{\color{purple}1}\mathclose{{\color{purple})}}}^{(\lambda,\lambda)}\right]\right\} ,\nonumber 
\end{align}
where~ 
\begin{equation}
\kappa=\frac{e^{2}e_{f}^{2}}{\pi}\sqrt{z_{0}\,z_{1}}.
\end{equation}
As we discussed in the previous section, the subindices \textcolor{purple}{(1)}
and \textcolor{purple}{(2)} distinguish contributions from the charge
conjugate diagrams, with photon emission from the quark and antiquark
respectively. We may join the two contributions and simplify analysis,
if we replace the dummy integration variable $z_{0}$ with a new variable
\begin{equation}
\zeta=\frac{k_{0}^{+}}{k_{0}^{+}+k_{1}^{+}}=\frac{z_{0}}{z_{0}+z_{1}}
\end{equation}
in the the terms which include functions $I,J$ with subscript index
\textcolor{purple}{(1)}, and with 
\begin{equation}
\zeta=\frac{k_{1}^{+}}{k_{0}^{+}+k_{1}^{+}}=1-\frac{z_{0}}{z_{0}+z_{1}}
\end{equation}
for the remaining terms with subscript index \textcolor{purple}{(2)}.
Physically, the new variable $\zeta$ equals a fraction of the incoming
photon's momentum carried by active fermion before emission of the
secondary photon, and thus cannot be less than $k_{\gamma}^{+}/q^{+}=\bar{\alpha}_{\eta_{c}}$.
Similar to our previous findings~(\ref{eq:APP-1}-\ref{eq:AMM-1}),
the result for amplitudes simplifies significantly in the helicity
basis and gets a form 
\begin{align}
\mathcal{A}_{2}^{(+,+)} & =\bar{\kappa}\int_{\bar{\alpha}_{\eta_{c}}}^{1}d\zeta\,d^{2}\boldsymbol{r}_{10}\,\,d^{2}\boldsymbol{b}_{10}\,\,d^{2}\boldsymbol{r}_{2}\,\,\mathcal{N}\left(x,\,\boldsymbol{r}_{10},\,\,\boldsymbol{b}_{10}\right)\exp\left[-i\boldsymbol{p}_{\perp}^{\eta_{c}}\cdot\left(\boldsymbol{b}_{10}-\frac{\bar{\alpha}_{\eta_{c}}}{\alpha_{\eta_{c}}}\boldsymbol{r}_{2}\right)-i\boldsymbol{k}_{\gamma}^{\perp}\cdot\left(\boldsymbol{r}_{2}+\boldsymbol{b}_{10}\right)\right]\times\label{eq:APP-3}\\
 & \times\left\{ -im\frac{K_{0}\left(mr_{10}\right)\sqrt{\bar{\zeta}}}{\left(\zeta-\bar{\alpha}_{\eta_{c}}\right)^{3/2}}\left[-2m^{2}\bar{\alpha}_{\eta_{c}}^{2}I_{\mathopen{{\color{purple}(}}{\color{purple}{\rm 1}}\mathclose{{\color{purple})}}}+\zeta\,\bar{\alpha}_{\eta_{c}}J_{\mathopen{{\color{purple}(}}{\color{purple}1}\mathclose{{\color{purple})}}}\right]+2m^{2}\frac{\zeta\bar{\alpha}_{\eta_{c}}^{2}K_{1}\left(mr_{10}\right)\sqrt{\bar{\zeta}}}{\left(\zeta-\bar{\alpha}_{\eta_{c}}\right)^{3/2}}e^{i\phi_{10}}\hat{I}_{\mathopen{{\color{purple}(}}{\color{purple}{\rm 1}}\mathclose{{\color{purple})}}}^{(-)}\right\} ,\nonumber 
\end{align}
\begin{align}
\mathcal{A}_{2}^{(+,-)} & =\bar{\kappa}\int_{\bar{\alpha}_{\eta_{c}}}^{1}d\zeta\,d^{2}\boldsymbol{r}_{10}\,\,d^{2}\boldsymbol{b}_{10}\,\,d^{2}\boldsymbol{r}_{2}\,\,\mathcal{N}\left(x,\,\boldsymbol{r}_{10},\,\,\boldsymbol{b}_{10}\right)\exp\left[-i\boldsymbol{p}_{\perp}^{\eta_{c}}\cdot\left(\boldsymbol{b}_{10}-\frac{\bar{\alpha}_{\eta_{c}}}{\alpha_{\eta_{c}}}\boldsymbol{r}_{2}\right)-i\boldsymbol{k}_{\gamma}^{\perp}\cdot\left(\boldsymbol{r}_{2}+\boldsymbol{b}_{10}\right)\right]\times\label{eq:AMM-3}\\
 & \times2m^{2}\frac{\zeta\bar{\alpha}_{\eta_{c}}^{2}K_{1}\left(mr_{10}\right)e^{-i\phi_{10}}\sqrt{\bar{\zeta}}}{\left(\zeta-\bar{\alpha}_{\eta_{c}}\right)^{3/2}}\hat{I}_{\mathopen{{\color{purple}(}}{\color{purple}{\rm 1}}\mathclose{{\color{purple})}}}^{(-)},\nonumber 
\end{align}
where $\bar{\kappa}$ was defined in the text under~(\ref{eq:AMM-1}),
$\phi_{10}$ is the angle which characterizes the azimuthal orientation
of the vector $\boldsymbol{r}_{10}$, and the arguments of the functions
$I,J$ are given by 
\begin{align}
 & \boldsymbol{\mathfrak{b}}=-\frac{\alpha_{\eta_{c}}\zeta\boldsymbol{r}_{10}+\bar{\alpha}_{\eta_{c}}\boldsymbol{r}_{2}}{\zeta-\bar{\alpha}_{\eta_{c}}},\qquad\mathfrak{\boldsymbol{r}}=\boldsymbol{r}_{2}+\bar{\zeta}\,\boldsymbol{r}_{10},\quad\lambda_{\mathopen{{\color{purple}(}}{\color{purple}1}\mathclose{{\color{purple})}}}=-\frac{\bar{\zeta}\,\bar{\alpha}_{\eta_{c}}}{\alpha_{\eta_{c}}\zeta},\quad\omega_{\mathopen{{\color{purple}(}}{\color{purple}1}\mathclose{{\color{purple})}}}=-\frac{\zeta\,\alpha_{\eta_{c}}\bar{\alpha}_{\eta_{c}}}{\bar{\zeta}\left(\zeta-\bar{\alpha}_{\eta_{c}}\right)^{2}}.\label{eq:Omega_2-1}
\end{align}

\subsection{Evaluation in the momentum space}

The results~(\ref{eq:APP-1},\ref{eq:AMM-1},\ref{eq:APP-3},\ref{eq:AMM-3})
allow to rewrite the full amplitude~(\ref{eq:ASum}) in a very compact
form in helicity basis,

\begin{align}
\mathcal{A}_{\gamma p\to\eta_{c}\gamma p}^{(+,+)} & =\int_{\bar{\alpha}_{\eta_{c}}}^{1}d\zeta\,d^{2}\boldsymbol{r}_{10}\,\,d^{2}\boldsymbol{b}_{10}\,\,d^{2}\boldsymbol{r}_{2}\,\,\mathcal{N}\left(x,\,\boldsymbol{r}_{10},\,\,\boldsymbol{b}_{10}\right)\exp\left[-i\boldsymbol{p}_{\perp}^{\eta_{c}}\cdot\left(\boldsymbol{b}_{10}-\frac{\bar{\alpha}_{\eta_{c}}}{\alpha_{\eta_{c}}}\boldsymbol{r}_{2}\right)-i\boldsymbol{k}_{\gamma}^{\perp}\cdot\left(\boldsymbol{r}_{2}+\boldsymbol{b}_{10}\right)\right]\times\label{eq:APP-4}\\
 & \times\left\{ -i\bar{\kappa}m\frac{K_{0}\left(mr_{10}\right)\sqrt{\bar{\zeta}}}{\left(\zeta-\bar{\alpha}_{\eta_{c}}\right)^{3/2}}\left[-2m^{2}\bar{\alpha}_{\eta_{c}}^{2}I_{\mathopen{{\color{purple}(}}{\color{purple}{\rm 1}}\mathclose{{\color{purple})}}}+\zeta\,\bar{\alpha}_{\eta_{c}}J_{\mathopen{{\color{purple}(}}{\color{purple}1}\mathclose{{\color{purple})}}}\right]+2\bar{\kappa}m^{2}\frac{\zeta\bar{\alpha}_{\eta_{c}}^{2}K_{1}\left(mr_{10}\right)\sqrt{\bar{\zeta}}}{\left(\zeta-\bar{\alpha}_{\eta_{c}}\right)^{3/2}}e^{i\phi_{10}}\hat{I}_{\mathopen{{\color{purple}(}}{\color{purple}{\rm a}}\mathclose{{\color{purple})}}}^{(-)}\right.\nonumber \\
 & +\frac{8i\pi m\bar{\kappa}\,\alpha_{\eta_{c}}^{2}}{\sqrt{\bar{\zeta}\left(\zeta-\bar{\alpha}_{\eta_{c}}\right)}}\left[-\frac{\left(\zeta-\bar{\alpha}_{\eta_{c}}\right)\bar{\alpha}_{\eta_{c}}}{2\zeta^{2}}\mathcal{J}_{\mathopen{{\color{purple}(}}{\color{purple}{\rm 1}}\mathclose{{\color{purple})}}}+m_{c}^{2}\bar{\alpha}_{\eta_{c}}^{2}\frac{\mathcal{I}_{\mathopen{{\color{purple}(}}{\color{purple}{\rm 1}}\mathclose{{\color{purple})}}}}{\zeta^{2}}\right]\Phi_{\eta_{c}}\left(\frac{\zeta-\bar{\alpha}_{\eta_{c}}}{\alpha_{\eta_{c}}},\,\boldsymbol{r}_{10}\right)+\nonumber \\
 & +\left.\frac{8\pi m\bar{\kappa}\,\alpha_{\eta_{c}}^{2}}{\sqrt{\bar{\zeta}\left(\zeta-\bar{\alpha}_{\eta_{c}}\right)}}\frac{\,\bar{\alpha}_{\eta_{c}}^{2}\hat{\mathcal{I}}_{\mathopen{{\color{purple}(}}{\color{purple}{\rm 1}}\mathclose{{\color{purple})}}}^{(+)}e^{-i\phi_{10}}}{\zeta}\partial_{r}\Phi_{\eta_{c}}\left(\frac{\zeta-\bar{\alpha}_{\eta_{c}}}{\alpha_{\eta_{c}}},\,\boldsymbol{r}_{10}\right)\right\} ,\nonumber 
\end{align}
\begin{align}
\mathcal{A}_{\gamma p\to\eta_{c}\gamma p}^{(+,-)} & =\int_{\bar{\alpha}_{\eta_{c}}}^{1}d\zeta\,d^{2}\boldsymbol{r}_{10}\,\,d^{2}\boldsymbol{b}_{10}\,\,d^{2}\boldsymbol{r}_{2}\,\,\mathcal{N}\left(x,\,\boldsymbol{r}_{10},\,\,\boldsymbol{b}_{10}\right)\exp\left[-i\boldsymbol{p}_{\perp}^{\eta_{c}}\cdot\left(\boldsymbol{b}_{10}-\frac{\bar{\alpha}_{\eta_{c}}}{\alpha_{\eta_{c}}}\boldsymbol{r}_{2}\right)-i\boldsymbol{k}_{\gamma}^{\perp}\cdot\left(\boldsymbol{r}_{2}+\boldsymbol{b}_{10}\right)\right]\times\label{eq:AMM-4}\\
 & \times\left\{ 2\bar{\kappa}m^{2}\frac{\zeta\bar{\alpha}_{\eta_{c}}^{2}K_{1}\left(mr_{10}\right)e^{-i\phi_{10}}\sqrt{\bar{\zeta}}}{\left(\zeta-\bar{\alpha}_{\eta_{c}}\right)^{3/2}}\hat{I}_{\mathopen{{\color{purple}(}}{\color{purple}{\rm 1}}\mathclose{{\color{purple})}}}^{(-)}\right.+\left.\frac{8\pi m_{c}\bar{\kappa}\,\alpha_{\eta_{c}}^{2}\sqrt{\bar{\zeta}}}{\zeta^{2}\sqrt{\zeta-\bar{\alpha}_{\eta_{c}}}}e^{-i\,\phi_{10}}\,\hat{\mathcal{I}}_{\mathopen{{\color{purple}(}}{\color{purple}{\rm 1}}\mathclose{{\color{purple})}}}^{(-)}\,\partial_{r}\Phi_{\eta_{c}}\left(\frac{\zeta-\bar{\alpha}_{\eta_{c}}}{\alpha_{\eta_{c}}},\,\boldsymbol{r}_{10}\right)\right\} .\nonumber 
\end{align}
However, a numerical evaluation of the amplitudes using relations~(\ref{eq:APP-4},~\ref{eq:AMM-4})
is technically challenging because it involves a 7-dimensional numerical
integration of oscillating functions. Furthermore, the functions $I,\,J,\,\mathcal{I},\,\mathcal{J}$
themselves are defined via multidimensional integrals, and in view
of a large number of arguments on which they depend, it is not feasible
to use any kind of caching (interpolation of pre-evaluated values)
in order to speed up the evaluations. As we will show below, all these
technical complications can be avoided if we perform integration in
the momentum (Fourier) space. Let's introduce the new functions (Fourier
images) $n_{\Phi_{\eta_{c}}},\,n_{\Phi_{\eta_{c}}}^{(\pm)},\,n_{0},\,n_{1}^{(\pm)}$
defined as~ 
\begin{equation}
n_{\Phi_{\eta_{c}}}\left(x,\boldsymbol{\ell},\,\boldsymbol{s},\,z\right)=\int d^{2}\boldsymbol{b}\,d^{2}\boldsymbol{r}e^{-i\boldsymbol{\ell}\cdot\boldsymbol{r}}e^{-i\boldsymbol{s}\cdot\boldsymbol{b}}\mathcal{N}\left(x,\boldsymbol{r},\boldsymbol{b}\right)\Phi_{\eta_{c}}\left(\boldsymbol{r},\,z\right),\label{eq:nPhi}
\end{equation}
\begin{equation}
n_{\Phi_{\eta_{c}}}^{(\pm)}\left(x,\,\boldsymbol{\ell},\,\boldsymbol{s},\,z\right)=\int d^{2}\boldsymbol{b}\,d^{2}\boldsymbol{r}e^{-i\boldsymbol{\ell}\cdot\boldsymbol{r}}e^{-i\boldsymbol{s}\cdot\boldsymbol{b}}\mathcal{N}\left(x,\,\boldsymbol{r},\,\,\boldsymbol{b}\right)e^{\pm i\phi_{r}}\partial_{r}\Phi_{\eta_{c}}\left(\boldsymbol{r},\,z\right),\label{eq:nPhi1}
\end{equation}
\begin{equation}
n_{0}\left(x,\,\boldsymbol{\ell},\,\,\boldsymbol{s}\right)=\int d^{2}\boldsymbol{b}\,d^{2}\boldsymbol{r}e^{-i\boldsymbol{\ell}\cdot\boldsymbol{r}}e^{-i\boldsymbol{s}\cdot\boldsymbol{b}}\mathcal{N}\left(x,\,\boldsymbol{r},\,\,\boldsymbol{b}\right)K_{0}\left(mr\right),
\end{equation}

\begin{equation}
n_{1}^{(\pm)}\left(x,\,\boldsymbol{\ell},\,\,\boldsymbol{s}\right)=\int d^{2}\boldsymbol{b}\,d^{2}\boldsymbol{r}e^{-i\boldsymbol{\ell}\cdot\boldsymbol{r}}e^{-i\boldsymbol{s}\cdot\boldsymbol{b}}\mathcal{N}\left(x,\,\boldsymbol{r},\,\,\boldsymbol{b}\right)K_{1}\left(mr\right)e^{\pm i\phi_{r}},\label{eq:n1}
\end{equation}
where the angle $\phi_{r}={\rm arg}(r_{x}+ir_{y})$ characterizes
the azimuthal orientation of the vector $\boldsymbol{r}$, and all
the $n$-functions have a mild dependence on Bjorken variable~$x$.

The dependence on the arguments $\left(\boldsymbol{\ell},\,\,\boldsymbol{s}\right)$
in functions $n_{\Phi_{\eta_{c}}},\,n_{\Phi_{\eta_{c}}}^{(\pm)},\,n_{0},\,n_{1}^{(\pm)}$
is relatively smooth and decreases rapidly at large $\ell,\,s$, for
this reason in numerical evaluations it is possible to approximate
them with interpolation of the cached (pre-evaluated) values on a
modestly sized grid in variables ($\ell,s$). The functions $n_{\Phi_{\eta_{c}}},\,n_{\Phi_{\eta_{c}}}^{(\pm)}$also
depend on variable $z$, however this dependence has a relatively
simple shape, and largely repeats the $z$-dependence of the $\eta_{c}$
wave function. Applying the inverse Fourier transform to~(\ref{eq:nPhi}-\ref{eq:n1}),
we may obtain 
\begin{align}
\mathcal{N}\left(x,\,\boldsymbol{r},\,\,\boldsymbol{b}\right) & \Phi_{\eta_{c}}\left(\boldsymbol{r},\,z_{\eta_{c}}\right)=\int\frac{d^{2}\ell}{(2\pi)^{2}}\,\frac{d^{2}s}{(2\pi)^{2}}\,n_{\Phi_{\eta_{c}}}\left(x,\,\boldsymbol{\ell},\,\boldsymbol{s},\,z_{\eta_{c}}\right)e^{i\boldsymbol{\ell}\cdot\boldsymbol{r}}e^{i\boldsymbol{s}\cdot\boldsymbol{b}},\\
\mathcal{N}\left(x,\,\boldsymbol{r},\,\,\boldsymbol{b}\right)e^{\pm i\phi_{r}} & \partial_{r}\Phi_{\eta_{c}}\left(\boldsymbol{r},\,z_{\eta_{c}}\right)=\int\frac{d^{2}\ell}{(2\pi)^{2}}\,\frac{d^{2}s}{(2\pi)^{2}}\,n_{\Phi_{\eta_{c}}}^{(\pm)}\left(x,\,\boldsymbol{\ell},\,\,\boldsymbol{s},\,z_{\eta_{c}}\right)e^{i\boldsymbol{\ell}\cdot\boldsymbol{r}}e^{i\boldsymbol{s}\cdot\boldsymbol{b}},\\
\mathcal{N}\left(x,\,\boldsymbol{r},\,\,\boldsymbol{b}\right)K_{0}\left(mr\right) & =\int\frac{d^{2}\ell}{(2\pi)^{2}}\,\frac{d^{2}s}{(2\pi)^{2}}\,n_{0}\left(x,\,\boldsymbol{\ell},\,\,\boldsymbol{s}\right)e^{i\boldsymbol{\ell}\cdot\boldsymbol{r}}e^{i\boldsymbol{s}\cdot\boldsymbol{b}},\\
\mathcal{N}\left(x,\,\boldsymbol{r},\,\,\boldsymbol{b}\right)K_{1}\left(mr\right)e^{\pm i\phi_{r}} & =\int\frac{d^{2}\ell}{(2\pi)^{2}}\,\frac{d^{2}s}{(2\pi)^{2}}\,n_{1}^{(\pm)}\left(x,\,\boldsymbol{\ell},\,\,\boldsymbol{s}\right)e^{i\boldsymbol{\ell}\cdot\boldsymbol{r}}e^{i\boldsymbol{s}\cdot\boldsymbol{b}}.
\end{align}

The corresponding expressions for the amplitudes in terms of these
new functions have a form 
\begin{align}
\mathcal{A}_{\gamma p\to\eta_{c}\gamma p}^{(+,+)} & =\bar{\kappa}\int_{\bar{\alpha}_{\eta_{c}}}^{1}d\zeta\,d^{2}\boldsymbol{r}_{10}\,\,d^{2}\boldsymbol{b}_{10}\,\,d^{2}\boldsymbol{r}_{2}\int\frac{d^{2}\ell}{(2\pi)^{2}}\,\frac{d^{2}s}{(2\pi)^{2}}\,e^{i\boldsymbol{\ell}\cdot\boldsymbol{r}_{10}}e^{i\boldsymbol{s}\cdot\boldsymbol{b}_{10}}\int\frac{d^{2}\boldsymbol{P}}{(2\pi)^{2}}\,\frac{d^{2}\boldsymbol{K}}{(2\pi)^{2}}\,\times\label{eq:APP-2}\\
 & \,\,\times\exp\left[-i\boldsymbol{p}_{\perp}^{\eta_{c}}\cdot\left(\boldsymbol{b}_{10}-\frac{\bar{\alpha}_{\eta_{c}}}{\alpha_{\eta_{c}}}\boldsymbol{r}_{2}\right)-i\boldsymbol{k}_{\gamma}^{\perp}\cdot\left(\boldsymbol{r}_{2}+\boldsymbol{b}_{10}\right)\right]\times\nonumber \\
 & \times\left\{ m\sqrt{\bar{\zeta}}\frac{2m\zeta\bar{\alpha}_{\eta_{c}}^{2}n_{1}^{(+)}\left(x,\,\boldsymbol{\ell},\,\,\boldsymbol{s}\right)\,\left(P_{x}-iP_{y}\right)-in_{0}\left(x,\,\boldsymbol{\ell},\,\,\boldsymbol{s}\right)\left[-2m^{2}\bar{\alpha}_{\eta_{c}}^{2}+\zeta\,\bar{\alpha}_{\eta_{c}}\left(\boldsymbol{P}^{2}+\frac{\left(\alpha_{\eta_{c}}-2\bar{\zeta}\right)^{2}m^{2}}{\alpha_{\eta_{c}}^{2}}\right)\right]}{\left(\zeta-\bar{\alpha}_{\eta_{c}}\right)^{3/2}\left(\boldsymbol{K}^{2}-\frac{\zeta\,\alpha_{\eta_{c}}\bar{\alpha}_{\eta_{c}}}{\bar{\zeta}\left(\zeta-\bar{\alpha}_{\eta_{c}}\right)^{2}}\boldsymbol{P}^{2}-\frac{\bar{\alpha}_{\eta_{c}}\left(\alpha_{\eta_{c}}-4\zeta\,\bar{\zeta}\right)m^{2}}{\alpha_{\eta_{c}}\bar{\zeta}\,\left(\alpha_{\eta_{c}}-\bar{\zeta}\right)}\right)}\right.\nonumber \\
 & \quad\times\Phi_{\eta_{c}}\left(z_{\eta_{c}},\,\boldsymbol{P}\right)e^{i\left[\boldsymbol{K}\cdot\left(\boldsymbol{r}_{2}+\frac{\bar{\zeta}}{\alpha_{\eta_{c}}}\boldsymbol{r}_{10}\right)+\frac{\boldsymbol{P}}{\zeta}\cdot\left(-\frac{\zeta-\bar{\alpha}_{\eta_{c}}}{\alpha_{\eta_{c}}}\boldsymbol{r}_{10}+\bar{\alpha}_{\eta_{c}}\boldsymbol{r}_{2}\right)\right]}\nonumber \\
 & \quad+\frac{8i\pi m\,\alpha_{\eta_{c}}^{2}e^{i\left[\boldsymbol{K}\cdot\left(\boldsymbol{r}_{2}+\bar{\zeta}\,\boldsymbol{r}_{10}\right)-\frac{\boldsymbol{P}}{\zeta-\bar{\alpha}_{\eta_{c}}}\left(\alpha_{\eta_{c}}\zeta\boldsymbol{r}_{10}+\bar{\alpha}_{\eta_{c}}\boldsymbol{r}_{2}\right)\right]}}{\sqrt{\bar{\zeta}\left(\zeta-\bar{\alpha}_{\eta_{c}}\right)}\left(\boldsymbol{P}^{2}+m^{2}\right)\left(\boldsymbol{K}^{2}+\frac{\left(\zeta-\bar{\alpha}_{\eta_{c}}\right)\,\bar{\alpha}_{\eta_{c}}}{\bar{\zeta}\,\zeta^{2}}\boldsymbol{P}^{2}+\frac{\alpha_{\eta_{c}}\bar{\alpha}_{\eta_{c}}m^{2}}{\zeta\,\bar{\zeta}\,}\right)}\times\nonumber \\
 & \quad\times\left[-\frac{\left(\zeta-\bar{\alpha}_{\eta_{c}}\right)\bar{\alpha}_{\eta_{c}}}{2\zeta^{2}}\left(\boldsymbol{P}^{2}+m^{2}\right)+m^{2}\bar{\alpha}_{\eta_{c}}^{2}\frac{1}{\zeta^{2}}\right]n_{\Phi_{\eta_{c}}}\left(x,\,\boldsymbol{\ell},\,\,\boldsymbol{s},\,z_{\eta_{c}}\right)\nonumber \\
 & \quad\left.+\frac{8\pi m\,\alpha_{\eta_{c}}^{2}}{\sqrt{\bar{\zeta}\left(\zeta-\bar{\alpha}_{\eta_{c}}\right)}}\frac{\,\bar{\alpha}_{\eta_{c}}^{2}}{\zeta}\frac{n_{\Phi_{\eta_{c}}}^{(-)}\left(x,\,\boldsymbol{\ell},\,\,\boldsymbol{s},\,z_{\eta_{c}}\right)\left(P_{x}+iP_{y}\right)e^{i\left[\boldsymbol{K}\cdot\left(\boldsymbol{r}_{2}+\bar{\zeta}\,\boldsymbol{r}_{10}\right)-\frac{\boldsymbol{P}}{\zeta-\bar{\alpha}_{\eta_{c}}}\left(\alpha_{\eta_{c}}\zeta\boldsymbol{r}_{10}+\bar{\alpha}_{\eta_{c}}\boldsymbol{r}_{2}\right)\right]}}{\left(\boldsymbol{P}^{2}+m^{2}\right)\left(\boldsymbol{K}^{2}+\frac{\left(\zeta-\bar{\alpha}_{\eta_{c}}\right)\,\bar{\alpha}_{\eta_{c}}}{\bar{\zeta}\,\zeta^{2}}\boldsymbol{P}^{2}+\frac{\alpha_{\eta_{c}}\bar{\alpha}_{\eta_{c}}m^{2}}{\zeta\,\bar{\zeta}\,}\right)}\right\} _{z_{\eta_{c}}=\frac{\zeta-\bar{\alpha}_{\eta_{c}}}{\alpha_{\eta_{c}}}}.\nonumber 
\end{align}

\begin{align}
\mathcal{A}_{\gamma p\to\eta_{c}\gamma p}^{(+,-)} & =\bar{\kappa}\int\frac{d^{2}\ell}{(2\pi)^{2}}\,\frac{d^{2}s}{(2\pi)^{2}}\,e^{i\boldsymbol{\ell}\cdot\boldsymbol{r}_{10}}e^{i\boldsymbol{s}\cdot\boldsymbol{b}_{10}}\times\label{eq:AMM-2}\\
 & \times\int_{\bar{\alpha}_{\eta_{c}}}^{1}d\zeta\,d^{2}\boldsymbol{r}_{10}\,\,d^{2}\boldsymbol{b}_{10}\,\,d^{2}\boldsymbol{r}_{2}\,\,\exp\left[-i\boldsymbol{p}_{\perp}^{\eta_{c}}\cdot\left(\boldsymbol{b}_{10}-\frac{\bar{\alpha}_{\eta_{c}}}{\alpha_{\eta_{c}}}\boldsymbol{r}_{2}\right)-i\boldsymbol{k}_{\gamma}^{\perp}\cdot\left(\boldsymbol{r}_{2}+\boldsymbol{b}_{10}\right)\right]\times\nonumber \\
 & \times\left\{ 2m^{2}\frac{\zeta\bar{\alpha}_{\eta_{c}}^{2}n_{1}^{(-)}\left(x,\,\boldsymbol{\ell},\,\,\boldsymbol{s}\right)\sqrt{\bar{\zeta}}}{\left(\zeta-\bar{\alpha}_{\eta_{c}}\right)^{3/2}}\frac{\left(P_{x}-iP_{y}\right)}{\left(\boldsymbol{K}^{2}-\frac{\zeta\,\alpha_{\eta_{c}}\bar{\alpha}_{\eta_{c}}}{\bar{\zeta}\left(\zeta-\bar{\alpha}_{\eta_{c}}\right)^{2}}\boldsymbol{P}^{2}-\frac{\bar{\alpha}_{\eta_{c}}\left(\alpha_{\eta_{c}}-4\zeta\,\bar{\zeta}\right)m^{2}}{\alpha_{\eta_{c}}\bar{\zeta}\,\left(\alpha_{\eta_{c}}-\bar{\zeta}\right)}\right)}\right.\times\nonumber \\
 & \qquad\times\Phi_{\eta_{c}}\left(z_{\eta_{c}},\,\boldsymbol{P}\right)e^{i\left[\boldsymbol{K}\cdot\left(\boldsymbol{r}_{2}+\frac{\bar{\zeta}}{\alpha_{\eta_{c}}}\boldsymbol{r}_{10}\right)+\frac{\boldsymbol{P}}{\zeta}\left(-\frac{\zeta-\bar{\alpha}_{\eta_{c}}}{\alpha_{\eta_{c}}}\boldsymbol{r}_{10}+\bar{\alpha}_{\eta_{c}}\boldsymbol{r}_{2}\right)\right]}\nonumber \\
 & \quad+\left.\frac{8\pi m\alpha_{\eta_{c}}^{2}\sqrt{\bar{\zeta}}}{\zeta^{2}\sqrt{\zeta-\bar{\alpha}_{\eta_{c}}}}\,\frac{\,n_{\Phi_{\eta_{c}}}^{(-)}\left(x,\,\boldsymbol{\ell},\,\,\boldsymbol{s},\,z_{\eta_{c}}\right)\left(P_{x}-iP_{y}\right)e^{i\left[\boldsymbol{K}\cdot\left(\boldsymbol{r}_{2}+\bar{\zeta}\,\boldsymbol{r}_{10}\right)-\frac{\boldsymbol{P}}{\zeta-\bar{\alpha}_{\eta_{c}}}\left(\alpha_{\eta_{c}}\zeta\boldsymbol{r}_{10}+\bar{\alpha}_{\eta_{c}}\boldsymbol{r}_{2}\right)\right]}}{\left(\boldsymbol{P}^{2}+m^{2}\right)\left(\boldsymbol{K}^{2}+\frac{\left(\zeta-\bar{\alpha}_{\eta_{c}}\right)\,\bar{\alpha}_{\eta_{c}}}{\bar{\zeta}\,\zeta^{2}}\boldsymbol{P}^{2}+\frac{\alpha_{\eta_{c}}\bar{\alpha}_{\eta_{c}}m^{2}}{\zeta\,\bar{\zeta}\,}\right)}\right\} _{z_{\eta_{c}}=\frac{\zeta-\bar{\alpha}_{\eta_{c}}}{\alpha_{\eta_{c}}}}.\nonumber 
\end{align}
After taking integrals over the coordinates $\boldsymbol{b}_{10},\,\boldsymbol{r}_{10},\,\boldsymbol{r}_{2}$,
we may get 
\begin{equation}
\left(2\pi\right)^{6}\delta\left(\boldsymbol{s}-\boldsymbol{p}_{\perp}^{\eta_{c}}-\boldsymbol{k}_{\gamma}^{\perp}\right)\delta\left(\boldsymbol{\ell}+\boldsymbol{K}\cdot\frac{\bar{\zeta}}{\alpha_{\eta_{c}}}-\frac{\boldsymbol{P}}{\zeta}\frac{\zeta-\bar{\alpha}_{\eta_{c}}}{\alpha_{\eta_{c}}}\right)\delta\left(\boldsymbol{p}_{\perp}^{\eta_{c}}\frac{\bar{\alpha}_{\eta_{c}}}{\alpha_{\eta_{c}}}-\boldsymbol{k}_{\gamma}^{\perp}+\boldsymbol{K}+\frac{\boldsymbol{P}\bar{\alpha}_{\eta_{c}}}{\zeta}\right)
\end{equation}
for the terms which include functions $n_{0},\,n_{1}^{(\pm)}$, and

\begin{equation}
\left(2\pi\right)^{6}\delta\left(\boldsymbol{s}-\boldsymbol{p}_{\perp}^{\eta_{c}}-\boldsymbol{k}_{\gamma}^{\perp}\right)\delta\left(\boldsymbol{\ell}+\boldsymbol{K}\bar{\zeta}\,-\frac{\alpha_{\eta_{c}}\zeta\,\boldsymbol{P}}{\zeta-\bar{\alpha}_{\eta_{c}}}\right)\delta\left(\boldsymbol{p}_{\perp}^{\eta_{c}}\frac{\bar{\alpha}_{\eta_{c}}}{\alpha_{\eta_{c}}}-\boldsymbol{k}_{\gamma}^{\perp}+\boldsymbol{K}-\frac{\boldsymbol{P}\bar{\alpha}_{\eta_{c}}}{\zeta-\bar{\alpha}_{\eta_{c}}}\right)
\end{equation}
for the terms which include $n_{\Phi_{\eta_{c}}},\,n_{\Phi_{\eta_{c}}}^{(\pm)}$.
These $\delta$-functions allow to get rid of the integrals over $\boldsymbol{K},\,\boldsymbol{P}$
and obtain the amplitudes~

\begin{align}
\mathcal{A}_{\gamma p\to\eta_{c}\gamma p}^{(+,+)} & =\bar{\kappa}m\int_{\bar{\alpha}_{\eta_{c}}}^{1}d\zeta\,\int\frac{d^{2}\ell}{(2\pi)^{2}}\,\times\label{eq:Amplitude_PP}\\
 & \times\left\{ \frac{-in_{0}\left(x,\,\boldsymbol{\ell},\,\,-\boldsymbol{\Delta}_{\perp}\right)\bar{\zeta}\sqrt{\bar{\zeta}\left(\zeta-\bar{\alpha}_{\eta_{c}}\right)}\left[\zeta\,\bar{\alpha}_{\eta_{c}}\left(\alpha_{\eta_{c}}^{2}\boldsymbol{P}_{(1)}^{2}+\left(\alpha_{\eta_{c}}-2\bar{\zeta}\right)^{2}m^{2}\right)-2m^{2}\alpha_{\eta_{c}}^{2}\bar{\alpha}_{\eta_{c}}^{2}\right]\Phi_{\eta_{c}}\left(z_{\eta_{c}},\,\boldsymbol{P}_{(1)}\right)}{\alpha_{\eta_{c}}^{2}\left(\bar{\zeta}\left(\zeta-\bar{\alpha}_{\eta_{c}}\right)^{2}\boldsymbol{K}_{(1)}^{2}-\zeta\,\alpha_{\eta_{c}}\bar{\alpha}_{\eta_{c}}\boldsymbol{P}_{(1)}^{2}-\frac{\bar{\alpha}_{\eta_{c}}}{\alpha_{\eta_{c}}}\left(\zeta-\bar{\alpha}_{\eta_{c}}\right)\left(\alpha_{\eta_{c}}-4\zeta\,\bar{\zeta}\right)m^{2}\right)}\right.+\nonumber \\
 & \quad+\frac{2m\zeta\bar{\zeta}\sqrt{\bar{\zeta}\left(\zeta-\bar{\alpha}_{\eta_{c}}\right)}\bar{\alpha}_{\eta_{c}}^{2}n_{1}^{(+)}\left(x,\,\boldsymbol{\ell},\,\,-\boldsymbol{\Delta}_{\perp}\right)\Phi_{\eta_{c}}\left(z,\,\boldsymbol{P}_{(1)}\right)\,\left(P_{(1)x}-iP_{(1)y}\right)}{\left(\bar{\zeta}\left(\zeta-\bar{\alpha}_{\eta_{c}}\right)^{2}\boldsymbol{K}_{(1)}^{2}-\zeta\,\alpha_{\eta_{c}}\bar{\alpha}_{\eta_{c}}\boldsymbol{P}_{(1)}^{2}-\frac{\bar{\alpha}_{\eta_{c}}}{\alpha_{\eta_{c}}}\left(\zeta-\bar{\alpha}_{\eta_{c}}\right)\left(\alpha_{\eta_{c}}-4\zeta\,\bar{\zeta}\right)m^{2}\right)}\nonumber \\
 & \quad+\sqrt{\frac{\bar{\zeta}}{\zeta-\bar{\alpha}_{\eta_{c}}}}\frac{4i\pi\,\alpha_{\eta_{c}}^{2}n_{\Phi_{\eta_{c}}}\left(x,\,\boldsymbol{\ell},\,\,-\boldsymbol{\Delta}_{\perp},\,z_{\eta_{c}}\right)\left[2m^{2}\bar{\alpha}_{\eta_{c}}^{2}-\,\left(\zeta-\bar{\alpha}_{\eta_{c}}\right)\bar{\alpha}_{\eta_{c}}\left(\boldsymbol{P}_{(2)}^{2}+m^{2}\right)\right]}{\left(\boldsymbol{P}_{(2)}^{2}+m^{2}\right)\left(\bar{\zeta}\,\zeta^{2}\boldsymbol{K}_{(2)}^{2}+\left(\zeta-\bar{\alpha}_{\eta_{c}}\right)\,\bar{\alpha}_{\eta_{c}}\boldsymbol{P}_{(2)}^{2}+\alpha_{\eta_{c}}\bar{\alpha}_{\eta_{c}}\zeta m^{2}\right)}\nonumber \\
 & \quad\left.+\sqrt{\frac{\bar{\zeta}}{\zeta-\bar{\alpha}_{\eta_{c}}}}\frac{8\pi\,\alpha_{\eta_{c}}^{2}\bar{\alpha}_{\eta_{c}}^{2}\zeta n_{\Phi_{\eta_{c}}}^{(-)}\left(x,\,\boldsymbol{\ell},\,\,-\boldsymbol{\Delta}_{\perp},\,z_{\eta_{c}}\right)\left(P_{(2)x}+iP_{(2)y}\right)}{\left(\boldsymbol{P}_{(2)}^{2}+m^{2}\right)\left(\bar{\zeta}\,\zeta^{2}\boldsymbol{K}_{(2)}^{2}+\left(\zeta-\bar{\alpha}_{\eta_{c}}\right)\,\bar{\alpha}_{\eta_{c}}\boldsymbol{P}_{(2)}^{2}+\alpha_{\eta_{c}}\bar{\alpha}_{\eta_{c}}\zeta m^{2}\right)}\right\} _{z_{\eta_{c}}=\frac{\zeta-\bar{\alpha}_{\eta_{c}}}{\alpha_{\eta_{c}}}},\nonumber 
\end{align}
\begin{align}
\mathcal{A}_{\gamma p\to\eta_{c}\gamma p}^{(+,-)} & =\bar{\kappa}m\int_{\bar{\alpha}_{\eta_{c}}}^{1}d\zeta\,\int\frac{d^{2}\ell}{(2\pi)^{2}}\,\times\label{eq:Amplitude_PM}\\
 & \times\left\{ \frac{2m\,\zeta\bar{\zeta}\bar{\alpha}_{\eta_{c}}^{2}\sqrt{\bar{\zeta}\left(\zeta-\bar{\alpha}_{\eta_{c}}\right)}n_{1}^{(-)}\left(x,\,\boldsymbol{\ell},\,\,-\boldsymbol{\Delta}_{\perp}\right)\Phi_{\eta_{c}}\left(z_{\eta_{c}},\,\boldsymbol{P}_{(1)}\right)\left(P_{(1)x}-iP_{(1)y}\right)}{\left(\bar{\zeta}\left(\zeta-\bar{\alpha}_{\eta_{c}}\right)^{2}\boldsymbol{K}_{(1)}^{2}-\zeta\,\alpha_{\eta_{c}}\bar{\alpha}_{\eta_{c}}\boldsymbol{P}_{(1)}^{2}-\frac{\bar{\alpha}_{\eta_{c}}}{\alpha_{\eta_{c}}}\left(\zeta-\bar{\alpha}_{\eta_{c}}\right)\left(\alpha_{\eta_{c}}-4\zeta\,\bar{\zeta}\right)m^{2}\right)}\right.+\nonumber \\
 & \quad+\left.\sqrt{\frac{\bar{\zeta}}{\zeta-\bar{\alpha}_{\eta_{c}}}}\,\frac{8\pi\alpha_{\eta_{c}}^{2}\bar{\zeta}\,\,n_{\Phi_{\eta_{c}}}^{(-)}\left(x,\,\boldsymbol{\ell},\,\,-\boldsymbol{\Delta}_{\perp},\,z_{\eta_{c}}\right)\left(P_{(2)x}-iP_{(2)y}\right)}{\left(\boldsymbol{P}_{(2)}^{2}+m^{2}\right)\left(\bar{\zeta}\,\zeta^{2}\boldsymbol{K}_{(2)}^{2}+\left(\zeta-\bar{\alpha}_{\eta_{c}}\right)\,\bar{\alpha}_{\eta_{c}}\boldsymbol{P}_{(2)}^{2}+\alpha_{\eta_{c}}\bar{\alpha}_{\eta_{c}}\zeta m^{2}\right)}\right\} _{z_{\eta_{c}}=\frac{\zeta-\bar{\alpha}_{\eta_{c}}}{\alpha_{\eta_{c}}}},\nonumber 
\end{align}
where $\Phi_{\eta_{c}}\left(z,\,\boldsymbol{k}_{{\rm rel}}\right)$
is the momentum-dependent wave function of the $\eta_{c}$ meson,
and we introduced shorthand notations 
\begin{align}
 & \boldsymbol{K}_{(1)}=-\frac{\boldsymbol{\ell}\bar{\alpha}_{\eta_{c}}}{\zeta}+\frac{\left(\zeta-\bar{\alpha}_{\eta_{c}}\right)}{\alpha_{\eta_{c}}^{2}\zeta}\left(\alpha_{\eta_{c}}\boldsymbol{k}_{\gamma}^{\perp}-\bar{\alpha}_{\eta_{c}}\boldsymbol{p}_{\perp}^{\eta_{c}}\right)=-\frac{\boldsymbol{P}_{(1)}\bar{\alpha}_{\eta_{c}}}{\zeta}+\frac{\left(\alpha_{\eta_{c}}\boldsymbol{k}_{\gamma}^{\perp}-\bar{\alpha}_{\eta_{c}}\boldsymbol{p}_{\perp}^{\eta_{c}}\right)}{\alpha_{\eta_{c}}},\\
 & \boldsymbol{P}_{(1)}=\boldsymbol{\ell}+\frac{\bar{\zeta}}{\alpha_{\eta_{c}}^{2}}\left(\alpha_{\eta_{c}}\boldsymbol{k}_{\gamma}^{\perp}-\bar{\alpha}_{\eta_{c}}\boldsymbol{p}_{\perp}^{\eta_{c}}\right),
\end{align}
\begin{align}
\boldsymbol{K}_{(2)} & =\frac{\boldsymbol{\ell}\bar{\alpha}_{\eta_{c}}+\zeta\left(\alpha_{\eta_{c}}\boldsymbol{k}_{\gamma}^{\perp}-\bar{\alpha}_{\eta_{c}}\boldsymbol{p}_{\perp}^{\eta_{c}}\right)}{\zeta-\bar{\alpha}_{\eta_{c}}}=\frac{\boldsymbol{P}_{(2)}\bar{\alpha}_{\eta_{c}}}{\zeta-\bar{\alpha}_{\eta_{c}}}+\frac{\left(\alpha_{\eta_{c}}\boldsymbol{k}_{\gamma}^{\perp}-\bar{\alpha}_{\eta_{c}}\boldsymbol{p}_{\perp}^{\eta_{c}}\right)}{\alpha_{\eta_{c}}},\\
\boldsymbol{P}_{(2)} & =\boldsymbol{\ell}+\frac{\bar{\zeta}}{\alpha_{\eta_{c}}}\left(\alpha_{\eta_{c}}\boldsymbol{k}_{\gamma}^{\perp}-\bar{\alpha}_{\eta_{c}}\boldsymbol{p}_{\perp}^{\eta_{c}}\right).
\end{align}
All the vectors $\boldsymbol{K}_{({\rm 1})},\,\boldsymbol{K}_{({\rm 2})},\,\boldsymbol{P}_{({\rm 1})},\,\boldsymbol{P}_{(2)}$,
depend only on the dummy integration variable $\boldsymbol{\ell}$
and a combination of the external momenta

\begin{equation}
\boldsymbol{L}=\left(\alpha_{\eta_{c}}\boldsymbol{k}_{\gamma}^{\perp}-\bar{\alpha}_{\eta_{c}}\boldsymbol{p}_{\perp}^{\eta_{c}}\right)=\boldsymbol{p}_{\perp}+\boldsymbol{\Delta}_{\perp}\left(\frac{1}{2}-\alpha_{\eta_{c}}\right),\label{eq:LVector}
\end{equation}
but not on the individual vectors $\boldsymbol{k}_{\gamma}^{\perp},\,\boldsymbol{p}_{\perp}^{\eta_{c}}$.
In the special but very important case when the dipole amplitude $N\left(x,\,\boldsymbol{r},\,\boldsymbol{b}\right)$
does not depend on relative orientation of the vectors $\boldsymbol{r},\,\boldsymbol{b}$,
the Fourier images ($n$-functions) depend only on the absolute values
of the vectors $\boldsymbol{\ell},\,\boldsymbol{\Delta}.$ Since the
vector $\Delta$ appears only in the argument of the $n$-functions
in~(\ref{eq:Amplitude_PP},\ref{eq:Amplitude_PM}), after integration
over the momentum $\boldsymbol{\ell}$ the final result will be sensitive
to the absolute values of the vectors $\boldsymbol{L}$ and $\boldsymbol{\Delta}$,
but not their relative orientation. This unique property may be used
to test possible angular correlations between the vectors $\boldsymbol{r},\,\boldsymbol{b}$
in a dipole amplitude.

\section{Numerical estimates}

\label{sec:Numer}

\subsection{Differential cross-sections}

\label{subsec:diff}In what follows for the sake of definiteness,
we use for our estimates the bCGC parametrization of the color dipole
amplitude with fit parameters from~\cite{RESH} corresponding to
a charm mass $m_{c}\approx1.4$~GeV and will take into account corrections
due to skewedness and real part of the amplitude, as given by~(\ref{eq:reA},~\ref{eq:Rg}).
For the wave function $\Phi_{\eta_{c}}$ we use the LC-Gauss parametrization~\cite{Dosch:1996ss,Benic:2023,RESH},
\begin{equation}
\Phi_{\eta_{c}}\left(z,\,\boldsymbol{r}\right)=\mathcal{N}_{p}\,\,z\bar{z}\exp\left(-\frac{m_{c}^{2}\mathcal{R}_{p}^{2}}{8z\bar{z}}-\frac{2z\bar{z}}{\mathcal{R}_{p}^{2}}r^{2}+\frac{1}{2}m_{c}^{2}\mathcal{R}_{p}^{2}\right),
\end{equation}
with $\mathcal{R}_{p}^{2}=2.48\,{\rm GeV}^{-2}$ and $\mathcal{N}_{p}=0.547$.

\begin{figure}
\includegraphics[width=6cm]{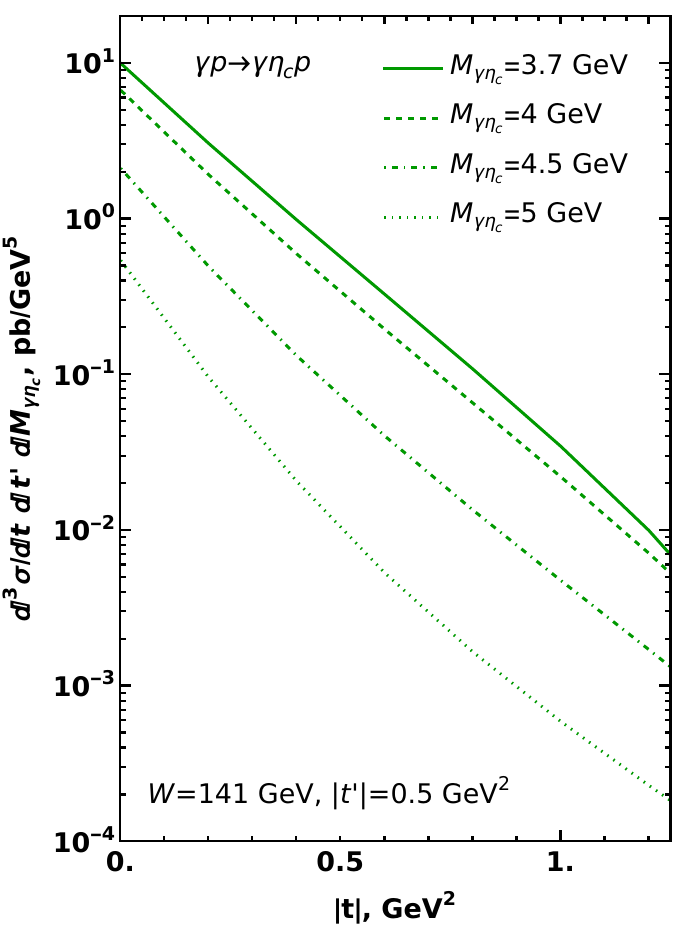}\includegraphics[width=6cm]{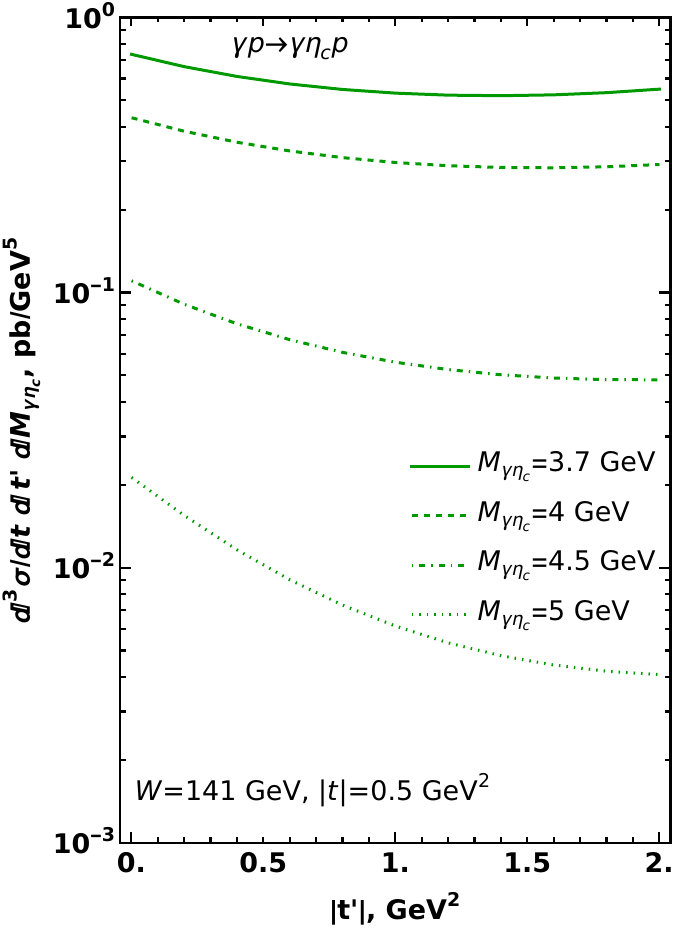}\includegraphics[width=6cm]{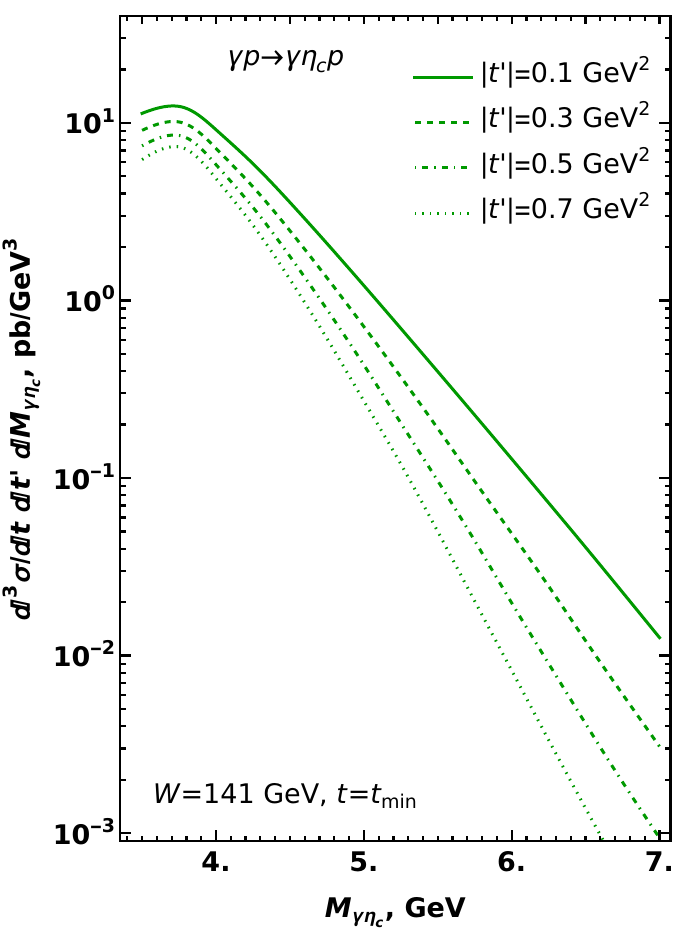}

\caption{Dependence of the photoproduction cross-section~(\ref{eq:Photo})
on the invariant momentum transfer $t$ to the target (left), on the
variable $|t'|$ defined in~~(\ref{eq:Photo}) (center), and on
the invariant masses $M_{\gamma\eta_{c}}$ (right). For other energies,
the $t,\,t'$-dependence has similar shape.}\label{fig:tDep}
\end{figure}

We would like to start the presentation of our results with discussion
of the threefold differential cross-section~(\ref{eq:Photo}) and
its dependence on various kinematic variables shown in the Figure~\ref{fig:tDep}:
the invariant momentum transfer $t$ to the target, the variable $|t'|$
defined in~~(\ref{eq:Photo}), and the invariant mass $M_{\gamma\eta_{c}}$.
In the high energy kinematics, the value of the parameter $\xi$ is
very small, for this reason we can approximate $t$ as $t\approx-\boldsymbol{\Delta}_{\perp}^{2}$.
In our evaluations, the pronounced $t$-dependence largely stems from
the $t$-dependence of the $n$-functions. While we also have $\boldsymbol{\Delta}_{\perp}$
in the momentum space coefficient functions, it always contributes
multiplied by a small factor $\left(1-2\alpha_{\eta_{c}}\right)$
and in combination with momentum $\boldsymbol{p}_{\perp}$. For this
reason, the $t$-dependence largely reflects the impact parameter
dependence of the dipole forward scattering amplitude. A sharply decreasing
$t$-dependence is common to many exclusive processes, and for the
$\eta_{c}\gamma$ photoproduction implies that photon and $\eta_{c}$
predominantly are produced with oppositely directed transverse momenta
$\boldsymbol{p}_{\eta_{c}}^{\perp},\,\boldsymbol{p}_{\eta_{c}}$.
In view of a simple and well-understood dependence on $t$, in what
follows we will tacitly assume that $|t|=|t_{{\rm min}}|$, or consider
the observables in which the dependence on $t$ is integrated out.
The dependence on the variable $t'$ at fixed $t,\,M_{\gamma\eta_{c}}$,
which appears in the central panel of the Figure~\ref{fig:tDep}
is much milder than the dependence on other variables and effectively
reflects the dependence on the variable $\alpha_{\eta_{c}}$, which
is kinematically constrained to the region $M_{\eta_{c}}^{2}/M_{\gamma\eta_{c}}^{2}\le\alpha_{\eta_{c}}\le1$.
For very small values of $t'$, as could be seen from~(\ref{eq:KinApprox}),
the variable $\boldsymbol{p}_{\perp}$ is suppressed, which leads
to increase of the cross-section. Finally, in the right panel of the
Figure~\ref{fig:tDep} we show the dependence on invariant mass $M_{\gamma\eta_{c}}$.
This variable, together with quarkonium mass $M_{\eta_{c}}$, plays
the role of the hard scale which determines the scale of all the transverse
momenta, and especially the parameter $\boldsymbol{L}$. As we can
see from~(\ref{eq:Amplitude_PP},\ref{eq:Amplitude_PM}), the amplitudes
of the process has a form of the convolution of the $n$-functions~(\ref{eq:nPhi}-\ref{eq:n1})
and the quarkonia wave functions in momentum space. The distance between
the maximums of these functions grows with $M_{\gamma\eta_{c}}$,
and in the implemented parametrizations both functions decrease exponentially
at large values of the transverse momenta in their arguments, for
this reason we obtained a pronounced \emph{exponential} suppression
of the cross-section at large $M_{\gamma\eta_{c}}$. We need to mention
that this result differs significantly from our previous findings
in the collinear factorization framework: in that approach the transverse
momentum dependence is effectively disregarded (integrated out) both
in proton and in quarkonium distributions, whereas for longitudinal
components of momenta we can no longer assume that $\xi\ll1$.

\begin{figure}
\includegraphics[width=9cm]{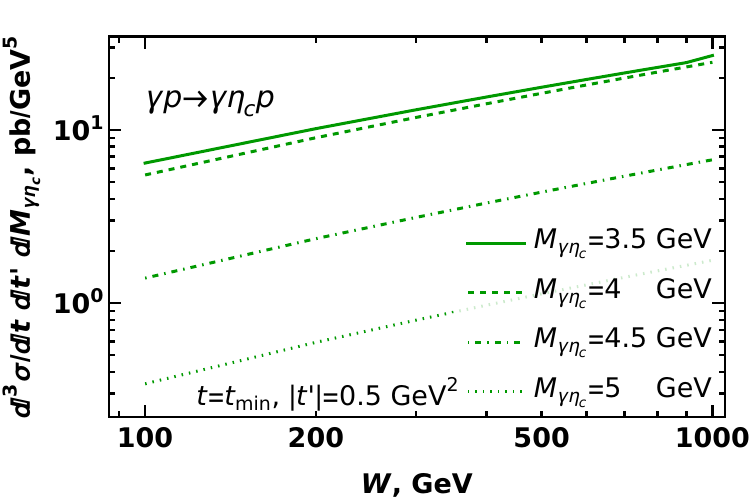}\includegraphics[width=9cm]{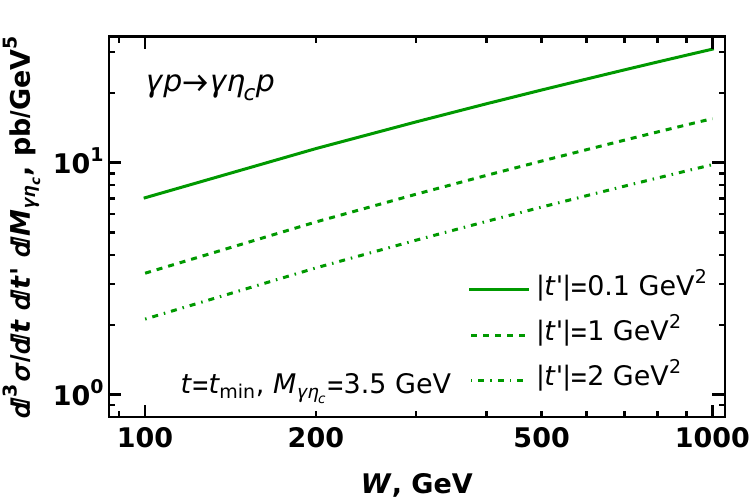}\caption{ Dependence of the cross-section on the invariant collision energy
$W$ at different values of the variable $|t'|$ (left) and invariant
mass $M_{\gamma\eta_{c}}$ (right). In both cases the energy dependence
can be approximated as $\sim W^{2\alpha}$, where the constant $\alpha$
has a mild dependence on other kinematic variables.}\label{fig:WDep}
\end{figure}

In the Figure~\ref{fig:WDep} we show the dependence of the cross-section
on the invariant energy $W$ at fixed $M_{\gamma\eta_{c}}$ and $|t'|$.
In both panels we can observe the power law growth $d\sigma(W)\sim W^{2\alpha}$,
where the constant $\alpha$ grows mildly as a function of $M_{\gamma\eta_{c}},$
and in the range which could present interest for experimental studies,
it takes values in the range (0.3,~0.35). Such behavior is a consequence
of a mild dependence of the slope of the $x$-dependence on the dipole
size in the implemented phenomenological parametrization of dipole
cross-section. The observed $W$-dependence is compatible with results
found in collinear factorization framework~\cite{Siddikov:2024blb},
though the value of the slope $\alpha$ is slightly smaller due to
different $x$-dependence in the implemented generalized parton distribution
and the dipole amplitudes.

\begin{figure}
\includegraphics[width=9cm]{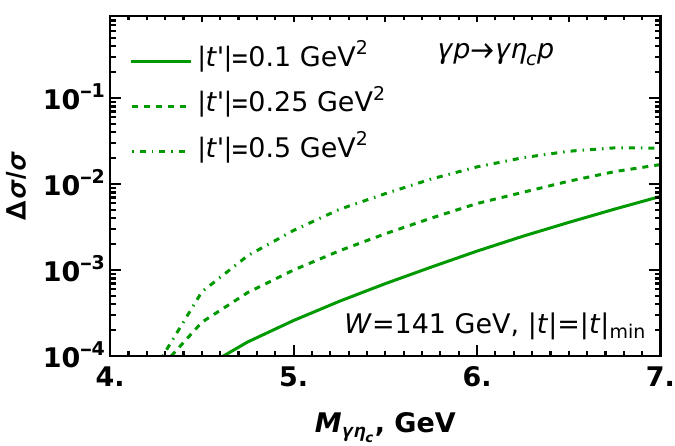}\includegraphics[width=9cm]{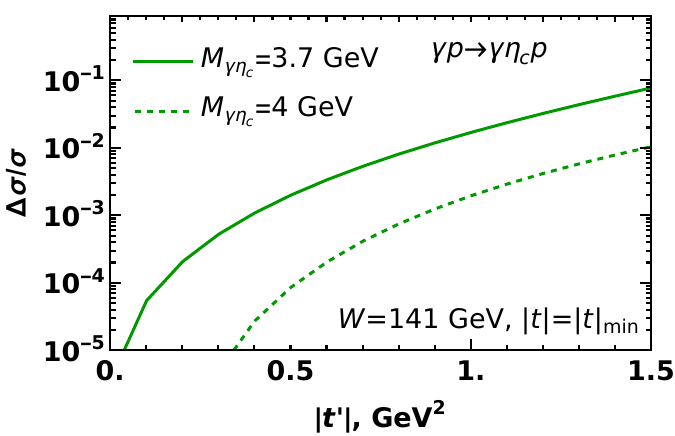}

\caption{The ratio of the cross-sections with and without photon helicity
flip, as a function of the invariant mass of the $\eta_{c}\gamma$
pair and the momentum transfer $|t'|$ to the photon. The ratio remains
small (below 10\%) in the whole range analyzed in this paper.}\label{fig:Polarized}
\end{figure}

Finally, we in the Figure~\ref{fig:Polarized} we plotted the ratio
of the cross-sections with and without the helicity flip of the final-state
photon. The ratio increases as a function of the variables $|t'|$
(momentum transfer to the photon) and $M_{\gamma\eta_{c}}$. However,
it remains small (a few per cent or below) in the whole kinematical
range where the cross-section is sufficiently large for experimental
studies. For this reason, we may safely assume that the scattered
(final-state) and incoming photons have the same polarization.

\subsection{Integrated cross-sections and counting rates}

The threefold differential cross-sections considered in the previous
section present the cleanest probes of the basic objects which are
well suited for theoretical studies. Unfortunately, it is difficult
to measure such small cross-sections because of insufficient statistics,
and for this reason now we will provide predictions for the yields
integrated over some or all kinematic variables. However, the numerical
evaluation of the cross-sections using explicit expressions~(\ref{eq:Amplitude_PP},~\ref{eq:Amplitude_PM})
for the amplitudes is relatively slow, and for this reason could even
make impossible such evaluations. Fortunately, in the heavy quark
mass limit it is possible to slightly simplify the amplitudes~(\ref{eq:Amplitude_PP},~\ref{eq:Amplitude_PM})
in order to be able to evaluate the integrated cross-sections. Precisely,
the wave function of $\eta_{c}$, which appears in~(\ref{eq:Amplitude_PP},~\ref{eq:Amplitude_PM}),
takes into account a relative motion of the quarks inside the $\eta_{c}$-charmonium,
which is formally suppressed as $\alpha_{s}\left(m_{c}\right)$ in
the heavy quark mass limit, and potentially can be disregarded altogether,
as is conventionally done in NRQCD framework. As was discussed in
detail in~\cite{Ma:2006hc,Wang:2013ywc,Wang:2017bgv}, the transverse
momentum integrated wave function (``distribution amplitude'') in
this limit can be approximated as 
\begin{equation}
\Phi_{\eta_{c}}\left(z_{\eta_{c}}\right)=\hat{\Phi}_{\eta_{c}}\left(z_{\eta_{c}}\right)\frac{\left\langle 0\left|\hat{\mathcal{O}}_{\eta_{c}}\right|\eta_{c}(p)\right\rangle }{2\sqrt{m_{c}}}\left(1+\mathcal{O}\left(v^{2}\right)\right),\label{eq:F}
\end{equation}
where $\left\langle 0\left|\hat{\mathcal{O}}_{\eta_{c}}\right|\eta_{c}(p)\right\rangle $
is the color singlet NRQCD long distance matrix element (LDME), and
\begin{equation}
\hat{\Phi}_{\eta_{c}}\left(z_{\eta_{c}}\right)\sim\delta\left(z_{\eta_{c}}-\frac{1}{2}\right)+\mathcal{O}\left(v^{2}\right).\label{eq:PhiHat}
\end{equation}
is the perturbative (partonic-level) distribution amplitude. For the
transverse momentum dependent wave function $\Phi_{\eta_{c}}\left(z_{\eta_{c}},\boldsymbol{k}_{\perp}^{({\rm rel})}\right)$
it is appealing to assume that the dependence on $\boldsymbol{k}_{\perp}^{({\rm rel})}$
is given by $\sim\delta^{2}\left(\boldsymbol{k}_{\perp}^{({\rm rel})}\right)$.
However, in this limit the functions $n_{\Phi_{\eta_{c}}},n_{\Phi_{\eta_{c}}}^{(\pm)}$
are not well-defined because, as can be seen from definitions~(\ref{eq:nPhi},\ref{eq:nPhi1}),
the dipole amplitude $\mathcal{N}\left(\boldsymbol{r},\,\,\boldsymbol{b}\right)$
has non-commuting limits $r\to\infty$ and $b\to\infty$. For this
reason we will not use heavy quark mass limit for transverse momentum
and assume that the wave function $\Phi_{\eta_{c}}$ can be approximated
as 
\begin{equation}
\Phi_{\eta_{c}}\left(z_{\eta},\,\boldsymbol{k}_{\perp}^{({\rm rel})}\right)=\delta\left(z_{\eta_{c}}-\frac{1}{2}\right)\overline{\Phi_{\eta_{c}}}\left(\boldsymbol{k}_{\perp}^{({\rm rel})}\right),
\end{equation}
where we defined 
\begin{equation}
\overline{\Phi_{\eta_{c}}}\left(\boldsymbol{k}_{\perp}^{({\rm rel})}\right)=\int_{0}^{1}dz_{\eta_{c}}\Phi_{\eta_{c}}\left(z_{\eta},\,\boldsymbol{k}_{\perp}^{({\rm rel})}\right).
\end{equation}

For the amplitudes in this approximation we may get

\begin{align}
\mathcal{A}_{\gamma p\to\eta_{c}\gamma p}^{(+,+)} & =\bar{\kappa}m\int\frac{d^{2}\ell}{(2\pi)^{2}}\,\left\{ \alpha_{\eta_{c}}\bar{\alpha}_{\eta_{c}}\overline{\Phi_{\eta_{c}}\left(\boldsymbol{P}_{(1)}\right)}\right.\times\label{eq:APP_Approx}\\
 & \quad\times\frac{-in_{0}\left(x,\,\boldsymbol{\ell},\,\,-\boldsymbol{\Delta}_{\perp}\right)\left[\left(2-\alpha_{\eta_{c}}\right)\,\boldsymbol{P}_{(1)}^{2}-4m^{2}\bar{\alpha}_{\eta_{c}}\right]+2m\left(2-\alpha_{\eta_{c}}\right)\bar{\alpha}_{\eta_{c}}n_{1}^{(+)}\left(x,\,\boldsymbol{\ell},\,\,-\boldsymbol{\Delta}_{\perp}\right)\,\left(P_{(1)x}-iP_{(1)y}\right)}{\left(\alpha_{\eta_{c}}^{2}\boldsymbol{K}_{(1)}^{2}-4\left(2-\alpha_{\eta_{c}}\right)\,\bar{\alpha}_{\eta_{c}}\boldsymbol{P}_{(1)}^{2}+4\bar{\alpha}_{\eta_{c}}^{2}m^{2}\right)}\nonumber \\
 & +\frac{16i\pi\,\alpha_{\eta_{c}}\bar{\alpha}_{\eta_{c}}n_{\Phi_{\eta_{c}}}\left(x,\,\boldsymbol{\ell},\,\,-\boldsymbol{\Delta}_{\perp},\,\frac{1}{2}\right)\left[-\alpha_{\eta_{c}}\left(\boldsymbol{P}_{(2)}^{2}+m^{2}\right)+4m^{2}\bar{\alpha}_{\eta_{c}}\right]}{\left(\boldsymbol{P}_{(2)}^{2}+m^{2}\right)\left(\left(2-\alpha_{\eta_{c}}\right)^{2}\boldsymbol{K}_{(2)}^{2}+4\bar{\alpha}_{\eta_{c}}\boldsymbol{P}_{(2)}^{2}+4\bar{\alpha}_{\eta_{c}}m^{2}\left(2-\alpha_{\eta_{c}}\right)\right)}+\nonumber \\
 & \left.+\frac{32\pi\,\alpha_{\eta_{c}}\,\bar{\alpha}_{\eta_{c}}^{2}\left(2-\alpha_{\eta_{c}}\right)n_{\Phi_{\eta_{c}}}^{(-)}\left(x,\,\boldsymbol{\ell},\,\,\boldsymbol{-\boldsymbol{\Delta}_{\perp}},\,\frac{1}{2}\right)\left(P_{(2)x}+iP_{(2)y}\right)}{\left(\boldsymbol{P}_{(2)}^{2}+m^{2}\right)\left(\left(2-\alpha_{\eta_{c}}\right)^{2}\boldsymbol{K}_{(2)}^{2}+4\bar{\alpha}_{\eta_{c}}\boldsymbol{P}_{(2)}^{2}+4\bar{\alpha}_{\eta_{c}}m^{2}\left(2-\alpha_{\eta_{c}}\right)\right)}\right\} ,\nonumber 
\end{align}
\begin{align}
\mathcal{A}_{\gamma p\to\eta_{c}\gamma p}^{(+,-)} & =\bar{\kappa}m\,\int\frac{d^{2}\ell}{(2\pi)^{2}}\,\left\{ \frac{2m\alpha_{\eta_{c}}\left(2-\alpha_{\eta_{c}}\right)\bar{\alpha}_{\eta_{c}}^{2}n_{1}^{(-)}\left(x,\,\boldsymbol{\ell},\,-\boldsymbol{\Delta}_{\perp}\right)\left(P_{(1)x}-iP_{(1)y}\right)\overline{\Phi_{\eta_{c}}\left(\boldsymbol{P}_{(1)}\right)}}{\left(\alpha_{\eta_{c}}^{2}\boldsymbol{K}_{(1)}^{2}-4\left(2-\alpha_{\eta_{c}}\right)\,\bar{\alpha}_{\eta_{c}}\boldsymbol{P}_{(1)}^{2}+4\bar{\alpha}_{\eta_{c}}^{2}m^{2}\right)}\right.+\label{eq:APM_Approx}\\
 & +\left.\frac{32\pi\alpha_{\eta_{c}}^{2}\,n_{\Phi_{\eta_{c}}}^{(-)}\left(x,\,\boldsymbol{\ell},\,-\boldsymbol{\Delta}_{\perp},\,\frac{1}{2}\right)\left(P_{(2)x}-iP_{(2)y}\right)}{\left(\boldsymbol{P}_{(2)}^{2}+m^{2}\right)\left(\left(2-\alpha_{\eta_{c}}\right)^{2}\boldsymbol{K}_{(2)}^{2}+4\bar{\alpha}_{\eta_{c}}\boldsymbol{P}_{(2)}^{2}+4\bar{\alpha}_{\eta_{c}}m^{2}\left(2-\alpha_{\eta_{c}}\right)\right)}\right\} .\nonumber 
\end{align}

The vectors $\boldsymbol{K}_{(1)},\,\boldsymbol{K}_{(2)},\,\boldsymbol{P}_{(1)},\,\boldsymbol{P}_{(2)}$
in this limit simplify as

\begin{align}
\boldsymbol{K}_{(1)} & =\frac{\boldsymbol{L}-2\boldsymbol{\ell}\alpha_{\eta_{c}}\bar{\alpha}_{\eta_{c}}}{\alpha_{\eta_{c}}\left(2-\alpha_{\eta_{c}}\right)}=\frac{\boldsymbol{p}_{\perp}}{2-\alpha_{\eta_{c}}}\frac{1}{\alpha_{\eta_{c}}}+\frac{\boldsymbol{\Delta}_{\perp}}{2-\alpha_{\eta_{c}}}\frac{1/2-\alpha_{\eta_{c}}}{\alpha_{\eta_{c}}}-\frac{2\bar{\alpha}_{\eta_{c}}}{2-\alpha_{\eta_{c}}}\boldsymbol{\ell},\\
\boldsymbol{P}_{(1)} & =\boldsymbol{\ell}+\frac{1}{2\alpha_{\eta_{c}}}\boldsymbol{L}=\boldsymbol{\ell}+\frac{\boldsymbol{p}_{\perp}}{2\alpha_{\eta_{c}}}+\frac{1-2\alpha_{\eta_{c}}}{2\alpha_{\eta_{c}}}\left(\frac{\boldsymbol{\Delta}_{\perp}}{2}\right),
\end{align}

\begin{align}
 & \boldsymbol{K}_{(2)}=\frac{2\boldsymbol{\ell}\bar{\alpha}_{\eta_{c}}+\left(2-\alpha_{\eta_{c}}\right)\boldsymbol{L}}{\alpha_{\eta_{c}}}=\frac{2\bar{\alpha}_{\eta_{c}}}{\alpha_{\eta_{c}}}\boldsymbol{\ell}+\left(2-\alpha_{\eta_{c}}\right)\frac{\boldsymbol{p}_{\perp}}{\alpha_{\eta_{c}}}-\frac{\left(2-\alpha_{\eta_{c}}\right)\left(1-2\alpha_{\eta_{c}}\right)}{2\alpha_{\eta_{c}}}\boldsymbol{\Delta}_{\perp},\\
 & \boldsymbol{P}_{(2)}=\boldsymbol{\ell}+\frac{1}{2}\boldsymbol{L}=\boldsymbol{\ell}+\frac{\boldsymbol{p}_{\perp}}{2}+\frac{1-2\alpha_{\eta_{c}}}{4}\boldsymbol{\Delta}_{\perp}.
\end{align}

\begin{figure}
\includegraphics[width=9cm]{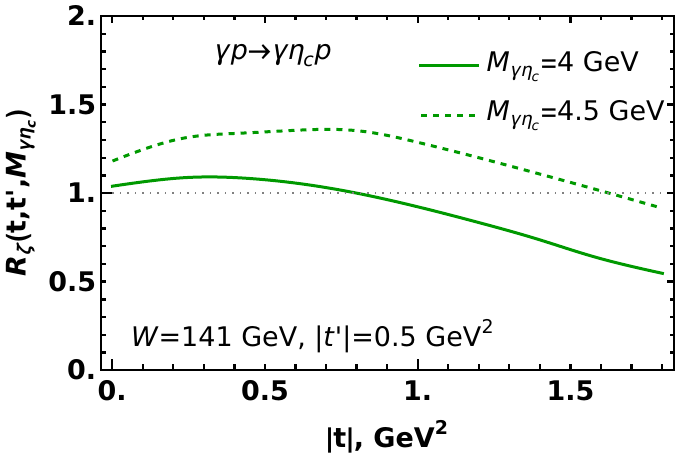}\includegraphics[width=9cm]{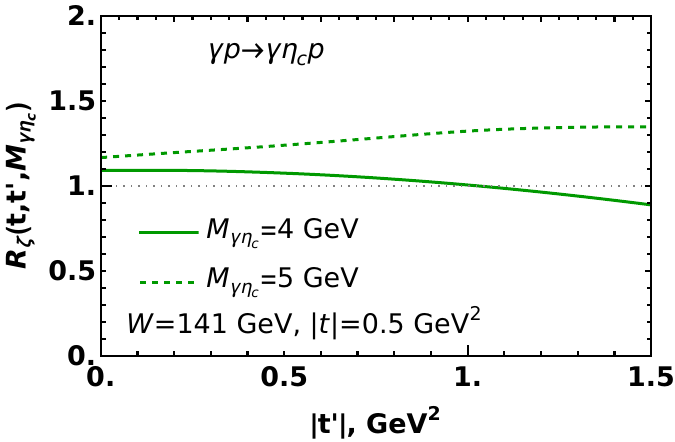}

\caption{The ratio of the threefold cross-sections evaluated with approximate
and exact expressions for the amplitudes, as defined in~(\ref{eq:RZeta}).
The ratio remains small in the kinematics which gives the dominant
contribution to the integrated cross-sections.}\label{fig:Approx}
\end{figure}

In order to estimate the precision of this approximation, in the Figure~\ref{fig:Approx}
we have shown the ratio of the threefold cross-sections evaluated
using approximate and exact expressions, 
\begin{equation}
R_{\zeta}\left(t,t',M_{\gamma\eta_{c}}\right)=\frac{\left(d^{3}\sigma/dt\,dt'\,dM_{\gamma\eta_{c}}\right)_{[{\rm Eqs}.\,\ref{eq:APP_Approx},\ref{eq:APM_Approx}]}}{\left(d^{3}\sigma/dt\,dt'\,dM_{\gamma\eta_{c}}\right)_{[{\rm Eqs}.\,\ref{eq:Amplitude_PP},\ref{eq:Amplitude_PM}]}}.\label{eq:RZeta}
\end{equation}
In agreement with expectations based on heavy quark mass limit, the
ratio remains small (below $\alpha_{s}\left(m_{c}\right)\sim1/3$)
in the kinematical range which gives the dominant contribution to
the integrated cross-sections. The ratio grows at larger $|t|,\,|t'|$,
however the cross-section is suppressed in that kinematics and gives
negligible contribution to the $t,t'$-integrated observables. For
this reason in what follows we will use this approximation for numerical
estimates.

\begin{figure}
\includegraphics[totalheight=6cm]{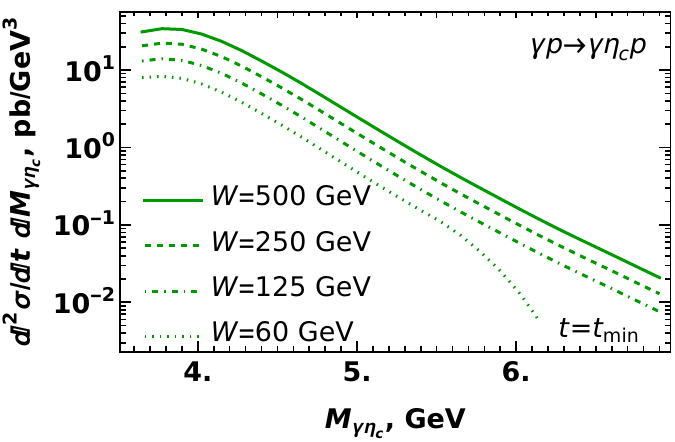}\includegraphics[totalheight=6cm]{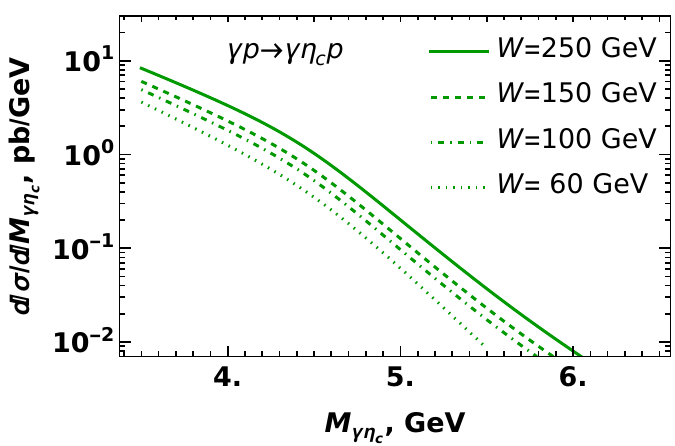}

\caption{The double- and single-differential cross-sections $d\sigma/dt\,dM_{\gamma\eta_{c}}$,
$d\sigma/dM_{\gamma\eta_{c}}$ as a function of the invariant mass
of $\eta_{c}\gamma$ pair at different collision energies $W$. For
the curve $W=60$~GeV we observe partial suppression of the phase
space due to implemented cut $x_{B}\lesssim10^{-2}$.}\label{fig:M12}
\end{figure}

In the left panel of the Figure~\ref{fig:M12} we have shown the
cross-sections $d\sigma/dt\,dM_{\gamma\eta_{c}}$ and $d\sigma/dM_{\gamma\eta_{c}}$
for different energies. As we discussed in the previous section, for
the threefold cross-section the dependence on $W,\,t,\,M_{\gamma\eta_{c}}$
largely factorizes, for this reason after integration we observe a
similar shape.

\begin{figure}
\includegraphics[width=6cm]{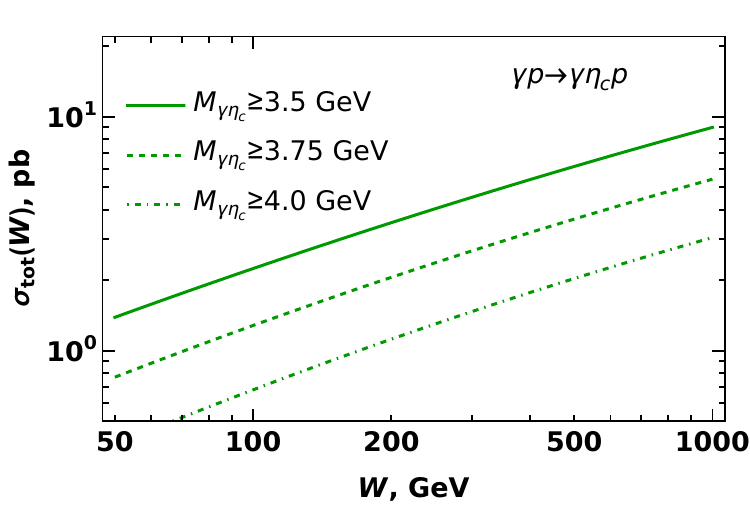}\includegraphics[width=6cm]{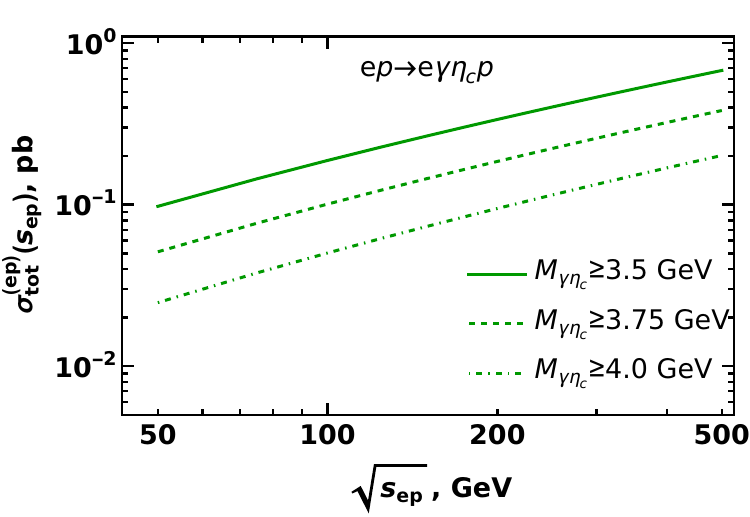}\includegraphics[width=6cm]{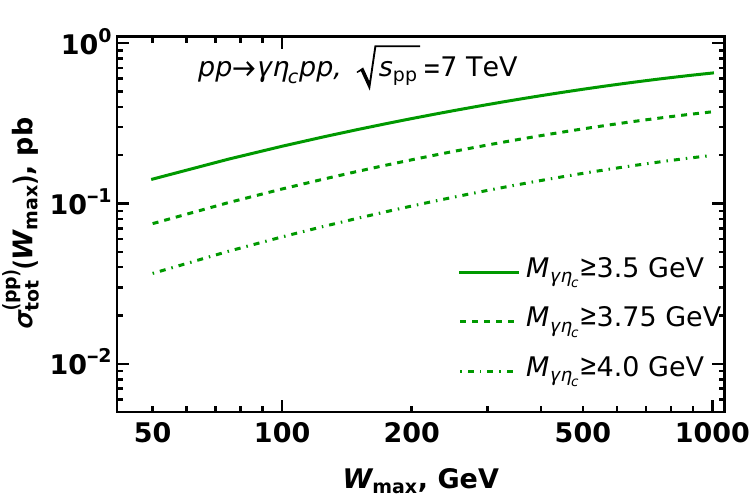}

\caption{Predictions for the total (\textquotedblleft fiducial\textquotedblright )
production cross-section of of $\eta_{c}\gamma$ in photoproduction
(left), electroproduction (center) and ultraperipheral production
in $pp$ collisions (right) as a function of energy $W$ for different
cutoffs on the invariant mass of $\eta_{c}\gamma$ pairs. For $ep$
and $pp$ collisions we integrated the flux over the photon virtuality
up to $Q_{{\rm max}}^{2}\approx1\,{\rm GeV^{2}}$. For $pp$ collisions,
we have shown the result as a function of maximal photon-proton energy
$W_{{\rm max}}$ at fixed $\sqrt{s_{pp}}$.}\label{fig:Total}
\end{figure}

In the left panel of the Figure~\ref{fig:Total} we provide the energy
dependence of the total (fully integrated) cross-section $\sigma_{{\rm tot}}(W)$
for the photoproduction of $\eta_{c}\gamma$. As we discussed earlier,
the suggested mechanism may be applied only in the kinematics where
the invariant mass $M_{\gamma\eta_{c}}$ of the photon-meson pair
is sufficiently large in order to avoid the feed-down contributions
from radiative decays of the excited quarkonia states. The choice
of the minimal value $\left(M_{\gamma\eta_{c}}\right)_{{\rm min}}$
is somewhat arbitrary, for this reason we have shown the results for
several possible cutoffs. For all cutoffs, the total cross-section
has a power law growth with energy $W$, namely 
\begin{equation}
\sigma_{{\rm tot}}\left(W,\,M_{\gamma\eta_{c}}\ge3.5\,{\rm GeV}\right)\approx2.2\,{\rm pb}\,\left(\frac{W}{100\,{\rm GeV}}\right)^{0.6},
\end{equation}
in agreement with our earlier findings in Section~\ref{subsec:diff}
for the energy dependence of the \textit{differential} cross-sections.
In order to facilitate feasibility studies of this channel, in the
central and right panel of the same Figure~\ref{fig:Total} we have
shown our estimates for the fully integrated cross-section of the
electroproduction $ep\to e\gamma\eta_{c}p$ and ultraperipheral production
$pp\to pp\gamma\eta_{c}$ as a function of the collision energy. In
these estimates we implemented an additional cut $Q^{2}\lesssim1\,{\rm GeV^{2}}$
for the virtuality of the incident photon when integrating over the
kinematics of the scattered projectile and assumed that the cross-section
remains nearly flat as a function of $Q^{2}$. This assumption is
justified in the heavy quark mass limit, because the characteristic
scale which controls transition from photoproduction to Bjorken regime
is $M_{\eta_{c}}$. Both $ep$ and $pp$ cross-sections are comparable
by magnitude; the smallness of the cross-section is due to $\sim\alpha_{{\rm em}}$
in the leptonic prefactor (see~(\ref{eq:LTSep})) and a very steep
slope of the $t$-dependence in the exclusive process. For the typical
EIC energy $\sqrt{s_{ep}}\approx100\,$GeV the corresponding cross-section
is 
\begin{equation}
\sigma_{{\rm tot}}^{({\rm ep})}\left(\sqrt{s_{ep}}=100\,{\rm GeV},\,M_{\gamma\eta_{c}}\ge3.5\,{\rm GeV}\right)\approx208\,{\rm fb}.\label{eq:XSecEIC}
\end{equation}
For the instantaneous luminosity $\mathcal{L}=10^{34}\,{\rm cm^{-2}s^{-1}}=0.864\,{\rm fb}^{-1}{\rm day}^{-1}$
at the future Electron Ion Collider~\cite{Accardi:2012qut,AbdulKhalek:2021gbh}
the cross-section~(\ref{eq:XSecEIC}) gives a production rate $dN/dt\approx$180
events/day, with approximately $N=2.1\times10^{4}$ produced $\eta_{c}\gamma$
pairs per each $\int dt\,\mathcal{L}=100\,{\rm fb^{-1}}$ of integrated
luminosity. Since the $\eta_{c}$ meson is not detected directly but
rather via its decays into light hadrons, for analysis of feasibility
it is also interesting to know the counting rate $dN_{d}/dt$ and
the total number of detected events $N_{d}$ for a chosen decay mode.
Technically, these quantities may be found multiplying $dN/dt$ and
$N$ by the branching fraction of $\eta_{c}$ to the chosen decay
mode. In experimental studies the $\eta_{c}$-meson is frequently
identified via its decays to pions and kaons, e.g.: $\eta_{c}(1S)\to K_{S}^{0}K^{+}\pi^{-}$,
for which the corresponding branching fraction is~\cite{BESIII:2019eyx,Navas:2024X}
\begin{equation}
{\rm Br}_{\eta_{c}}={\rm Br}\left(\eta_{c}(1S)\to K_{S}^{0}K^{+}\pi^{-}\right)=2.6\%.
\end{equation}
This translates into detection (counting) rate $dN_{d}/dt\approx$139
events/month, with $N_{d}=$540 detected events per $100{\rm fb^{-1}}$
of integrated luminosity.

\subsection{Comparison with $\eta_{c}$ photoproduction}

\label{subsec:odderon}For a long time the exclusive $\eta_{c}$ photoproduction
process $\gamma p\to\eta_{c}p$ has been considered as one of the
most promising channels for studies of odderons: the $C$-odd 3-gluon
exchanges in $t$-channel predicted in~\cite{Odd2,Odd3} and extensively
studied in ~\cite{Odd1,Bartels:2001hw,Odd4,Odd5,Odd6,Odd7}. While
the existence of the odderons has never been questioned, for a long
time the magnitude of the odderon-mediated processes remained largely
unknown because it is controlled by a completely new nonperturbative
amplitude (see~\cite{Odd7,Benic:2023} for a short overview). Only
recently the experimental measurements could confirm nonzero contribution
of odderons from comparison of $pp$ and $p\bar{p}$ elastic cross-sections
measured at LHC and Tevatron~\cite{OddTotem1,OddTotem2}. Since that
analysis potentially could include sizable uncertainties, the searches
of odderons shifted towards channels which require the $C$-odd $t$-channel
exchanges. The photoproduction $\gamma p\to\eta_{c}p$ is a rather
clean channel for study of odderons using perturbative methods, and
for this reason it will remain in focus of future experimental studies,
both at HL-LHC and at the future EIC.

The $\eta_{c}\gamma$ photoproduction in this context deserves a lot
of interest because it could constitute a sizable background to $\eta_{c}$
photoproduction. Indeed, the $\eta_{c}\gamma$ photoproduction does
not require small $C$-odd exhanges in $t$-channel, and the latter
fact potentially could compensate the expected $\mathcal{O}\left(\alpha_{{\rm em}}\right)$-suppression
of its cross-section. Since the acceptance of modern detectors for
photons is less than one, potentially the $\eta_{c}\gamma$ photoproduction
with undetected final-state photons could be misinterpreted as $\gamma p\to\eta_{c}p$
subprocess. Furthermore, if the emitted photon is not detected, it
becomes impossible to impose constraints (cuts) on the invariant mass
$M_{\gamma\eta_{c}}$ and for this reason potentially a sizable contribution
could be obtained from the feed-down contributions (radiative decays
of heavier charmonia). The accurate estimate of the background in
general requires detailed knowledge of the detector's geometry and
acceptance. For the sake of simplicity we will assume that \emph{all}
photons are undetected, and will discuss the cross-section $d\sigma/dt$,
integrating over the phase space of the produced photon. Such approach
provides an upper estimate for the background.

\begin{figure}
\includegraphics[width=9cm]{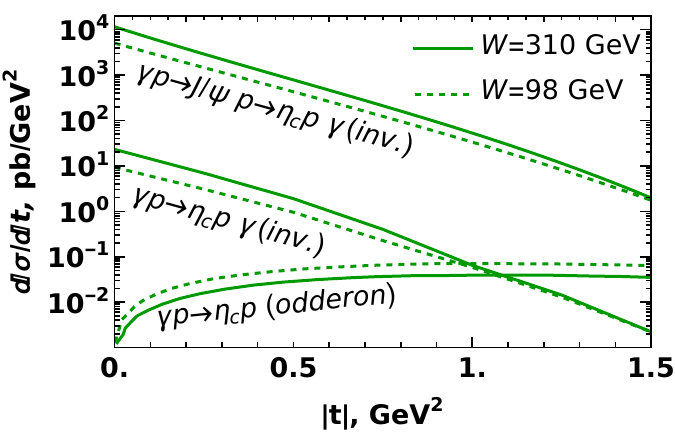}\caption{ Comparison of the $\eta_{c}$ photoproduction cross-sections via
different mechanisms: upper curve corresponds to production via radiative
decay of $J/\psi$ meson (this mechanism dominates in absence of cuts
for invariant mass $M_{\gamma\eta_{c}}$), middle curve corresponds
to $\gamma p\to\eta_{c}\gamma p$ mechanism discussed in this paper,
and lower curve corresponds to odderon-mediated photoproduction $\gamma p\to\eta_{c}p$,
whose cross-section is taken from~\cite{Odd7,Benic:2023} (curves
$x=10^{-2}$ and $x=10^{-3}$, the values of $W$ are restored assuming
the relation $x=M_{\eta_{c}}^{2}/W^{2}$).}\label{fig:Odd}
\end{figure}

In the Figure~(\ref{fig:Odd}) we compare the cross-sections of the
$\eta_{c}$ and $\eta_{c}\gamma$ with undetected (integrated out)
photon in the final state. We can see that the largest contribution
comes from the mechanism mediated by radiative decay $J/\psi\to\eta_{c}\gamma$
decay, which we estimated using the CGC (color dipole) framework~\cite{RESH}
and the branching ratio ${\rm Br}\left(J/\psi\to\eta_{c}\gamma\right)=1.7\pm0.4\,\%$~\cite{Navas:2024X}.
At present we can't estimate the feed-down contributions from other
excited quarkonia due to significant uncertainties in their wave functions,
however we believe that these contributions are significantly smaller
due to smallness of their production cross-sections and smaller branchings.

The suggested mechanism of non-resonant production of $\eta_{c}\gamma$
is the second largest and exceeds significantly the contribution of
the odderon-mediated $\eta_{c}$ production in the kinematics of small-$t$
($|t|\lesssim1\,{\rm GeV}^{2}$). However, this contribution is suppressed
exponentially at higher $t$ as explained in previous section. The
cross-section of $\eta_{c}\gamma$ grows faster with energy and thus
eventually will exceed the $\eta_{c}$ cross-section at any $t$~\footnote{For comparison with detection rates from the previous section, the
odderon-mediated exclusive photoproduction of $\eta_{c}$ yields detection
rates $dN_{d}/dt=$93 and 52 events/month at energies $W=98$ GeV
and $W=310$ GeV respectively.}. In BFKL language, the difference of energy dependencies of $\eta_{c}\gamma$
and $\eta_{c}$ is a consequence of the fact that pomeron has larger
intercept than odderon.

The existence of other mechanisms which can supersede the odderon
contribution in the small-$t$ kinematics has been discussed earlier
in~\cite{Benic:2023}. However, the corrections discussed in that
paper are significantly less important for processes on neutrons,
namely for $\gamma n\to\eta_{c}n$ subprocess. In contrast, the $\eta_{c}\gamma$
production with undetected photon gives the same contribution for
proton and neutron targets.

We need to mention that potentially it is possible to separate different
production mechanisms mentioned in this section, even when the final
photon is not detected, imposing appropriate cuts on the variable
$\left(q-\Delta\right)^{2}$, where $q$ is the momentum of the \textit{incoming}
photon, and $\Delta$ is the momentum transfer to the recoil proton.
This variable corresponds to $M_{\eta_{c}}^{2}$ in case of odderon-mediated
production, equals $M_{J/\psi}^{2}$ in case of $\eta_{c}\gamma$
production via radiative decays of $J/\psi$, and coincides with invariant
mass $M_{\gamma\eta_{c}}^{2}$ of $\eta_{c}\gamma$ pair in case of
non-resonant $\eta_{c}\gamma$ production. However, separation of
the odderon contribution using this method would require measurements
of the momenta of recoil proton and scattered electron with outstanding
precision.

\section{Conclusions}

\label{sec:Conclusions}In this manuscript we studied the exclusive
$\eta_{c}\gamma$ photoproduction in the Color Glass Condensate framework.
We found that the amplitude of the process may be represented in terms
of the forward dipole scattering amplitude, and a target-independent
partonic amplitude. The evaluation simplifies drastically in the momentum
space in the heavy quark mass limit: the amplitude of the process
becomes closely related to Fourier images of the dipole scattering
amplitude (multiplied by known functions), and in this way can be
used for phenomenological analysis of this fundamental nonperturbative
object. We also made numerical estimates in the kinematics of small-$x_{B}\ll1$
which can be studied in the kinematics of the the ongoing experiments
at LHC in ultraperipheral kinematics and the future EIC. In the region
of moderate energies ($W\sim100\,{\rm GeV},$ $x_{B}\sim10^{-2}$)
we found that the evaluations in dipole approach agree (up to a factor
of two) with similar evaluations in the collinear factorization framework
from~\cite{Siddikov:2024blb} and are comparable by order of magnitude
for other $2\to3$ channels discussed in the literature~\cite{GPD2x3:9,GPD2x3:8,Siddikov:2024blb,GPD2x3:7,GPD2x3:6,GPD2x3:5,GPD2x3:4,GPD2x3:3,GPD2x3:2,GPD2x3:1,Duplancic:2022wqn,Qiu:2024mny,Qiu:2023mrm,Deja:2023ahc,Siddikov:2022bku,Siddikov:2023qbd}.
We also analyzed feasibility to study experimentally this channel
in the kinematics of ongoing ultraperipheral collisions experiments
at LHC and the highest-energy $ep$ collisions at future Electron
Ion Collider. We found that in both setups this channel gives the
production rate of a few thousands of $\eta_{c}\gamma$ pairs and
detection rate of a few hundreds of events per each 100 ${\rm fb}^{-1}$
of integrated luminosity, assuming that $\eta_{c}$ is detected via
its decays to $\eta_{c}(1S)\to K_{S}^{0}K^{+}\pi^{-}$. We also observed
that the suggested channel (together with radiative decays of $J/\psi\,\to\eta_{c}\gamma$)
gives a large background to exclusive production of $\eta_{c}$ mesons,
which is being considered as a golden channel for study of the odderons,
and gives the dominant contribution in the kinematics of small momenta
transfer $|t|$.

\section*{Acknowledgments}

We thank our colleagues at UTFSM university for encouraging discussions.
This research was partially supported by Proyecto ANID PIA/APOYO AFB220004
(Chile), and ANID grants Fondecyt Regular \textnumero 1220242 and
Fondecyt Postdoctoral \textnumero 3230699. I. Zemlyakov also acknowledges
support of the scholarship of Direccion de Postgrado, Universidad
Tecnica Federico Santa Maria. \textquotedbl Powered@NLHPC: This research
was partially supported by the supercomputing infrastructure of the
NLHPC (ECM-02)\textquotedbl .

\appendix

\section{Wave functions and their overlaps}

\label{sec:WFsAndOverlaps}As we discussed in Section~\ref{subsec:Derivation},
the amplitude of the $\eta_{c}\gamma$ photoproduction can be represented
in terms of convolutions which include the wave function of $c\bar{c}\gamma$
Fock state of the photon (amplitude of the $\gamma\to\bar{c}c\gamma$
subprocess) or the light-cone amplitude of the $\eta_{c}\gamma\to\bar{c}c$
subprocess in the mixed (light-cone) notations. For the former, the
leading order contribution requires evaluation of the diagrams shown
in the Figure~\ref{fig:Diags-2}. This evaluation has been discussed
in detail in~\cite{Lappi:2016oup,Hanninen:2017ddy,Beuf:2021qqa,Beuf:2022ndu}
for the $\gamma\to\bar{c}cg$ subprocess (see e.g. the diagrams $j,k,\ell,m$
in~~\cite{Beuf:2022ndu}) and can be trivially extended to $\gamma\to c\bar{c}\gamma$
adjusting the color factors.

\begin{figure}
\includegraphics[width=6cm]{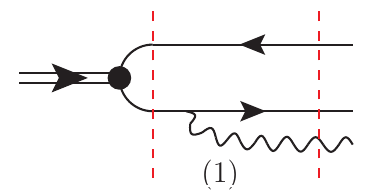}\includegraphics[width=6cm]{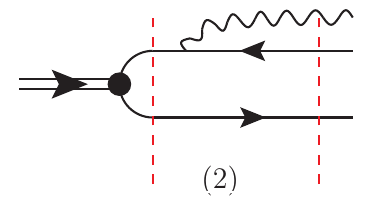}

\includegraphics[width=6cm]{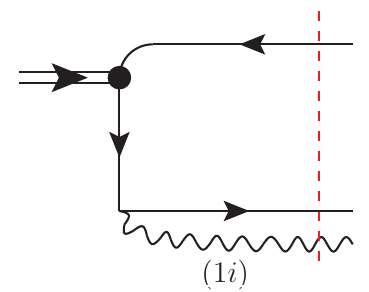}\includegraphics[width=6cm]{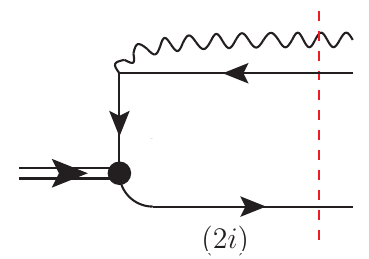}

\caption{The leading order diagrams which correspond to the $\eta_{c}\to\gamma q\bar{q}$
subprocess. As explained in the text, they are related to the amplitudes
of the quarkonium photodissociation subprocess $\gamma\eta_{c}\to q\bar{q}$
shown in the Figure~\ref{fig:Diags} only by change of a sign of
the photon momentum $k_{\gamma}^{\mu}$. The vertical dashed lines
denote the light-cone denominators of the corresponding wave functions
in momentum space.}\label{fig:Diags-1}
\end{figure}

The evaluation of the amplitude of the subprocess $\eta_{c}\gamma\to\bar{c}c$
can be reduced to a form which closely resembles the evaluation of
the above-mentioned $\gamma\to c\bar{c}\gamma$ subprocess. Indeed,
in the leading order of the light-cone perturbation theory, due to
crossing symmetry the amplitudes of $\eta_{c}\gamma\to\bar{c}c$ and
$\eta_{c}\to\gamma\bar{c}c$ are related to each other by inversion
of the photon's 4-momentum (with proper adjustment of its polarization
vector), as can be seen from the Figure~\ref{fig:Diags-1}. On the
other hand, the evaluation of the $\eta_{c}\to\gamma\bar{c}c$ amplitude
technically is very similar to $\gamma\to c\bar{c}\gamma$ and differs
only in minor details. For this reason, we will focus on evaluation
of the former process, and at the end of this Appendix we'll briefly
discuss the adjustments which are required in order to reproduce the
results for $\gamma\to c\bar{c}\gamma$.

In what follows we will use the standard rules of the light-cone perturbation
theory~\cite{Lepage:1980fj,Brodsky:1997de}. Since the dominant Fock
state in $|\eta_{c}\rangle$ is the color singlet pseudoscalar $\bar{Q}Q$,
conventionally it is assumed that the coupling of $\eta_{c}$ to heavy
quarks is given by~\cite{Benic:2023,Dosch:1996ss} 
\begin{equation}
\hat{V}_{\eta_{c}}^{(h,\bar{h})}\left(z_{\eta_{c}},\,\boldsymbol{k}_{\perp}^{({\rm rel})}\right)=-i\phi_{\eta_{c}}\left(z_{\eta_{c}},\,\boldsymbol{k}_{\perp}^{({\rm rel})}\right)\frac{\bar{u}_{h}\left(z_{\eta_{c}},\,\boldsymbol{k}_{\perp}^{({\rm rel})}\right)}{\sqrt{z_{\eta_{c}}}}\gamma_{5}\frac{v_{\bar{h}}\left(1-z_{\eta_{c}},\,-\boldsymbol{k}_{\perp}^{({\rm rel})}\right)}{\sqrt{1-z_{\eta_{c}}}},\label{eq:VEta}
\end{equation}
where $z_{\eta_{c}}$ is the light-cone fraction carried by the quark,
$\boldsymbol{k}_{\perp}^{({\rm rel})}$ is the relative transverse
momentum, and the wave function $\phi_{\eta_{c}}\left(z_{\eta_{c}},\,\boldsymbol{k}_{\perp}^{({\rm rel})}\right)$
is essentially nonperturbative object which can be fixed either from
potential models or from phenomenological fits. The function $\phi_{\eta_{c}}\left(z_{\eta},\,\boldsymbol{k}_{\perp}^{({\rm rel})}\right)$
is related to the light-cone wave function $\Phi_{\eta_{c}}\left(z_{\eta_{c}},\,\boldsymbol{r}_{\perp}\right)$
by a Fourier transform over the transverse moments.

In what follows we assume that the momenta of outgoing $c,\bar{c},\gamma$
are given by 
\begin{equation}
k_{c}^{\mu}=\left(k_{0}^{+},\,\frac{\boldsymbol{k}_{0}^{2}+m_{c}^{2}}{2k_{0}^{+}},\,\boldsymbol{k}_{0}\right),\quad k_{\bar{c}}^{\mu}=\left(k_{1}^{+},\frac{\boldsymbol{k}_{1}^{2}+m_{c}^{2}}{2k_{1}^{+}},\,\boldsymbol{k}_{1}\right),\qquad k_{\gamma}^{\mu}\equiv\left(k_{2}^{+},\frac{\boldsymbol{k}_{2}^{2}}{2k_{2}^{+}},\,\boldsymbol{k}_{2}\right),
\end{equation}
respectively. The 4-momentum of $\eta_{c}$ is given by 
\begin{align}
P_{\eta_{c}}^{\mu} & =\left(p_{\eta_{c}}^{+},\,\,\frac{\boldsymbol{p}_{\eta_{c}}+m_{\eta_{c}}^{2}}{2p_{\eta_{c}}^{+}}\,\,,\boldsymbol{p}_{\perp}^{\eta_{c}}\right),\qquad p_{\eta_{c}}^{+}=k_{0}^{+}+k_{1}^{+}+k_{2}^{+},\qquad\boldsymbol{p}_{\perp}^{\eta_{c}}=\boldsymbol{k}_{0}+\boldsymbol{k}_{1}+\boldsymbol{k}_{2},
\end{align}
The evaluation of the diagram 1 in the first row of the Figure~\ref{fig:Diags-1}
yields 
\begin{align}
\mathcal{\psi}_{1}\left(Z_{0},\,\boldsymbol{r}_{0},\,Z_{1},\,\boldsymbol{r}_{1},\,Z_{2},\,\boldsymbol{r}_{2}\right) & =\int\prod_{a=1}^{3}\frac{d^{2}k_{a}}{\left(2\pi\right)^{2}}\exp\left(i\sum_{a=1}^{3}\boldsymbol{k}_{a}\cdot\boldsymbol{r}_{a}\right)\delta\left(\boldsymbol{k}_{0}+\boldsymbol{k}_{1}+\boldsymbol{k}_{2}-\boldsymbol{p}_{\perp}^{\eta_{c}}\right)\times\label{eq:WF}\\
 & \times\sum_{\mathfrak{h}}\frac{\bar{u}_{h}\left(Z_{0},\,\boldsymbol{k}_{0}\right)\hat{\varepsilon}_{\gamma}\left(k_{2}\right)u_{\mathfrak{h}}\left(Z_{0'},\,\boldsymbol{k}_{0'}\right)\hat{V}_{\eta_{c}}^{(\mathfrak{h},\,\bar{h})}\left(z_{\eta_{c}},\,\boldsymbol{k}_{\perp}^{({\rm rel})}\right)}{2k_{0'}^{+}\,D_{11}D_{12}},\nonumber 
\end{align}
where $Z_{0'}=Z_{0}+Z_{2}$, $\boldsymbol{k}_{0'}=\boldsymbol{k}_{0}+\boldsymbol{k}_{2}$,
the factor $2k_{0'}^{+}$ in denominator stems from the conventional
light-cone rules for (internal) quark-gluon vertices, and the energy
denominators $D_{11},\,D_{12}$ are given by~\footnote{In what follows we consider that the first subindex is diagram number,
and the second one enumerates the number of cut} 
\begin{align}
D_{11} & =\frac{\left(\boldsymbol{k}_{0}+\boldsymbol{k}_{1}+\boldsymbol{k}_{2}\right)^{2}+m_{\eta_{c}}^{2}}{2\left(k_{0}^{+}+k_{1}^{+}+k_{2}^{+}\right)}-\frac{\left(\boldsymbol{k}_{0}+\boldsymbol{k}_{2}\right)^{2}+m_{c}^{2}}{2\left(k_{0}^{+}+k_{2}^{+}\right)}-\frac{\left(\boldsymbol{k}_{1}\right)^{2}+m_{c}^{2}}{2k_{1}^{+}}=-\frac{\boldsymbol{P}^{2}+m^{2}\left(1-2Z_{1}\right)^{2}}{p_{\eta_{c}}^{+}Z_{1}\bar{Z}_{1}},\label{eq:D11}\\
D_{12} & =\frac{\left(\boldsymbol{k}_{0}+\boldsymbol{k}_{1}+\boldsymbol{k}_{2}\right)^{2}+m_{\eta_{c}}^{2}}{2\left(k_{0}^{+}+k_{1}^{+}+k_{2}^{+}\right)}-\frac{\left(\boldsymbol{k}_{0}\right)^{2}+m_{c}^{2}}{2k_{0}^{+}}-\frac{\left(\boldsymbol{k}_{2}\right)^{2}}{2k_{2}^{+}}-\frac{\left(\boldsymbol{k}_{1}\right)^{2}+m_{c}^{2}}{2k_{1}^{+}}=-\frac{\boldsymbol{P}^{2}}{p_{\eta_{c}}^{+}Z_{1}\bar{Z}_{1}}-\frac{\boldsymbol{K}^{2}\bar{Z}_{1}}{p_{\eta_{c}}^{+}Z_{0}Z_{2}}+\frac{m^{2}\left(Z_{2}+4Z_{0}Z_{1}-1\right)}{p_{\eta_{c}}^{+}Z_{0}Z_{1}}.\label{eq:D12}
\end{align}
The vectors $\boldsymbol{P},\,\boldsymbol{K}$ in~(\ref{eq:D11},\ref{eq:D12})
are some linear combinations of vectors $\boldsymbol{k}_{i}$ defined
as 
\begin{align}
\boldsymbol{P} & =-\boldsymbol{k}_{1}+Z_{1}\boldsymbol{p}_{\perp}^{\eta_{c}}=Z_{1}\left(\boldsymbol{k}_{0}+\boldsymbol{k}_{2}\right)-\left(1-Z_{1}\right)\boldsymbol{k}_{1},\label{eq:P1Def}\\
\boldsymbol{K} & =\frac{Z_{0}}{Z_{0}+Z_{2}}\left(\boldsymbol{k}_{2}-\frac{Z_{2}}{Z_{0}}\boldsymbol{k}_{0}\right)=\frac{Z_{0}\boldsymbol{k}_{2}-Z_{2}\boldsymbol{k}_{0}}{Z_{0}+Z_{2}}.\label{eq:K1Def}
\end{align}
Physically, these vectors can be interpreted as relative transverse
momentum of the $\bar{c}c$ pair inside $\eta_{c}$ and the relative
transverse momentum of the quark and photon after the emission. The
definitions~(\ref{eq:P1Def}, \ref{eq:K1Def}) can be inverted and
rewritten in the form 
\begin{equation}
\boldsymbol{k}_{0}=-\boldsymbol{K}+\frac{Z_{0}}{Z_{0}+Z_{2}}\left(\boldsymbol{P}+\bar{z}_{1}\boldsymbol{p}_{\perp}^{\eta_{c}}\right),\quad\boldsymbol{k}_{1}=-\boldsymbol{P}+Z_{1}\boldsymbol{p}_{\perp}^{\eta_{c}},\quad\boldsymbol{k}_{2}=\boldsymbol{K}+\frac{Z_{2}}{Z_{0}+Z_{2}}\left(\boldsymbol{P}+\bar{Z}_{1}\boldsymbol{p}_{\perp}^{\eta_{c}}\right).
\end{equation}
The argument of the exponent in the first line of~(\ref{eq:WF})
in terms of these vectors can be rewritten as 
\begin{equation}
\sum_{a=1}^{3}\boldsymbol{k}_{a}\cdot\boldsymbol{r}_{a}=\boldsymbol{K}\cdot(\boldsymbol{r}_{2}-\boldsymbol{r}_{0})+\boldsymbol{P}\cdot\left(\frac{Z_{0}\boldsymbol{r}_{0}+Z_{2}\boldsymbol{r}_{2}}{Z_{0}+Z_{2}}-\boldsymbol{r}_{1}\right)+\boldsymbol{q}\cdot\left(Z_{0}\boldsymbol{r}_{0}+Z_{1}\boldsymbol{r}_{1}+Z_{2}\boldsymbol{r}_{2}\right),\label{eq:phase}
\end{equation}
which shows that $\boldsymbol{K}$ is a Fourier conjugate to the distance
between the photon and the quark after emission, and $\boldsymbol{P}$
is a Fourier conjugate of the distance between the antiquark and the
quark before emission (the center of mass of quark-photon system).
If we treat $Z_{0},Z_{1},Z_{2}$ as light-cone masses (as was suggested
in~\cite{Bjorken:1970ah}), then we may see that effectively constants
in front of $\boldsymbol{K},\,\boldsymbol{P},\,\boldsymbol{q}$ in~(\ref{eq:phase})
coincide with the so-called Jacobi coordinates for the three-body
problem.

The numerator in~(\ref{eq:WF}) can be evaluated using conventional
(anti)commutation rules for Dirac matrices, as was done in~\cite{Lappi:2016oup,Hanninen:2017ddy,Beuf:2021qqa,Beuf:2022ndu}
for $\gamma\to\gamma\bar{c}c$, however this leads to a lengthy and
tedious procedure. For this reason, we performed the evaluations with
\emph{Mathematica}, using explicit form of the spinors and Dirac matrices
from~\cite{Lepage:1980fj,Brodsky:1997de} and contracting them directly
in helicity basis. The final result has a remarkably simple form and
is given by contributions proportional to $I_{(1)},\,I_{(1)}^{(\pm)},\hat{I}_{(1)}^{(\pm)},\,I_{(1)}^{(\pm,\pm)}$
in~(\ref{eq:PsiQQEtac}). The instantaneous contribution shown in
the diagram ($1i$) of the Figure~~\ref{fig:Diags-1} requires to
replace the internal spinors and summation over the internal helicity
index $\mathfrak{h}$ with instantaneous quark propagator $\gamma^{+}/2k_{0'}^{+}$,
viz: 
\begin{align}
\mathcal{\psi}_{1i}\left(Z_{0},\,\boldsymbol{r}_{0},\,Z_{1},\,\boldsymbol{r}_{1},\,Z_{2},\,\boldsymbol{r}_{2}\right) & =\int\prod_{a=1}^{3}\frac{d^{2}k_{a}}{\left(2\pi\right)^{2}}\exp\left(i\sum_{a=1}^{3}\boldsymbol{k}_{a}\cdot\boldsymbol{r}_{a}\right)\delta\left(\boldsymbol{k}_{0}+\boldsymbol{k}_{1}+\boldsymbol{k}_{2}-\boldsymbol{p}_{\perp}^{\eta_{c}}\right)\times\label{eq:WF-2}\\
 & \times\frac{\bar{u}_{h}\left(Z_{0},\,\boldsymbol{k}_{0}\right)\hat{\varepsilon}_{\gamma}\left(k_{2}\right)\gamma^{+}\gamma_{5}v_{h}\left(Z_{1},\,\boldsymbol{k}_{1}\right)\phi_{\eta_{c}}\left(z_{\eta_{c}},\,\boldsymbol{k}_{\perp}^{({\rm rel})}\right)}{2k_{0'}^{+}D_{12}}.\nonumber 
\end{align}
Afterwards, repeating literally the steps used for evaluation of the
diagram (1), we may obtain the contributions given by $J_{(1)}$ in~(\ref{eq:PsiQQEtac}).
The evaluation of the diagram (2) and instantaneous correction shown
in diagram ($2i$) in the Figure~\ref{fig:Diags-1} largely follows
the same procedure and yields 
\begin{align}
\mathcal{\psi}_{2}\left(Z_{0},\,\boldsymbol{r}_{0},\,Z_{1},\,\boldsymbol{r}_{1},\,Z_{2},\,\boldsymbol{r}_{2}\right) & =\int\prod_{a=1}^{3}\frac{d^{2}k_{a}}{\left(2\pi\right)^{2}}\exp\left(i\sum_{a=1}^{3}\boldsymbol{k}_{a}\cdot\boldsymbol{r}_{a}\right)\delta\left(\boldsymbol{k}_{0}+\boldsymbol{k}_{1}+\boldsymbol{k}_{2}-\boldsymbol{p}_{\perp}^{\eta_{c}}\right)\times\label{eq:WF-1}\\
 & \times\sum_{\mathfrak{h}}\frac{\hat{V}_{\eta_{c}}^{(h,\,\mathfrak{h})}\left(z_{\eta_{c}},\,\boldsymbol{k}_{\perp}^{({\rm rel})}\right)\bar{v}_{\mathfrak{h}}\left(Z_{1'},\,\boldsymbol{k}_{1'}\right)\hat{\varepsilon}_{\gamma}\left(k_{2}\right)v_{h}\left(Z_{0'},\,\boldsymbol{k}_{0'}\right)}{2k_{1'}^{+}\,D_{21}D_{22}},\nonumber 
\end{align}
\begin{align}
\mathcal{\psi}_{2i}\left(Z_{0},\,\boldsymbol{r}_{0},\,Z_{1},\,\boldsymbol{r}_{1},\,Z_{2},\,\boldsymbol{r}_{2}\right) & =\int\prod_{a=1}^{3}\frac{d^{2}k_{a}}{\left(2\pi\right)^{2}}\exp\left(i\sum_{a=1}^{3}\boldsymbol{k}_{a}\cdot\boldsymbol{r}_{a}\right)\delta\left(\boldsymbol{k}_{0}+\boldsymbol{k}_{1}+\boldsymbol{k}_{2}-\boldsymbol{p}_{\perp}^{\eta_{c}}\right)\times\label{eq:WF-1-1}\\
 & \times\frac{\bar{u}_{h}\left(Z_{0},\,\boldsymbol{k}_{0}\right)\gamma_{5}\gamma^{+}\hat{\varepsilon}_{\gamma}\left(k_{2}\right)v_{\bar{h}}\left(Z_{1},\,\boldsymbol{k}_{1}\right)\phi_{\eta_{c}}\left(z_{\eta_{c}},\,\boldsymbol{k}_{\perp}^{({\rm rel})}\right)}{2k_{1'}^{+}\,D_{22}},\nonumber 
\end{align}
where $Z_{1'}=Z_{1}+Z_{2}$, $\boldsymbol{k}_{1'}=\boldsymbol{k}_{1}+\boldsymbol{k}_{2}$,
and the light-cone energy denominators $D_{21},\,D_{22}$ are given
by 
\begin{align}
D_{21} & =\frac{\left(\boldsymbol{k}_{0}+\boldsymbol{k}_{1}+\boldsymbol{k}_{2}\right)^{2}+m_{\eta_{c}}^{2}}{2\left(k_{0}^{+}+k_{1}^{+}+k_{2}^{+}\right)}-\frac{\left(\boldsymbol{k}_{1}+\boldsymbol{k}_{2}\right)^{2}+m_{c}^{2}}{2\left(k_{1}^{+}+k_{2}^{+}\right)}-\frac{\left(\boldsymbol{k}_{0}\right)^{2}+m_{c}^{2}}{2k_{0}^{+}}=-\frac{\boldsymbol{P}^{2}+m^{2}\left(1-2Z_{0}\right)^{2}}{2p_{\eta_{c}}^{+}Z_{0}\bar{Z}_{0}},\label{eq:D21}\\
D_{22} & =\frac{\left(\boldsymbol{k}_{0}+\boldsymbol{k}_{1}+\boldsymbol{k}_{2}\right)^{2}+m_{\eta_{c}}^{2}}{2\left(k_{0}^{+}+k_{1}^{+}+k_{2}^{+}\right)}-\frac{\left(\boldsymbol{k}_{1}\right)^{2}+m_{c}^{2}}{2k_{1}^{+}}-\frac{\left(\boldsymbol{k}_{2}\right)^{2}}{2k_{2}^{+}}-\frac{\left(\boldsymbol{k}_{0}\right)^{2}+m_{c}^{2}}{2k_{0}^{+}}=-\frac{\boldsymbol{P}^{2}}{p_{\eta_{c}}^{+}Z_{0}\bar{Z}_{0}}-\frac{\boldsymbol{K}^{2}\bar{Z}_{0}}{p_{\eta_{c}}^{+}Z_{1}Z_{2}}+\frac{m^{2}\left(Z_{2}+4Z_{0}Z_{1}-1\right)}{p_{\eta_{c}}^{+}Z_{0}Z_{1}},\label{eq:D22}
\end{align}
The vectors $\boldsymbol{P},\,\boldsymbol{K}$ in~(\ref{eq:D21},~\ref{eq:D22})
are defined as linear combinations of vectors $\boldsymbol{k}_{i}$
\begin{align}
\boldsymbol{P} & =\boldsymbol{k}_{0}-Z_{0}\boldsymbol{p}_{\perp}^{\eta_{c}}=\left(1-Z_{0}\right)\boldsymbol{k}_{0}-Z_{0}\left(\boldsymbol{k}_{1}+\boldsymbol{k}_{2}\right),\label{eq:P1Def-1}\\
\boldsymbol{K} & =\frac{Z_{1}}{Z_{1}+Z_{2}}\left(\boldsymbol{k}_{2}-\frac{Z_{2}}{Z_{1}}\boldsymbol{k}_{1}\right)=\frac{Z_{1}\boldsymbol{k}_{2}-Z_{2}\boldsymbol{k}_{1}}{Z_{1}+Z_{2}}.\label{eq:K1Def-1}
\end{align}
Similar to the previous case, these vectors can be interpreted as
relative transverse momentum of the $\bar{c}c$ pair inside $\eta_{c}$
and the relative transverse momentum of the antiquark and photon after
the emission. The definitions~(\ref{eq:P1Def-1}, \ref{eq:K1Def-1})
can be inverted and rewritten in the form 
\begin{equation}
\boldsymbol{k}_{0}=\boldsymbol{P}+Z_{0}\boldsymbol{p}_{\perp}^{\eta_{c}},\quad\boldsymbol{k}_{1}=-\boldsymbol{K}-\frac{Z_{1}}{Z_{1}+Z_{2}}\left(\boldsymbol{P}-\bar{Z}_{0}\boldsymbol{p}_{\perp}^{\eta_{c}}\right),\quad\boldsymbol{k}_{2}=\boldsymbol{K}-\frac{Z_{2}}{Z_{1}+Z_{2}}\left(\boldsymbol{P}-\bar{Z}_{0}\boldsymbol{p}_{\perp}^{\eta_{c}}\right),
\end{equation}
which allows to rewrite the argument of the exponent in~(\ref{eq:WF-1})
as

\begin{equation}
\sum_{a=1}^{3}\boldsymbol{k}_{a}\cdot\boldsymbol{r}_{a}=\boldsymbol{K}\cdot(\boldsymbol{r}_{2}-\boldsymbol{r}_{1})+\boldsymbol{P}\cdot\left(\boldsymbol{r}_{0}-\frac{Z_{1}\boldsymbol{r}_{1}+Z_{2}\boldsymbol{r}_{2}}{Z_{1}+Z_{2}}\right)+\boldsymbol{q}\cdot\left(Z_{0}\boldsymbol{r}_{0}+Z_{1}\boldsymbol{r}_{1}+Z_{2}\boldsymbol{r}_{2}\right).\label{eq:Phase1-1}
\end{equation}
The evaluation of $\mathcal{\psi}_{2}$ with \emph{Mathematica} in
helicity basis and subsequent Fourier transformation gives the contributions
proportional to $I_{(2)},\,I_{(2)}^{(\pm)},\hat{I}_{(2)}^{(\pm)},\,I_{(2)}^{(\pm,\pm)}$
in~(\ref{eq:PsiQQEtac}), and a similar evaluation of $\psi_{2i}$
yields contributions proportional to $J_{(2)}$ in~(\ref{eq:PsiQQEtac}),
respectively. 

Finally, we need to mention that the results for the amplitude of
the $\gamma\to c\bar{c}\gamma$ may be obtained in a similar way,
replacing the effective vertex~(\ref{eq:VEta}) with conventional
light-cone QED vertex from~\cite{Lepage:1980fj,Brodsky:1997de} and
adjusting $Z_{a}\to z_{a}$, $m_{\eta_{c}}^{2}\to-Q^{2}$, where $Q^{2}$
is the virtuality of the incoming photon ($Q=0$ for real photons
considered in this paper).

\end{document}